\documentclass[iop,numberedappendix,twocolappendix]{emulateapj}
\pdfoutput=1
\usepackage{amsmath,amssymb}
\usepackage{graphicx}
\usepackage{soul,xcolor}
\usepackage{tensor}

\newcommand{\fornax}{F{\textsc{ornax}}}
\setstcolor{red}

\begin{document}
\setlength{\abovedisplayskip}{7pt}
\setlength{\belowdisplayskip}{7pt}

\slugcomment{Received June 19, 2018; Accepted January 20, 2019}

\title{F{\sc{ornax}}: a Flexible Code for Multiphysics Astrophysical Simulations}

\author{M. Aaron Skinner\altaffilmark{1}}
\author{Joshua C. Dolence\altaffilmark{2}}
\author{Adam Burrows\altaffilmark{3}}
\author{David Radice\altaffilmark{3,4}}
\author{David Vartanyan\altaffilmark{3}}

\altaffiltext{1}{Livermore National Laboratory, 7000 East Ave., Livermore, CA 94550-9234; skinner15@llnl.gov}
\altaffiltext{2}{CCS-2, Los Alamos National Laboratory, P.O. Box 1663, Los Alamos, NM 87545; jdolence@lanl.gov}
\altaffiltext{3}{Department of Astrophysical Sciences, Princeton University, Princeton NJ 08544; burrows@astro.princeton.edu,dvartany@astro.princeton.edu}
\altaffiltext{4}{Schmidt Fellow, Institute for Advanced Study, 1 Einstein Drive, Princeton, NJ 08540; dradice@astro.princeton.edu}

\begin{abstract}
This paper describes the design and implementation of our new multi-group, multi-dimensional radiation hydrodynamics (RHD) code \fornax{} and provides a suite of code tests to validate its application in a wide range of physical regimes.  
Instead of focusing exclusively on tests of neutrino radiation hydrodynamics relevant to the 
core-collapse supernova problem for which \fornax{} is primarily intended, we present here classical 
and rigorous demonstrations of code performance relevant to a broad range of multi-dimensional 
hydrodynamic and multi-group radiation hydrodynamic problems. Our code solves the comoving-frame 
radiation moment equations using the $M1$ closure, utilizes conservative high-order reconstruction, 
employs semi-explicit matter and radiation transport via a high-order time stepping scheme, and is suitable 
for application to a wide range of astrophysical problems.
To this end, we first describe the philosophy, algorithms, and methodologies of \fornax{}
and then perform numerous stringent code tests, that collectively
and vigorously exercise the code, demonstrate the excellent numerical fidelity with 
which it captures the many physical effects of radiation hydrodynamics, 
and show excellent strong scaling well above 100k MPI tasks.

\end{abstract}


\section{Introduction}
\label{fornaxintro}

In recent years, there has been an explosion of techniques and codes to solve 
the equations of radiation hydrodynamics in astrophysical environments (Hubeny \& Burrows 2007; 
Krumholz et al. 2007; Swesty \& Myra 2009; Stone et al. 2008; Marek \& Janka 2009; Vaytet et al. 2011; 
M\"uller, Janka, \& Dimmelmeier 2010; Shibata et al. 2011; Kuroda, Kotake, \& Takiwaki 2012; 
Zhang et al. 2011,2013; Davis et al. 2012; Ott et al. 2012,2013; Kolb et al. 2013; Couch 2013; Couch \& O'Connor 2014; 
Teyssier 2015; Roberts et al. 2016; Bruenn et al. 2016; Nagakura et al. 2018; O'Connor \& Couch 2018).  
To address a broad class of radiation hydrodynamics problems, we have recently developed an entirely new multi-dimensional, multi-group
radiation hydrodynamic code, \fornax{}, primarily, but not exclusively, to study core-collapse
supernovae (Skinner, Burrows, \& Dolence 2016; Radice et al. 2017; Burrows et al. 2018; 
Vartanyan et al. 2018). However, to keep this paper manageable and focused, we
reserve for a later paper the explicit tests of the specifically neutrino radiation hydrodynamics
of relevance in the core-collapse supernovae problem and focus on tests relevant to 
generic radiation-hydrodynamic problems.

Most of \fornax{} is written in C, with only a few Fortran 95 routines 
for reading in microphysical data tables, and we use an MPI/OpenMP 
hybrid parallelism model now optimized for Cray and KNL architectures.
\fornax{} employs general orthogonal coordinates in one, two, and three spatial dimensions, solves
the comoving-frame, multi-group, two-moment, velocity-dependent transport equations to $\mathcal{O}(v/c)$,
and uses the $M1$ tensor closure for the second and third moments of the radiation
fields (Dubroca \& Feugeas 1999; Vaytet et al. 2011). \fornax{} can accommodate 
any orthogonal geometry, though for the core-collapse supernova (CCSN) problem we generally 
employ a spherical grid.  In this paper, we describe in detail the computational philosophy and methods of 
\fornax{} (\S\ref{general}), the formulation of the equations (\S\ref{sec:formulation}), the discretization approach (\S\ref{discrete}), 
specifics concerning reconstruction, solvers, and coupling (\S\ref{recon_method}), the algorithmic 
steps (\S\ref{steps}), the implementation details of the dendritic grid in 3-d spherical coordinates
(\S\ref{dendritic_grid}), the results of numerous standard hydrodynamic 
(\S\ref{hydro_tests}) and radiation (\S\ref{rad_tests}) test problems, and conclude 
with some general observations in \S\ref{conclusion}.

\section{General Description of Fornax}
\label{general}

The hydrodynamics in \fornax{} is based on a directionally-unsplit, Godunov-type finite-volume
method.  Fluxes at cell faces are computed with the fast and accurate HLLC approximate
Riemann solver based on left and right states reconstructed from the underlying volume-averages.  
The reconstruction is accomplished via a novel algorithm we developed specifically for \fornax{}
that uses moments of the coordinates within each cell and the volume-averaged variables to
reconstruct TVD-limited, high-order profiles.  The profiles always respect the cells' volume averages and,
in smooth parts of the solution away from extrema, yield third-order accurate states on the faces.  

The code is written in a generalized covariant/coordinate-independent fashion, 
and so can employ any coordinate mapping (see Appendix \S\ref{metric_tech}).  
This allows the use of an arbitrary orthogonal coordinate system and facilitates the artful distribution of zones in a given geometry.
To circumvent Courant limits due to converging angular 
zones when using spherical coordinates, the code can deresolve in both angles ($\theta$ 
and $\phi$) independently near the origin and polar axes as needed, conserving hydrodynamic and radiative fluxes in a
manner similar to the method employed in SMR (static-mesh-refinement) codes at refinement boundaries.
The use of such a ``dendritic grid" (discussed further in \S\ref{dendritic_grid}) allows 
us to avoid angular Courant limits near coordinate singularities, while maintaining accuracy and enabling one to
employ the useful spherical coordinate system natural for the supernova problem.
All components of the velocity vector are included in the hydrodynamics, enabling, e.g., the evolution of a rotating flow in 2-d axisymmetry. 

Importantly, the overheads for the calculations of various geometric quantities such as Christoffel symbols are minimal, 
since the code uses static mesh refinement, and hence the terms need
to be calculated only once (in the beginning).  Therefore, the overhead
associated with the covariant formulation is almost nonexistent.  In the context
of a multi-species, multi-group, radiation hydrodynamics calculation, the additional memory 
overhead is small (note that the radiation typicially requires hundreds of variables 
to be stored per zone).  The additional costs associated with occasionally
transforming between contravariant and covariant quantities and in the
evaluation of the geometric source terms are negligible, especially in the context of a
radiation hydrodynamics calculation.

The various radiation species (photons or neutrinos) are followed using
an explicit Godunov method applied to the radiation transport
operators, but an implicit method is used for the radiation source terms.
In this way, the radiative transport is handled locally, without the need for a global solution on the entire mesh.
This is also the recent approach taken by Just, Obergaulinger, \& Janka (2015), Roberts et al. (2016), and O'Connor \& Couch (2018),
though with some important differences. By addressing the transport operator with an explicit method,
we significantly reduce the computational complexity and communication overhead of traditional multi-dimensional
radiative transfer schemes by circumventing the need for global iterative solvers that have typically shown
to be slow and/or problematic beyond $\sim$10,000 cores. 
We have demonstrated excellent strong scaling in three dimensions well beyond 100,000 MPI tasks using \fornax{} on KNL and
Cray architectures (Appendix \S\ref{scaling}). 

The light-crossing time of a zone generally sets the
time step, but since the speed of light and the speed of sound in the inner core are
not far apart in the core-collapse problem after bounce, this numerical stability
constraint on the time step is similar to the CFL constraint of the explicit
hydrodynamics. For cases in which this near correspondence does not obtain,
we have implemented a reduced-speed-of-light approximation, {although we do not employ this approximation in our core-collapse simulations}.  Radiation quantities are reconstructed 
with linear profiles and an HLLE-like solver is used to determine numerical fluxes at zone boundaries.  These HLLE fluxes are corrected to
obtain the correct asymptotic behavior in the diffusion limit (Berthon, Charrier, \& Dubroca, 2007). The momentum and energy transfer
between the radiation and the gas are operator-split and addressed implicitly.

When using spherical coordinates in 2-d and 3-d, gravity is handled with a multipole solver (M\"uller \& Steinmetz 1995), where we
generally set the maximum spherical harmonic order necessary equal to twelve.
The monopole gravitational term is altered to approximate
general-relativistic (GR) gravity (Marek et al. 2006),
and we employ the metric terms, $g_{rr}$ and $g_{tt}$, derived from this potential
in the neutrino transport equations to incorporate general relativistic redshift
effects (in the manner of Rampp \& Janka 2002; see also Skinner et al. 2016).
For the CCSN problem, we have used for our recent 2-d and 3-d simulations the $K = 220\text{ MeV}$ equation of state (EOS) of
Lattimer \& Swesty (1991), the SFHo EOS (Steiner et al. 2013), and the DD2 EOS (Typel et al. 2010; Hempel \& Schaffner-Bielich 2010).

With gravity, energy conservation is excellent before and after core bounce (\S\ref{1d_conservation_hydro}).
However, as with all other supernova codes, at bounce the total energy as defined in integral form
experiences a glitch by $\ge 10^{49}$ erg,\footnote{In comparison, most
supernova codes experience a jump in the total energy at bounce in excess of 10$^{50}$ erg.} due to the fact that the gravitational terms are traditionally handled in the momentum and
energy equations as source terms, hence the equations are not solved in conservation form.

Though constructed originally for the CCSN problem,
\fornax{} is a flexible general radiation-hydrodynamics code 
that functions well in a variety of geometries and coordinate systems.
As already stated, in this paper we focus on this generic character as we demonstrate its strengths
and capabilities and do not here emphasize its implementation in its original CCSN
context.  This we leave to a future paper geared specifically to the use of \fornax{} in studying CCSNe. Again, the main advantages of the \fornax{} code are its
efficiency due to its use of explicit transport, its excellent strong scaling
to hundreds of thousands of cores (Appendix \S\ref{scaling}), its multi-dimensional transport capabilities,
and its interior static mesh derefinement near spherical coordinate singularities (\S\ref{dendritic_grid}).  

\section{Formulation of the Equations}
\label{sec:formulation}

In \fornax{}, we use a 
general covariant formalism based on an arbitrary orthogonal metric to write
our equations in a way that makes no reference to any particular geometry 
or set of coordinates.  Perhaps the biggest advantage of this approach is 
flexibility; as we discuss below, switching geometries and coordinates 
in \fornax{} is straightforward.

\subsection{Hydrodynamics}

As is standard practice, we denote contravariant components of a vector with 
raised indices, covariant components with lowered indices, covariant differentiation 
with a semicolon, partial differentiation with a comma, and make use of 
Einstein notation for summation over repeated indices.  Here and throughout 
this work, we adopt a coordinate basis.  In this notation, the equations of Newtonian 
hydrodynamics can be written
\begin{subequations}
\begin{align}
\rho_{,t} + (\rho v^i)_{;i}  &= 0 \label{eq:hydro_cont_start}\,, \\ 
(\rho v_j)_{,t} + (\rho v^i v_j + P \tensor{\delta}{^i_j})_{;i} &= S_j\,, \label{eq:mom_covderiv}\\
\left(\rho e\right)_{,t} + \left[\rho v^i \left(e + \frac{P}{\rho}\right)\right]_{;i} &= S_E\,, \\
(\rho X)_{,t} + (\rho X v^i)_{;i}  &= S_X\,, \label{eq:hydro_cont_stop}
\end{align}
\end{subequations}
where $e$ is the specific total energy of the gas, $X$ is an arbitrary 
scalar quantity that may represent, e.g., composition, and $S_j$, $S_E$, and $S_X$ 
are source terms that account for additional physics.  The contravariant components 
of the velocity are $v^i = dx^i/dt$, i.e., they are coordinate velocities.  In a 
coordinate basis, the covariant derivatives can be expanded to yield
\begin{subequations}
\begin{align}
\rho_{,t} + \frac{1}{\sqrt{g}}(\sqrt{g} \, \rho v^i)_{,i}  &= 0\,, \\
(\rho v_j)_{,t} + \frac{1}{\sqrt{g}}\left[\sqrt{g} \, \tensor{T}{^i_j} \right]_{,i} &= \Gamma^l_{jk} \tensor{T}{^k_l} + S_j\,, \label{eq:gas_mom}\\
\left(\rho e\right)_{,t} + \frac{1}{\sqrt{g}} \, \left[\sqrt{g} \rho v^i \left(e + \frac{P}{\rho}\right)\right]_{,i} &= S_E\,, \\
(\rho X)_{,t} + \frac{1}{\sqrt{g}} \Big( \sqrt{g} \, \rho X v^i \Big)_{,i} &= S_X\,, \label{eq:gas_comp}
\end{align}
\end{subequations}
where $g$ is the determinant of the metric, $\Gamma^l_{\;jk}$ are the 
Christoffel symbols, defined in terms of derivatives of the metric, 
and $\tensor{T}{^k_l} \equiv \rho v^i v_j + P\tensor{\delta}{^i_j}$ is the fluid stress tensor.

Note that we have chosen to express the momentum equation as a conservation 
law for the covariant compenents of the momentum.  There is good reason to 
do so.  Written this way, the geometric source terms, $\Gamma^l_{jk} \tensor{T}{^k_{l}}$, 
vanish identically for components associated with ignorable coordinates in 
the metric.  A good example is in spherical ($r$, $\theta$, $\phi$) coordinates 
where, since $\phi$ does not explicitly enter into the metric, the geometric 
source terms vanish for the $\rho v_\phi$ equation.  Physically, $\rho v_\phi$ is 
the angular momentum, so we are left with an explicit expression of angular 
momentum conservation that the numerics will satisfy to machine precision, 
rather than to the level of truncation error.  In general, the covariant 
expression of the momentum equation respects the geometry of the problem 
without special consideration or coordinate-specific modifications of the code.

\subsection{Radiation}

\fornax{} evolves the zeroth and first moments of the frequency-dependent comoving-frame 
radiation transport equation.  Keeping all terms to $\mathcal{O}(v/c)$ and dropping 
terms proportional to the fluid acceleration, the monochromatic radiation moment equations can be written
\begin{subequations}  \label{eq:rad_system}
\begin{align}
E_{\nu,t} + (F_{\nu}^i + v^i E_{\nu})_{;i} \qquad && \notag \\
+\, \tensor{v}{^i_{;j}} \left[ P_{\nu i}^j - \partial_\nu \left(\nu P_{\nu i}^j \right) \right] &= R_{\nu E}\,,\label{eq:rad_E} \\ 
F_{\nu j,t} + (c^2 P_{\nu j}^i + v^i F_{\nu j})_{;i} \qquad && \notag \\
+\, \tensor{v}{^i_{;j}} F_{\nu i}  - \tensor{v}{^i_{;k}} \,\partial_\nu \left( \nu Q^k_{\nu ji} \right) &= R_{\nu j}\,. \label{eq:rad_mom2}
\end{align}
\end{subequations}
In equations~\eqref{eq:rad_system}, $E_{\nu}$ and $F_{\nu j}$ denote the monochromatic energy density 
and flux of the radiation field at frequency $\nu$ in the comoving frame, 
$P^j_{\nu i}$ is the radiation pressure tensor ($2^{\rm nd}$ moment), 
$Q^k_{\nu j i}$ is the heat-flux tensor ($3^{\rm rd}$ moment), and 
$R_{\nu E}$ and $R_{\nu j}$ are source terms that 
account for interactions between the radiation and matter.  These interaction terms are written
\begin{subequations}
\begin{align}
R_{\nu E} &= j_\nu - c \kappa_\nu E_\nu\,, \\
R_{\nu j} &= -c(\kappa_\nu + \sigma_\nu) F_{\nu j}\,, \label{eq:rad_mom_src}
\end{align}
\end{subequations}
where $j_\nu$ is the emissivity, $\kappa_\nu$ is the absorption 
coefficient, and $\sigma_\nu$ is the scattering coefficient.  
Correspondingly, there are energy and momentum source terms in the fluid equations:
\begin{subequations}  \label{eq:radgas_source}
\begin{align}
S_j &= -\frac{1}{c^2} \int_0^\infty R_{\nu j} \,d\nu\,, \label{eq:radgas_E} \\ 
S_E &= -\int_0^\infty \left( R_{\nu E} + \frac{v^i}{c^2} R_{\nu i} \right) d\nu\,. \label{eq:radgas_mom}
\end{align}
\end{subequations}
As in the fluid sector, we rewrite these covariant derivatives as partial derivatives, 
introducing geometric source terms in the radiation momentum equation.

\fornax{} can treat either photon or neutrino radiation fields.  For neutrinos, 
Equations~\eqref{eq:rad_system} are solved separately for each 
species, and Equations~\eqref{eq:radgas_source} are summed over species.  
Additionlly, the electron fraction is evolved according to Equation~\ref{eq:gas_comp}, with $X=Y_e$ and
\begin{equation}
S_X = \sum_s \int_0^\infty \xi_{s\nu} (j_{s\nu} - c \kappa_{s\nu} E_{s\nu}) \,d\nu \,,
\end{equation}
where $s$ refers to the neutrino species and
\begin{equation}
\xi_{s\nu} = \begin{cases}
	-(N_A \nu)^{-1},&	\text{$s=\nu_e$}\\
	(N_A \nu)^{-1},&  \text{$s=\bar{\nu}_e$}\\
	0,&							  \text{$s=\nu_x$}
\end{cases}\,,
\end{equation}
where $N_A$ is Avogadro's number.

\section{Numerical Discretization}
\label{discrete}

To maintain conservation while admitting discontinuous shock solutions, we 
adopt a finite-volume discretization of the equations presented in 
Section~\ref{sec:formulation}.  Independent of geometry or coordinates ($x$), 
the volume element can always be expressed $dV=\sqrt{g} \,d^3x$.  Similarly, 
for arbitrary orthogonal coordinates the area element is $dA=\sqrt{g} \,d^2x$.  
Averaging the equations over control volumes (cells) leads to a set of exact 
equations describing the evolution of cell volume-averaged quantities.

Applying the divergence theorem, we express the divergence terms in each 
equation as a net flux through cell faces.  For example, the volume-averaged density in cell $(i,j,k)$ evolves according to
\begin{equation}
\begin{split}
\frac{d \rho_{i,j,k}}{dt} = -&\frac{1}{V_{i,j,k}} \\
\times[&(\rho v^0 A_0)_{i+1/2,j,k} - (\rho v^0 A_0)_{i-1/2,j,k} \\
+ &(\rho v^1 A_1)_{i,j+1/2,k} - (\rho v^1 A_1)_{i,j-1/2,k} \\
+ &(\rho v^2 A_2)_{i,j,k+1/2} - (\rho v^2 A_2)_{i,j,k-1/2}]\,.
\end{split} \label{eq:finite_vol}
\end{equation}
Two things are of note.  First, by integrating over cell volumes, we have 
transformed our set of partial differential equations into a large set of 
coupled ordinary differential equations.  Second, Equation~\eqref{eq:finite_vol} is exact, provided the fluxes on each face are taken to represent cell face area-averaged quantities.

Source terms appear in nearly all the evolution equations and, for consistency, 
must also be volume-averaged.  If these source terms were linear in the conserved 
variables, this volume averaging would be trivial.  Unfortunately, all the 
source terms are nonlinear, which requires that we make choices for how the 
volume averaging is carried out.  For example, the geometric source terms 
in Equation~\ref{eq:gas_mom} are typically treated as
\begin{equation}
\frac{1}{V} \int \sqrt{g}\ \Gamma^l_{jk} \tensor{T}{^k_l} \,d^3x \approx \langle \Gamma^l_{jk}\rangle \tensor{T}{^k_l}(\langle\mathbf{U}\rangle) \,,
\end{equation}
where angle brackets indicate a volume average over a given cell and $\tensor{T}{^k_l}(\mathbf{U})$ is 
stress tensor computed from the vector of volume-averaged conserved variables.  
It is straightforward to show that the error in this approximation is 
$\mathcal{O}(\Delta x^2)$, but this is not a unique second-order accurate 
expression.  Adopting alternative expressions for some geometric source terms 
can have desirable properties, such as exact conservation of angular momentum (see Appendix \S\ref{metric_tech}).

With space fully discretized, we now turn to the issue of temporal evolution.  Each of our equations can be written in the form
\begin{equation}
\frac{\partial Q}{\partial t} + (\mathcal{F}^i_Q)_{;i} = S_{\rm non-stiff} + S_{\rm stiff}\,,
\end{equation}
where $Q$ is some volume-averaged quantity evolved on the mesh, $\mathcal{F}^i_Q$ is the area-averaged flux of that quantity, 
and the volume-averaged source terms have been grouped together according to whether or not they are stiff.  
The only terms currently treated by \fornax{} that are stiff are the interaction 
terms that couple radiation and matter and the terms that handle reactions (nuclear or chemical).  
These stiff terms require an implicit treatment for numerical stability since 
the characteristic time scales for these interactions may be extremely short compared 
to the explicit Courant time step for the matter or radiation. All other terms are treated 
explicitly and are evolved together without operator splitting.  The explicit time integration 
is currently carried out using Shu \& Osher's (1988) optimal second-order TVD Runge-Kutta scheme.  

The frequency dependence of the radiation moments is represented in discrete
groups linearly or logarithmically spaced between a user-specified minimum and maximum frequency.\footnote{Alternatively, the user may specify the frequency range using equivalent energies in MeV, a standard unit for neutrino radiation.}  The number
of groups per species $N_{sg}$ is also set by the user and can be different
for each species.  Unless otherwise indicated, the radiation moments and interaction 
coefficients are integrated over each frequency group, appearing with a subscript $g$.  For example,
\begin{equation}
E_{sg} \equiv \int_{\nu_{g-1/2}}^{\nu_{g+1/2}} E_{s\nu} \,d\nu
\end{equation}
is the energy density of species $s$ in group $g$ which has units of energy density.  In integrating over frequency groups, the monochromatic radiation moment Equations~\eqref{eq:rad_system} become the group-integrated moment equations, given for each group $g$ (and for each species $s$) by
\begin{subequations}  \label{eq:rad_system2}
\begin{align}
E_{g,t} + (F_{g}^i + v^i E_{g})_{;i} \qquad && \notag \\
+\, \tensor{v}{^i_{;j}} \left[ P_{g i}^j - \int_{\nu_{g-1/2}}^{\nu_{g+1/2}} \partial_\nu \left(\nu P_{\nu i}^j \right)\,d\nu \right] &= R_{g E}\,, \\ 
F_{g j,t} + (c^2 P_{g j}^i + v^i F_{g j})_{;i} \qquad && \notag \\
+\, \tensor{v}{^i_{;j}} F_{g i}  - \tensor{v}{^i_{;k}} \int_{\nu_{g-1/2}}^{\nu_{g+1/2}} \partial_\nu \left( \nu Q^k_{\nu ji} \right)\,d\nu &= R_{g j}\,.
\end{align}
\end{subequations}

\section{Reconstruction, Solvers, and Interaction Terms}
\label{recon_method}

\subsection{Reconstruction}
\label{recon}

The hydrodynamics in \fornax{} is evolved using a directionally-unsplit, high-order
Godunov-type scheme.  Having already described the underlying spatial discretization
of the variables as well as the time-stepping scheme, the essential remaining
element is a method for computing fluxes on cell faces. Since \fornax{}
employs Runge-Kutta time stepping, there is no need for the characteristic
tracing step or transverse flux gradient corrections required in
single-step unsplit schemes.  This approach has the great advantage
of relative simplicity, especially in the multi-physics context.  \fornax{}
follows the standard approach used by similar codes, first reconstructing
the state profile at cell faces and then solving the resulting Riemann problem to obtain fluxes.

There are many potential approaches to reconstructing face data based on cell values.  
One particularly popular approach is to reconstruct the primitive variables 
($\rho$, $P$, and $v^i$) using the method described by Woodward \& Colella (1984), the so-called 
piecewise-parabolic method (PPM).  This method is based on the idea of reconstructing 
profiles of the volume-averaged data, with curvilinear coordinates incorporated by 
reconstructing the profiles in the volume coordinate rather than in the original coordinates themselves.  
This approach can be problematic in the vicinity of coordinate singularities.  
Blondin \& Lufkin (1993) suggest a generalized approach 
to PPM in cylindrical and spherical coordinates.  Unfortunately, it can be 
shown that their approach cannot be used generically and, 
in fact, fails even for standard spherical coordinates.  Here, we present an 
alternative approach that works in arbitrary geometries and coordinates, which 
contains PPM as a special case.

We begin by writing down an expression for a general $n^{\rm th}$-order polynomial:
\begin{equation}
p(x) = \sum_{j=0}^n c_j x^j\,,
\end{equation}
where $c_j$ is the coefficient of the $j^{\rm th}$-order monomial.  
As in PPM, we wish to construct a polynomial $p(x)$ consistent with $p_i$, 
the volume average of quantity $p$ in \mbox{cell $i$}.  In our 
formulation, the volume average of $p(x)$ must then satisfy a set of constraint equations of the form
\begin{align}
p_i &= \langle p \rangle_i \notag \\
    &= \frac{1}{V} \int_{x_{i-1/2}}^{x_{i+1/2}} p(x) \sqrt{g} \,dx \notag \\
    &= \sum_{j=0}^n c_j \langle x^j \rangle_i\,,  \label{eq:recon_constraints}
\end{align}
where $x_{i\pm 1/2} \equiv x_i \pm \Delta x/2$ and the quantities $\langle x^j \rangle_i$ can be precomputed for each cell either directly or via Romberg integration then stored in one-dimensional arrays.
For the case $n=0$, i.e., for piecewise-constant reconstruction, 
this yields the trivial result that $p(x)=p_i$.  For $n>0$, we must further constrain 
the reconstruction.  To proceed, we first distinguish between cases where $n$ is 
even and $n$ is odd.  For odd $n$, we form a symmetric stencil around each 
edge containing data in $n+1$ cells and require that $\langle p \rangle_i = p_i$ in each cell $i$.  This necessarily yields continuous left and right states at cell interfaces, but that continuity may be broken by the subsequent application of a slope limiter.  For even $n$, we similarly constrain 
the polynomial within a 
symmetric stencil of $n+1$ cells, this time centered on a given cell, yielding potentially
discontinuous data at cell faces even before the application of a slope limiter.

In \fornax{}, since we most often employ third-order piecewise-parabolic reconstruction ($n=2$), we 
now describe it in more detail.  The reconstruction of $p(x)$ in cell $i$ depends 
on the data values $p_{i-1}$, $p_i$, and $p_{i+1}$.  Note that this requires the same number 
of ghost cells as linear reconstruction, is of similar cost, and being of higher order, produces 
superior results for many problems, as we show below and in \S\ref{hydro_tests}.  

To begin, if $p_i$ is a local extremum of the data, then the reconstruction in cell $i$ is taken to be piecewise-constant.  Otherwise, three constraints in the form of Equation~\eqref{eq:recon_constraints} are used to form the linear system
\begin{equation}
\left[\begin{array}{lll}
1 & \langle x \rangle_{i-1} & \langle x^2 \rangle_{i-1} \\
1 & \langle x \rangle_{i  } & \langle x^2 \rangle_{i  } \\
1 & \langle x \rangle_{i+1} & \langle x^2 \rangle_{i+1}
\end{array}\right]
\left[\begin{array}{l}
c_0 \\
c_1 \\
c_2
\end{array}\right]
= \left[\begin{array}{l}
p_{i-1} \\
p_{i  } \\
p_{i+1}
\end{array}\right],  \label{eq:recon_system1}
\end{equation}
which is then solved for the unique interpolation coefficients $c_0$, $c_1$, and $c_2$.  The resulting polynomial is then used to determine $p_{R,i-1/2} \equiv p(x_{i-1/2})$, the right state at the lower face of cell $i$, and $p_{L,i+1/2} \equiv p(x_{i+1/2})$, the left state at the upper face of cell $i$.  If $p_{R,i-1/2}$ is sufficiently close to either $p_{i-1}$ or $p_i$ such that $p(x)$ has an extremum for $x \in (x_{i-1},x_{i})$, then $p_{R,i-1/2}$ is reset to ensure that $p(x)$ is monotone on this interval and has zero slope at $x_{i-1}$.  The value of $p_{L,i+1/2}$ is reset in an analogous manner as needed.  Next, $p(x)$ is reconstructed once more according to the solution of the linear system
\begin{equation}
\left[\begin{array}{lll}
1 & x_{i-1/2} & x^2_{i-1/2} \\
1 & \langle x \rangle_{i  } & \langle x^2 \rangle_{i  } \\
1 & x_{i+1/2} & x^2_{i+1/2}
\end{array}\right]
\left[\begin{array}{l}
c_0 \\
c_1 \\
c_2
\end{array}\right]
= \left[\begin{array}{l}
p_{R,i-1/2} \\
p_{i  } \\
p_{L,i+1/2}
\end{array}\right].  \label{eq:recon_system2}
\end{equation}
If neither state has been reset by prior monotonicity constraints, then the solutions of Equations~\eqref{eq:recon_system1} and~\eqref{eq:recon_system2} are identical.  Finally, if $p(x)$ has an extremum in cell $i$, then either $p_{R,i-1/2}$ or $p_{L,i+1/2}$---whichever state is nearest to the extremum---is reset to ensure that $p(x)$ is monotone in cell $i$ with zero slope at that face.  

For smooth data away from extrema, this process yields a third-order-accurate piecewise-parabolic reconstruction at the lower and upper faces of each cell.  Once the reconstruction step is completed in all cells and for all variables, the resulting left and right states at each face, which are derived from distinct interpolation parabolae, are then passed to a Riemann solver to obtain the numerical flux at that face as described in the next section.

\subsection{Riemann Solvers}
\label{riemann}

\fornax{} currently implements three choices for
computing fluxes as a function of the reconstructed left and right states:
the local Lax-Friedrichs, HLLE, and HLLC solvers.  The HLLC solver, incorporating
the most complete information on the modal structure of the equations, is
the least diffusive option and is the default choice.  Unfortunately, so-called
three-wave solvers are susceptible to the carbuncle or odd-even instability.
Many schemes have been proposed to inhibit the development of this numerical
 instability.  In \fornax{}, we tag interfaces determined to be within
shocks (by the pressure jump across the interface), and switch to the HLLE
solver in orthogonal directions for cells adjacent to the tagged interface.
In all our tests, this very simple approach has been successful at preventing
the instability, incurs essentially no cost, and is typically used so
sparingly that the additional diffusion introduced into the solution is likely negligible.

To $\mathcal{O}(\Delta x^2)$, we compute face-averaged fluxes between cells using
approximate Riemann solvers applied to reconstructed states at the area-centroid
of each face.  For the hydrodynamic fluxes, we use the HLLC solver of 
Batten et al. (1997) and Toro, Spruce, \& Speares (1994).  For the fluxes of the 
radiation moments (specifically the $F^i$ and $c^2 \tensor{P}{^i_j}$ terms),
we use the HLL solver as in Harten, Lax, \& van Leer (1983).  As has been pointed out by multiple authors,
the HLL solver applied to our radiation subsystem yields fluxes that fail to
preserve the asymptotic diffusion limit.  To recover this limit, we have
found it sufficient to introduce a correction to the energy fluxes of the form
\begin{equation}
\mathcal{F}^\mathrm{HLL}_{E,\mathrm{corrected}} = \frac{S_R F_L - S_L F_R + \epsilon S_R S_L (E_R - E_L)}{S_R - S_L}\,,  \label{eq:hll}
\end{equation}
with
\begin{equation}
\epsilon \equiv \min \left(1,\frac{1}{\tau_\mathrm{cell}}\right)\,.  \label{eq:epsilon}
\end{equation}
In Equation~\eqref{eq:epsilon}, $\tau_\mathrm{cell}$ is an approximation to the optical depth of a cell in the direction
normal to the face, computed using a simple arithmetic average of the total opacity
(absorption plus scattering) of the cells on either side of the face.  In Equation~\eqref{eq:hll}, $S_L$ and $S_R$ are wavespeed estimates for the fastest left-
and right-going waves, $F_L$ and $F_R$ are the normal components of the reconstructed radiation fluxes on the left and right side of the face, and $E_L$ and $E_R$ are the reconstructed
radiation energy densities on the left and right side of the face.  Note that when
$\tau_\mathrm{cell} \le 1$, the corrected numerical fluxes are identical to the standard HLL fluxes.  Note further
that from here on, we drop indices referring to radiation species
(photon or neutrino) and frequency group for clarity, though each of these
radiation terms are computed per species and per group.

Since we evolve the comoving-frame radiation moments, the intercell fluxes include
advective terms that depend on the fluid velocity component normal to the
faces.  For consistency and accuracy, we make use of the contact
(i.e., particle) velocity ($v^*$) computed during the construction
of the hydrodynamic fluxes in the HLLC Riemann solver.  At a given cell face, these advective
fluxes are then set to
\begin{subequations}
\begin{align}
(v E) &= \begin{cases}
					v^* E_{R}, & v^* < 0 \\
					v^* E_{L}, & v^* \ge 0
					\end{cases}\,, \\
(v \mathbf{F}) &= \begin{cases}
									v^* \mathbf{F}_{R}, & v^* < 0 \\
									v^* \mathbf{F}_{L}, & v^* \ge 0
									\end{cases}\,,
\end{align}
\end{subequations}
where the subscripts $L$ and $R$ indicate quantities reconstructed
to the left or right side of the cell face, respectively.
In other words, we use the reconstructed quantities in the upwind
direction, defined with respect to $v^*$, the particle velocity at a given cell face
computed by the Riemann solver for the hydrodynamics.

Several terms in the radiation moment equations depend upon
the gradient of the velocity field.  Ultimately, volume averages of each of
these velocity-gradient dependent terms are needed.  These are approximated, as in the calculation of the geometric source terms, by combining separately volume-averaged components as
\begin{equation}
\langle \tensor{v}{^i_{;j}}\rangle \approx \langle \tensor{v}{^i_{,j}} \rangle + \langle \Gamma^i_{j k} \rangle \langle v^k \rangle\, .
\end{equation}
Like before with the advective fluxes, we again make use of the particle velocities
computed during the Riemann solve for the hydrodynamic fluxes.  In
addition to the normal components, we also need the transverse
components; these are available in the construction
of the hydrodynamic fluxes as well.  Thus, given the velocity components on every face of a
cell, we assume a linear profile (in our generic coordinates $x^i$) for the
intracell velocities and compute the volume-averaged partial derivatives directly
via finite differences across the cell.  Finally, to construct the
volume-averaged velocities $\langle v^k \rangle$, we compute the volume-average
of the linear profile in each direction $k$.

For the terms involving frequency-space derivatives, we use an approach similar
to Vaytet et al. (2011), considering our multi-group treatment as a generalized
finite-volume discretization of frequency space.  The evolution equations for these
terms can be written as
\begin{subequations}  \label{eq:freq_adv_system}
\begin{align}
E_{g,t} - \tensor{v}{^i_{;j}} \left[ (\nu P^j_{\nu i})_{g+1/2} - (\nu P^j_{\nu i})_{g-1/2}\right] &= 0\,, \\ 
F_{g j,t} - \tensor{v}{^i_{;k}} \left[ (\nu Q^k_{\nu ji})_{g+1/2} - (\nu Q^k_{\nu ji})_{g-1/2}\right] &= 0\,.
\end{align}
\end{subequations}
As in Vaytet et al. (2011),
we adopt a simple upwind formulation of these terms, defining the intergroup
flux at a given group boundary $\nu_{g-1/2}$ as
\begin{subequations}
\begin{align}
(\nu P^j_{\nu i})_{g-1/2} &= \begin{cases}
                        \nu_{g-1/2} \,\langle P^j_{\nu i} \rangle_{L}, & \tensor{v}{^i_{;j}} < 0 \\
                        \nu_{g-1/2} \,\langle P^j_{\nu i} \rangle_{R}, & \tensor{v}{^i_{;j}} \ge 0
                     \end{cases}\,, \\
(\nu Q^k_{\nu ji})_{g-1/2} &= \begin{cases}
                            \nu_{g-1/2} \,\langle Q^k_{\nu ji} \rangle_{L}, & \tensor{v}{^i_{;k}} < 0 \\
                            \nu_{g-1/2} \,\langle Q^k_{\nu ji} \rangle_{R}, & \tensor{v}{^i_{;k}} \ge 0
                        \end{cases}\,,
\end{align}
\end{subequations}
where for a given frequency group $g$, $\langle P^j_{\nu i} \rangle_{g} = P^j_{gi}/\Delta \nu_g$ is the 
group-averaged spectral density, $\Delta \nu_g$ is the group width, and the subscripts $L$ and $R$ 
indicate the reconstruction of the spectral densities to the left or right side of the group boundary, respectively.

\subsection{Interaction Terms}
\label{interaction}

The interaction terms that couple the radiation and matter can introduce significant
stiffness into the system, requiring an implicit treatment to provide stability to their
numerical evolution on hydrodynamic time scales.
In \fornax{}, we use operator splitting to advance the system
\begin{subequations}
\begin{align}
(\rho v_j)_{,t} &= S_j\,,  \label{eq:Sj} \\
(\rho e)_{,t} &= S_E\,,  \label{eq:SE} \\
(\rho X)_{,t} &= S_X\,,  \label{eq:SX} \\
E_{sg,t} &= R_{sgE}\,,  \label{eq:RsgE} \\
F_{sgj,t} &= R_{sgj}\,,  \label{eq:Rsgj}
\end{align}
\end{subequations}
after the explicit transport update described above is applied.  It is essential
to treat these terms after the explicit update to capture stiff equilibria properly, 
as obtain in optically-thick environments. 

Equations~\eqref{eq:SE},~\eqref{eq:SX}, and~\eqref{eq:RsgE} treat the terms describing the emission and absorption of radiation and their coupling to the material energy and composition.  The essential task is to solve the system
\begin{subequations}  \label{eq:energy_source_system}
\begin{align}
\frac{u^{n+1} - u^{-}}{\Delta t} &= -\sum_s \sum_g (j_{sg}^{n+1} - c \kappa_{sg}^{n+1} E^{n+1}_{sg})\,, \\
\frac{X^{n+1} - X^{-}}{\Delta t} &= \sum_s \sum_g \xi_{sg} (j_{sg}^{n+1} - c \kappa_{sg}^{n+1} E_{sg}^{n+1})\,, \\
\frac{E^{n+1}_{sg} - E^{-}_{sg}}{\Delta t} &= j_{sg}^{n+1} - c \kappa_{sg}^{n+1} E^{n+1}_{sg}\,,
\end{align}
\end{subequations}
where $u$ denotes the material internal energy density and the $-$ superscript denotes the state just after the explicit update described above.
Although Equations~\eqref{eq:energy_source_system} are spatially decoupled, i.e., their solution depends entirely on local data, if solved directly, they would represent a large system of stiff nonlinear equations, especially in the multi-species, multi-group context.  Fortunately, the solution of this system can be made simpler by separating the updates into ``inner'' and ``outer'' parts.  In the inner update, the frequency groups are decoupled from one another and can be updated using a direct implicit scheme.  Then, in the outer update, all that remains is to find the roots of just a few equations, independent of the number of groups or species.  This can be accomplished using an iterative scheme.  In the photon radiation context, there is just a single equation, and in the core-collapse supernova context, there are two.  Without loss of generality, we continue our description for the core-collapse case, where $X$ represents the electron fraction, $Y_e$, and the opacities and emissivities depend on both $T$ and $Y_e$.

In the inner update, each individual frequency group is updated using a fully-implicit Backward-Euler scheme of the form
\begin{equation}
E^{k}_{sg} = \frac{E^{-}_{sg} + \Delta t \,j^k_{sg}}{1 + c \,\Delta t\, \kappa^k_{sg}}\,,  \label{eq:be_update}
\end{equation}
where the opacities and emissivities are computed using the values of $T^k$ and $Y_e^k$ at outer iteration $k$.  Equation~\eqref{eq:be_update} has the important property that for \mbox{$c\,\Delta t \,\kappa^k_{sg} \gg 1$}, i.e., for large optical depths, the updated energy, $E^{k}_{sg}$, approaches the thermal equilibrium solution, $j^k_{sg}/(c \,\kappa^k_{sg})$.

In the outer update, the hydrodynamic quantities $u$ and $\rho Y_e$ are updated implicitly according to the sum over groups (and of species) of the emission and absorption terms.  The integrated changes in energy and composition are given by
\begin{subequations}
\begin{align}
\Delta E^k &= \sum_s \sum_g (E_{sg}^k - E_{sg}^{-})\,, \\
\Delta X^k &= \sum_s \sum_g \xi_{sg} (E_{sg}^k - E_{sg}^{-})\,.
\end{align}
\end{subequations}
Finally, we compute the residuals
\begin{subequations}
\begin{align}
r^k_E &= u^k - u^{-} + \Delta E^k \,, \\
r^k_X &= \rho (Y_e^k - Y_e^{-}) - \Delta X^k \,,
\end{align}
\end{subequations}
where, like the opacities and emissivities, the internal energy density $u^k$ depends on both
$T^k$ and $Y_e^k$.  The solution to our implicit system is given by the values of $T$ and $Y_e$
for which $r^k_E = r^k_X = 0$.\footnote{More accurately, the numerical solution to our implicit system of two equations consists of the vector $[T, Y_e]^T$ that minimizes the norm of the relative residual vector $[r_E/u^-,r_X/(\rho Y_e^-)]^T$.}  To find these roots, we use a Newton-Raphson iteration, using finite differences to form the Jacobian
and employing a {backtracking} line search algorithm to improve robustness.\footnote{Occasionally, this procedure may fail to converge, in which case we resort to bisection.  For photon radiation, this is typically guaranteed to converge provided the matter temperature, $T$, can be bracketed.  For neutrino radiation, we alternately bisect in $T$ and $Y_e$ separately, holding the other variable fixed and, upon convergence, retrying the line search.  In the core-collapse context, we find this procedure to be quite robust.}  At each iteration,
a linear system is solved for $\delta T^{k+1} \equiv T^{k+1} - T^k$ and $\delta Ye \equiv Y_e^{k+1} - Y_e^k$, the changes in temperature and electron fraction, respectively,
between iterations $k$ and $k+1$.  It is in solving this linear system
that our reduction to two equations has obvious virtues.  Convergence
is checked by requiring that the relative change of the temperature and
composition between iterations is below a user-specfied tolerance,
typically $10^{-6}$.  In practice, this procedure yields converged solutions
within a few iterations under a wide range of conditions.

Finally, Equations~\mbox{\eqref{eq:Sj}, \eqref{eq:SE}, and \eqref{eq:Rsgj}} treat the momentum coupling
represented by the absorption and scattering of radiative flux.  For each frequency group, we first
update the $j$th component of the flux via the implicit Backward Euler scheme
\begin{equation}
F^{n+1}_{sgj} = \frac{F^{-}_{sgj}}{1 + c \,\Delta t (\kappa^{n+1}_{sg} + \sigma^{n+1}_{sg})} \,,
\end{equation}
which is crucial for maintaining the correct asymptotic behavior in the diffusion regime.
To see this, consider the relevant case of a near-equilibrium, static atmosphere with large cell optical depths,
as is approximately the case inside the proto-neutron star in the core-collapse
problem.  Under these conditions, $\|F\|/(cE) \ll 1$, so that the pressure tensor
under the $M1$ closure approaches the Eddington closure, $\tensor{P}{^i_j} = (E/3) \tensor{\delta}{^i_j}$.
The flux after the explicit transport update described above is then given by
\begin{equation}
F^{-}_{sgj} = F^{n}_{sgj} - \Delta t \frac{c^2}{3} E^{n}_{sg,j} \approx - \Delta t \frac{c^2}{3} E^{n}_{sg,j} \,,
\end{equation}
where the last approximate equality holds in a near-equilibrium, optically-thick atmosphere in which $F \sim cE/\tau \ll c^2 E_{,i}$.
In this case, our implicit update yields the updated flux
\begin{align}
F^{n+1}_{sgj} &= \frac{F^{-}_{sgj}}{1 + c \Delta t (\kappa^{n+1}_{sg} + \sigma^{n+1}_{sg})}\,, \notag \\
              &\approx - \left[ \frac{c}{3 (\kappa^{n+1}_{sg} + \sigma^{n+1}_{sg})} \right] E^{n}_{sg,j}  \,,
\end{align}
which is a temporally first-order accurate expression of Fick's law of diffusion.  In
other words, our operator-split implicit update of the flux reduces to Fick's law
in the asymptotic limit of high optical depth.  The corresponding explicit material momentum and energy updates \st{then follow from conservation, namely} {are then given by}
\begin{align}
(\rho v_j)^{n+1} &= (\rho v_j)^{-} - \frac{1}{c^2} \sum_s \sum_g \left( F^{n+1}_{sgj} - F^{-}_{sgj} \right) \,, \label{eq:radforce} \\
(\rho e)^{n+1} &= (\rho e)^{-} - \frac{1}{c^2} \sum_s \sum_g (v^j)^{-} \left(F^{n+1}_{sgj} - F^{-}_{sgj} \right) \,. \label{eq:vdotradforce}
\end{align}
{Note that equations~\eqref{eq:radforce} and~\eqref{eq:vdotradforce} are not momentum- and energy-conservative, respectively, because the equations themselves are not momentum- and energy-conservative as written in the comoving frame.  In section~\ref{multid_conservation_total}, we demonstrate how well \fornax{} conserves total energy in a realistic core-collapse problem for which the code was primarily intended.}
This completes the description of our treatment of the interaction terms.

\section{Algorithmic Steps}
\label{steps}

The following describes an Euler push sequence (i.e., first-order in time)
for the multi-species equations in the core-collapse context.  The generalization to a second-order Runge-Kutta scheme is straightforward.

\subsection{Step 1}
Advance the explicit transport subsystem (including advective flux terms and geometrical and gravitational source terms) by $\Delta t$:
\begin{subequations}
\begin{align}
\rho_{,t} + (\rho v^i)_{;i}  &= 0\,, \\
(\rho v_j)_{,t} + (\rho v^i v_j + P \tensor{\delta}{^i_j})_{;i} &= -\rho \phi_{,j}\,, \\
(\rho e)_{,t} + \left[\rho v^i \left(e + \frac{P}{\rho}\right)\right]_{;i} &= -\rho v^i \phi_{,i}\,, \\
(\rho Y_e)_{,t} + (\rho Y_e v^i)_{;i} &= 0\,, \\
E_{s\nu,t} + (F_{s\nu}^i + v^i E_{s\nu})_{;i} &= 0\,, \\
F_{s\nu j,t} + (c^2 P_{s\nu j}^i + v^i F_{s\nu j})_{;i} &\notag \\
+ v^i_{;j} F_{s\nu i} - v^i_{;k} Q^k_{s\nu ji} &= 0\,.
\end{align}
\end{subequations}

\subsection{Step 2}
Advance the frequency advection subsystem by $\Delta t$:
\begin{subequations}
\begin{align}
E_{s\nu,t} + v^i_{;j} \left[ P_{s\nu i}^j - \partial_\nu \left( \nu\,P_{s\nu i}^j \right) \right] &= 0\,, \\
F_{s\nu j,t} - v^i_{;k} \,\partial_\nu \left(\nu \,Q^k_{s\nu ji} \right) &= 0\,. \\
\end{align}
\end{subequations}

\subsection{Step 3}
Advance the radiation-matter energy interaction subsystem by $\Delta t$:
\begin{subequations}
\begin{align}
(\rho e)_{,t} &= -\sum_s \int_0^\infty \left(j_{s\nu} - c \kappa_{s\nu} E_{s\nu}\right) \,d\nu \,, \\
(\rho Y_e)_{,t} &= \sum_s \int_0^\infty \xi_{s\nu} (j_{s\nu} - c \kappa_{s\nu} E_{s\nu}) \,d\nu \,, \\
E_{s\varepsilon,t} &= j_{s\nu} - c \kappa_{s\nu} E_{s\nu} \,.
\end{align}
\end{subequations}

\subsection{Step 4}
Advance the radiation-matter momentum interaction subsystem by $\Delta t$:
\begin{subequations}
\begin{align}
(\rho v_j)_{,t} &= \frac{1}{c} \sum_s \int_0^\infty (\kappa_{s\nu} + \sigma_{s\nu}) F_{s\nu j} \,d\nu \,, \\
(\rho e)_{,t} &= \frac{v^i }{c} \sum_s \int_0^\infty (\kappa_{s\nu} + \sigma_{s\nu}) F_{s\nu i} \,d\nu \,, \\
F_{s\nu j,t} &= -c(\kappa_{s\nu} + \sigma_{s\nu}) F_{s\nu j} \,.
\end{align}
\end{subequations}

\begin{figure} [t]
\includegraphics[width=0.50\textwidth]{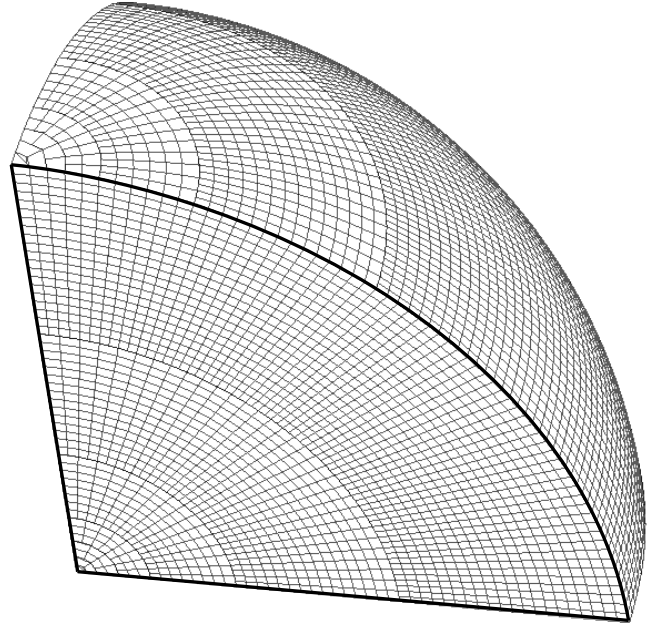}
\caption{\footnotesize
An example of the sort of dendritic grid one might employ for 3-d simulations.  Note that as both
the interior and the vertical axis are approached, the angular sizes increase independently.  This allows
us to maintain respectable resolution, while not being burdened by a severe Courant limit.
}
\label{dendritic}
\end{figure}

\section{Implementation of the Dendritic Grid in 3-d}
\label{dendritic_grid}

\subsection{Motivation}

The purpose of the spherical dendritic grid is to avoid the narrowing of zones 
in polar and azimuthal angular extent approaching the origin and the narrowing 
of zones in the azimuthal extent approaching the polar axes that is characteristic 
of the spherical polar coordinate mapping of a logically-Cartesian grid.  For 
example, a unigrid in ordinary spherical polar coordinates would have zones 
of size $\Delta r \times r\,\Delta \theta \times r\,\sin\theta\,\Delta\phi$.  
Even at moderate angular resolution, this narrowing of zones can place a severe 
constraint on the CFL time step.  One approach to circumvent this constraint is 
to coarsen the angular resolution as needed in order to maintain a roughly constant 
zone aspect ratio.  Thus, we coarsen the polar angle in order to keep $\Delta \theta \sim \Delta r/r$ as $r \to 0$, 
and similarly, we coarsen the azimuthal angle in order to keep $\Delta\phi \sim \Delta r/r\,\sin\theta$ as $\theta \to 0$ or $\theta \to \pi$. As a result, all zones on the grid have roughly equal aspect ratios, and near the 
origin where $\Delta r$ is approximately constant, all zones have roughly the same 
size.  This implies that the CFL time step is essentially constrained only by the 
minimum radial resolution, $\Delta r_\mathrm{min}$, i.e., the time step is not 
constrained by the resolution in either angular direction.
A sample arrangement of the dendritic grid in 3-d is given in Figure~\ref{dendritic}.

\subsection{Static Coarsening}

In \fornax{}, angular resolution is always coarsened by a factor of 2.  Thus, 
wherever the zone aspect ratio would otherwise deviate too far from 1, either $\Delta\theta$ or $\Delta\phi$ 
is coarsened as needed,\footnote{We do not allow both $\Delta\theta$ \emph{and} $\Delta\phi$ 
to be coarsened at the same radius, since that would represent 4-to-1 coarsening.} and 
the aspect ratio of the neighboring zone changes discontinuously by a factor of $\sim$2.  
It is optimal to coarsen in angle whenever a zone's aspect ratio would fall outside the 
range $[1/\sqrt{2}, \,\sqrt{2}]$, ensuring that any zone is at most $\sim$41\% longer 
in any direction than in the others.  It follows that the CFL time step constraint may 
be given by
\begin{equation}
\Delta t \le \frac{\Delta r_\mathrm{min}}{3\sqrt{2}\max_i\{v+c_s\}}\, ,
\end{equation}
where the $\max$ is taken over all zones $i$.  Note that the notion of coarsening 
does not here imply that the overall zone size is increasing.  On the contrary, 
in the portion of the grid where $\Delta r$ is constant, the zones are all roughly 
of equal size.  Therein, the resolution can be controlled simply by the value 
of $\Delta r_\mathrm{min}$.

To continue, let $\ell_\theta$ and $\ell_\phi$ denote the refinement levels in 
the $\theta$- and $\phi$-directions, respectively, where level $\ell_d = 0$ denotes 
the coarsest-resolution level in the $d$-direction.  We require a minimum of 2 
zones in the $\theta$-direction at the origin and a minimum of 4 zones in 
the $\phi$-direction at the polar axes, hence, the finest levels in each of 
these directions are given by $\ell_\theta = \log_2 N_\theta - 1$ and $\ell_\phi = \log_2 N_\phi - 2$, 
where $N_\theta$ and $N_\phi$ are the number of zones on the finest level 
in the $\theta$- and $\phi$-directions, respectively.  For example, with a 
resolution of $N_\theta \times N_\phi = 256 \times 512$ zones, we would have 
maximum levels $\ell_\theta = \ell_\phi = 7$.

The indexing scheme can be rather subtle. A given zone with global coordinates 
$(i,j,k)$ at levels $\ell_\theta(i)$ and $\ell_\phi(i,j)$ has a neighbor (or neighbors) 
in the $r$-direction at coordinates $(i',j',k')$ at levels $\ell_\theta(i')$ and 
$\ell_\phi(i',j')$.  The global coordinates are related by 
$j' = \lfloor n_\theta j \rfloor$ and $k' = \lfloor n_\phi k \rfloor$, where $n_\theta(i,i') \equiv 2^{\ell_\theta(i')-\ell_\theta(i)}$ and $n_\phi(i,i',j,j') \equiv 2^{\ell_\phi(i',j')-\ell_\phi(i,j)}$, and $\lfloor \cdot \rfloor$ 
denotes the integer floor operation.

\subsection{Three-Dimensional Reconstruction}
\label{recon_3d}

The general approach of \fornax{} towards reconstruction is described in \S\ref{recon_method}.
However, with our dendritic grid there are numerous special considerations. 
When performing reconstruction on the spherical dendritic grid in the $r$-direction 
in 2-d (and in both the $r$- and $\theta$-directions in 3-d), we must take special 
care at refinement boundaries to use a pencil of zones of constant coordinate volume.  
For a given zone at refinement level $\ell$, if its neighbor is at refinement 
level $\ell_d' > \ell_d$ in a given angular direction $d$, then we perform a fine-to-coarse 
restriction of the neighbor data to level $\ell_d$ using an appropriate volume-weighted 
average.  On the contrary, if its neighbor is at level $\ell_d' < \ell_d$, then we perform 
a coarse-to-fine prolongation of the data to level $\ell_d$ using an interpolation of 
the neighbor data based on monotonized-central-difference (MC) limited linear slopes in the transverse direction(s).

The restriction from level $\ell_d'$ down to level $\ell_d$ in the $r$-direction is then given by
\begin{equation}
\mathbf{U}_{ijk} = \frac{\sum_{p=j'}^{j'+n_\theta} \sum_{q=k'}^{k'+n_\phi} \Delta V_{ipq} \mathbf{U}_{ipq}}{\sum_{p=j'}^{j'+n_\theta} \sum_{q=k'}^{k'+n_\phi} \Delta V_{ipq}}\, .
\end{equation}
The analogous prolongation from level $\ell_d$ up to level $\ell_d'$ in the $r$-direction can be messy, but straightforward.
Prolongation and restriction are done in order to obtain a pencil of constant coordinate volume, which 
is required to perform a volume-coordinate-based reconstruction. 

\subsection{Domain Decomposition}

\fornax{} is parallelized with non-blocking point-to-point communication and global 
all-to-all reduction using an implementation of the MPI-3.0 standard.  Non-blocking 
communication allows us to hide most $-$ if not all $-$ of the communication latency behind a 
significant amount of local computation.  Due to the structure of the dendritic grid, domain 
decomposition becomes a non-trivial issue.  Unlike the case for a unigrid or for 
patch-based AMR/SMR, there is no known optimal domain decomposition vis-\`a-vis load 
balancing.  Therefore, we rely on a type of greedy algorithm to map the grid onto processors.  
In our decomposition algorithm, we attempt to drive the maximum processor load as close 
to the average load as possible, within reason.  At the same time, we attempt to minimize 
the total communication overhead required to exchange ghost zone data among neighboring grids.  
Since our decomposition algorithm scales as a power law in the number of processors, 
$N_\mathrm{proc}$, it can take minutes to hours to compute a single decomposition 
for a 3-d grid with a given resolution and spatial extent.  However, we need only to 
compute this once; the resulting decomposition can be saved to a file and read in at startup.

We typically require the maximum processor load be no more than 50\% larger than the average 
load.  This is not possible for every conceivable $N_\mathrm{proc}$, but for a given target, 
there is typically some nearby $N_\mathrm{proc}$ for which our algorithm yields sufficient 
load-balancing efficiency.  We need only search within some range of $N_\mathrm{proc}$ around 
the target $-$ a process that can be performed in parallel $-$ and select the value of 
$N_\mathrm{proc}$ giving the most evenly balanced decomposition.  Although our algorithm is 
likely suboptimal, we still achieve near-perfect strong scaling efficiency for 
$N_\mathrm{proc} \gtrsim \text{few} \times 10^5$ (Appendix \S\ref{scaling}).  This is due to the fact 
that the majority of zones belong to a single logically-Cartesian block that can be optimally 
decomposed in a trivial manner.  All that remains is to decompose the coarsened dendritic 
portion of the grid in a way that optimizes the overall load and communication overhead.

\begin{figure*}
\includegraphics[width=\columnwidth]{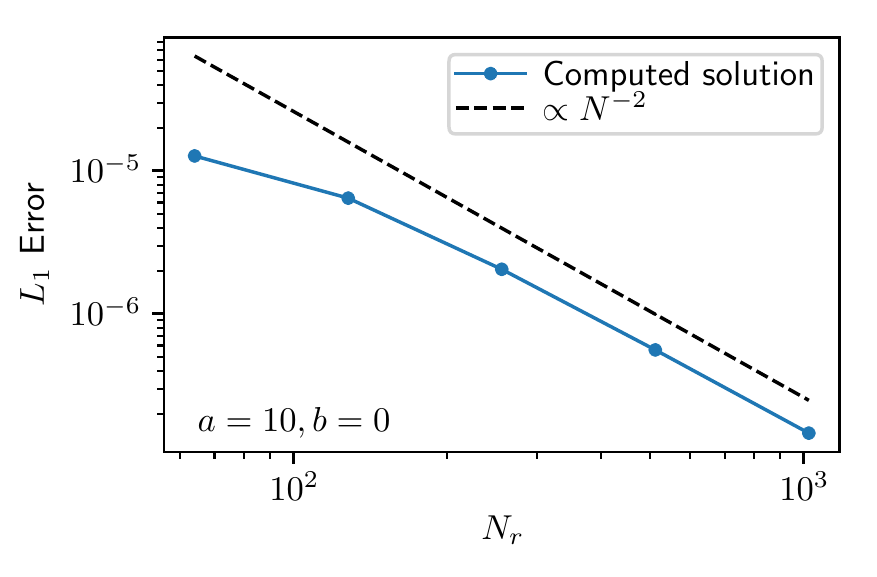}
\hfill
\includegraphics[width=\columnwidth]{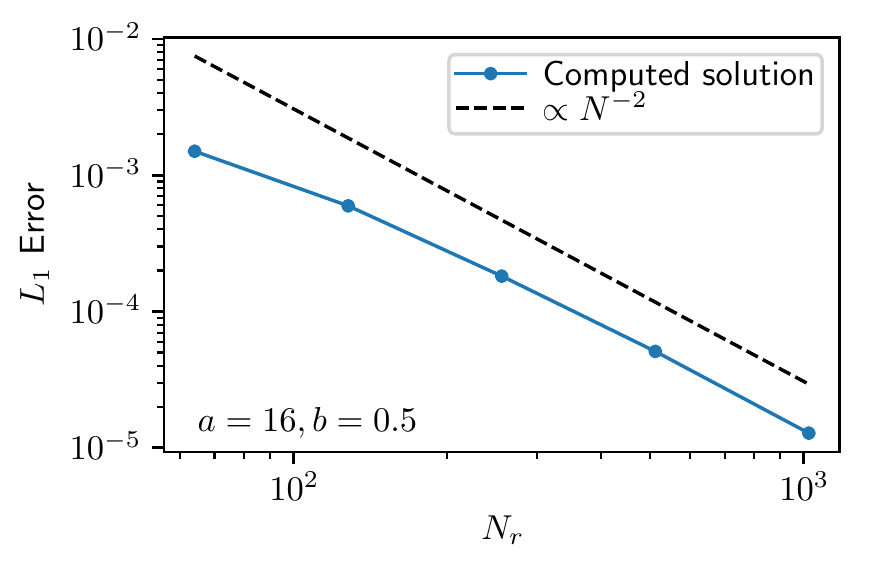}
\caption{Plotted are the $L_1$ error measurements for the spherical wind problem found in
Mignone (2014), as a function of number of radial zones ($N_r$). These resolution tests were performed
with two sets of parameters for the initial Gaussian density profiles used in Mignone,
with his parameter pairs $(a = 10; b = 0)$ (left panel) and $(a = 16; b = 0.5)$ (right panel)
(see Mignone 2014, his Eqs.~73 and~89). The solutions to this hydrodynamic wind problem
are analytic, facilitating the demonstration of the rate of convergence to the correct solution.
See text in \S\ref{mignone} for a discussion.}
\label{mignone1}
\end{figure*}

\subsection{Boundary Conditions}

In spherical coordinates, certain unit vectors change sign discontinuously across coordinate 
singularities, i.e., the $\hat{r}$ and $\hat{\phi}$ unit vectors change sign across the 
origin in the radial direction and the $\hat{\theta}$ and $\hat{\phi}$ unit vectors 
change sign across the axis in the polar direction. Following Baumgarte et al. (2013), we 
enforce parity conditions on the signs of the corresponding vector components in these 
ghost zones after copying the data from active zones.  In the azimuthal direction, 
the grid at $\phi=2\pi$ is mapped periodically to the grid at $\phi=0$, so that the data 
in active zone $(i,j,N_\phi-1)$ are mapped to ghost zone $(i,j,-1)$ and, similarly, the 
data in active zone $(i,j,0)$ are mapped to ghost zone $(i,j,N_\phi)$.  In the polar direction 
at the axes, the data are periodically shifted in the azimuthal direction by $N_\phi/2$ zones 
as they are copied from the active zones into the ghost zones.  Thus, the active zone $(i,0,k)$ 
is mapped to the ghost zone $(i,-1,k')$, where $k' = (k+N_\phi/2) \mod N_\phi$.  Finally, in 
the radial direction at the origin, the active-zone data are reflected in the polar direction 
and then periodically shifted as they are copied into the ghost zones.  Thus, the active 
zone $(0,j,k)$ is mapped to the ghost zone $(-1,j',k')$, where $j' = N_\theta - 1 - j$ and $k'$ is as before.  {At the outer radial boundary, \fornax{} can employ a variety of boundary conditions, including outflow conditions with or without a diode restriction on the mass or radiation flux, Dirichlet and Neumann conditions, and arbitrary user-specified conditions.}

\section{Hydrodynamic Tests}
\label{hydro_tests}

\subsection{Mignone Reconstruction Tests}
\label{mignone}

We have taken special care in treating the
reconstruction of quantities on the mesh and have formulated our
reconstruction methods to be agnostic to choices of coordinates and mesh
equidistance.  As in Mignone (2014), we do this by
forming our reconstruction expressions in terms of volume-averaged
quantities and moments of the coordinates in each cell.  The moments are
calculated at start up for the given mesh and coordinates.
In the radial direction, we have a uniform mesh in coordinate $x_1$, where
the normal spherical radius $r = r_t sinh(x_1/r_t)$, with $r_t$ a constant
parameter.  We reconstruct quantities in $x_1$, i.e., we locally compute $f(x_1)$
to specify $f$ on the faces of each cell.  The mesh in $x_1$ is entirely
uniform.  We have no need of, and do not construct, $f(r)$, where then
the samples of $f$ would be non-uniformly distributed in $r$.  The same is true in the
angular direction.  Thus, we have no ``non-equidistant" reconstruction
and, therefore, our reconstructions yield the same results as Mignone (2014).
For any choice of coordinates in any geometry, the method works the same and forms a
consistent parabolic reconstruction from volume-averaged inputs.

To demonstrate the scaling of \fornax{} under this reconstruction regime, we 
have performed at various resolutions the $L_1$ error radial/spherical wind test 
discussed in Mignone (2014).  This hydrodynamic problem has an analytic 
solution (Mignone, his equation 89) and we have conducted the test for the 
Gaussian density profile provided in his Equation~73, with his parameters 
$(a = 10; b = 0)$ and $(a = 16; b = 0.5)$. Figure~\ref{mignone1} depicts the 
associated convergence scaling for the two parameter sets, compared with 
quadratic ($\frac{1}{N^2}$) behavior, and should be compared with 
Mignone's Figure~10. Despite the fact this study was performed for a 
non-uniform radial grid, the expected scaling with resolution still emerges.


\begin{figure}
\includegraphics[width=0.48\columnwidth]{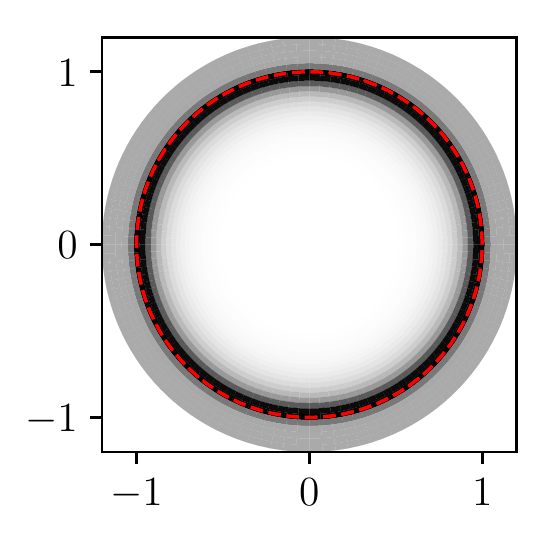}
\includegraphics[width=0.48\columnwidth]{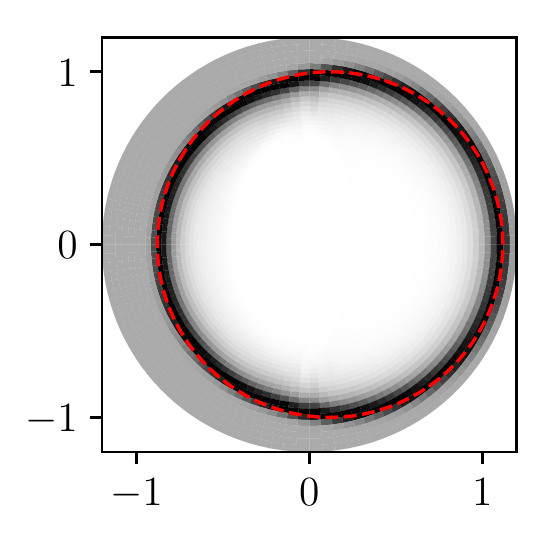}
\caption{Results of the three-dimensional Sedov blast wave test at time $t_\mathrm{final}=1$.  
We plot the density, $\rho$, for the on-axis (left) and off-axis (right) blast waves along a 
slice at $y=0$ using a linear gray color scale from $\rho=0$ (white) to $\rho=3$ (black).  
The analytic shock radius location, $r_\mathrm{sh}$, is overplotted (red dashed line).  
Small-scale axis artifacts are apparent in the off-axis version, but the solution remains 
fairly spherical even though the blast is nowhere aligned to the grid.}
\label{fig:sedov_rho}
\end{figure}
 
\begin{figure}
\includegraphics[width=0.48\columnwidth]{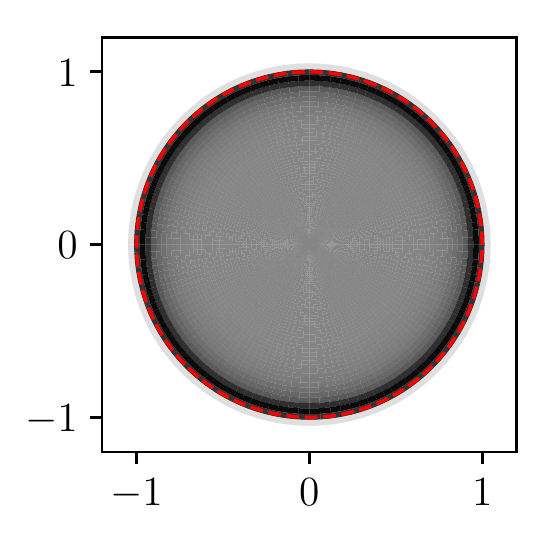}
\includegraphics[width=0.48\columnwidth]{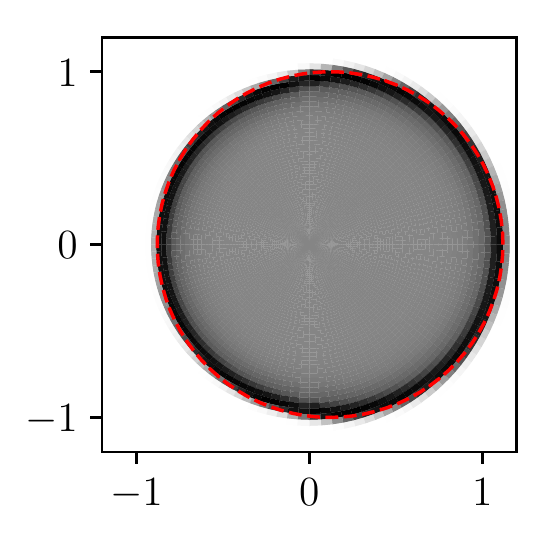}
\caption{Same as Figure~\ref{fig:sedov_rho} but for the internal energy density, $u$, which 
is plotted using a linear gray color scale from $u=0$ (white) to $u=0.25$ (black).  
Small-scale axis artifacts are less apparent, since pressure gradients tend to get 
smoothed out behind the shock front.}
\label{fig:sedov_u}
\end{figure}

\subsection{Centered and Off-Centered Sedov Blast Wave}
\label{sedov}

A standard test is the classical self-similar blast wave of Sedov (1959) describing, e.g., the 
initial energy-conserving phase of a supernova remnant.  The problem consists of a stationary 
background medium with uniform density $\rho_1$ and near-zero pressure $P_1$ into which a 
prescribed blast energy $E_1$ is deposited.  Since $P_1 \approx 0$, the solution depends on 
only the two parameters $\rho_1$ and $E_1$, it follows that the radius, $r$, and the time 
since explosion, $t$, are related by a dimensionless similarity variable, $\xi$.  Thus, it 
can be shown that the position of the shock radius is given by
\begin{equation}
r_\mathrm{sh}(t) = \xi_0 \left(\frac{E_1 t^2}{\rho_1}\right)^{1/5}\, ,
\end{equation}
where $\xi_0 \approx 1.033$ for an adiabatic equation of state with $\gamma=1.4$.  Following 
Kamm \& Timmes (2007), we set the dimensionless inputs to $\gamma=1.4$, $\rho_1=1$, $P_1=4\times10^{-13}$, 
and $E_1 = 0.851072$ such that the outer blast radius is at $r_\mathrm{sh}=1$ at time $t=1$.  We use 
a three-dimensional spherical dendritic grid with radial extent out to $r_\mathrm{max}=1.2$ and 
a resolution of $64\times 64 \times 128$ zones.  The radial grid spacing is constant in the 
interior with $\Delta r \approx 0.01$ out to $r\approx 0.3$, then smoothly transitions to logarithmic 
spacing out to $r_\mathrm{max}$.

For our first test, we deposit all of $E_1$ in the zones surrounding the origin.  As expected, 
the subsequent blast is aligned with the grid and maintains perfect spherical symmetry.  Next, 
we deposit the same energy in the zones closest to the $x$-axis at $r=0.12$ so that the blast 
is off-axis.  This time the blast is not aligned with the grid and must cross the polar axis as 
it expands.  Figures~\ref{fig:sedov_rho} and~\ref{fig:sedov_u} compare the density and internal 
energy of the on- and off-axis blast waves, respectively.  There are small-scale axis artifacts 
in the density in the off-axis version of the blast wave, but the blast wave remains fairly 
spherical despite being nowhere aligned with the spherical grid.  The scale of the artifacts is 
smaller in the internal energy, since pressure variations tend to be smoothed out behind the shock 
where the pressure is large.

Overall, this test demonstrates that the internal geometric boundary conditions at the origin and 
polar axes are correctly applied.  The artifacts at the axes are likely due to small errors in 
reconstruction in the azimuthal direction with our dendritic grid.  The direction of the polar 
and azimuthal coordinate vectors is divergent at the polar axis, hence angular variations in the 
data there can only be resolved by a correspondingly convergent grid.  In the limit of a locally 
``planar" flow across the poles, the rapid azimuthal variation comes entirely from the coordinates 
themselves, not from the data, although these variations must average to zero.  Our dendritic grid 
deresolves this rapid angular variation, introducing first-order errors due to monotonicity contraints.  
However, these artifacts are confined to a narrow polar axis region and do not adversely contaminate 
the solution elsewhere.

\subsection{Sod Shock Tube}
\label{noh_sod}

To demonstrate the shock-capturing capabilities of \fornax, we consider
the classical shock-tube test of Sod (1978). The initial data
consists of two states:
\begin{equation}
  (\rho, P) = \begin{cases}
    (1.0, 1.0), & \textrm{if } x < 0\,, \\
    (0.125, 0.1), & \textrm{otherwise}\,.
  \end{cases}
\end{equation}
The initial velocity is set to zero. For this test we adopt an adiabatic
equation of state with index \mbox{$\gamma = 1.4$}.

This test is not particularly challenging for modern HRSC codes, but it
is nevertheless interesting because it exhibits all fundamental
hydrodynamical waves. The exact solution consists of a left propagating
rarefaction wave, a right propagating contact discontinuity, and a right
propagating shock.

\begin{figure*}
  \includegraphics[width=\textwidth]{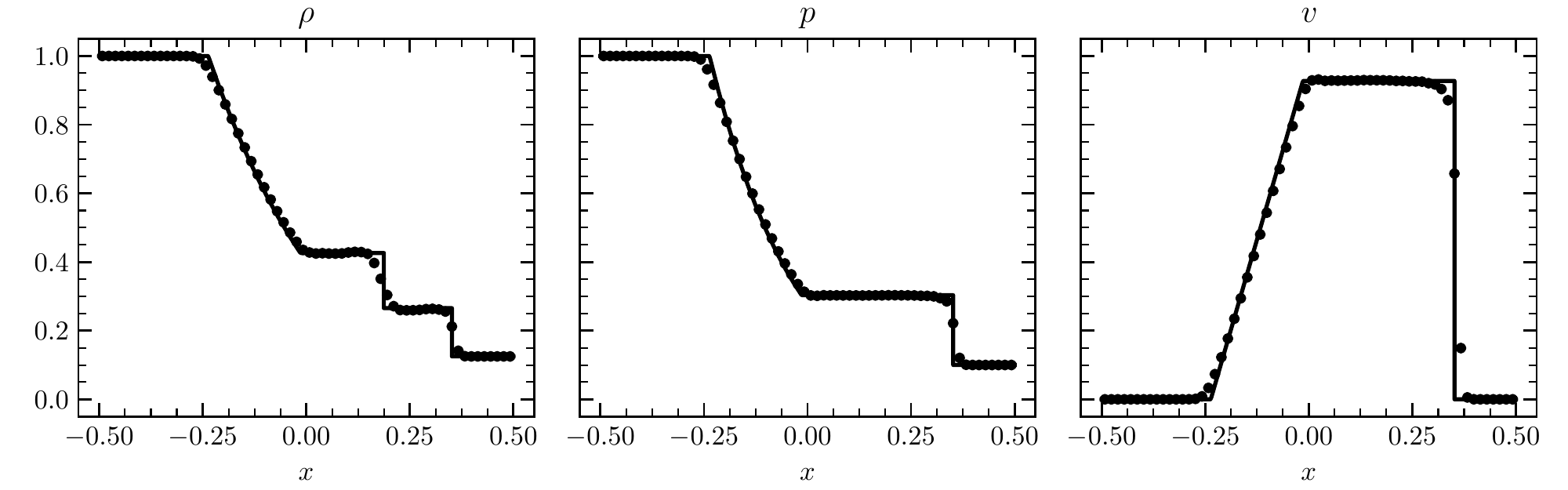}
  \caption{Density, pressure, and velocity (left, middle, and right
  panels respectively) for the Sod test. The solid line is the analytic
  solution, while the dots show the \fornax{} results at a resolution of
  $64$ grid points. All of the key features of the solution are
  correctly captured, but the contact wave is smeared over ${\sim}4$
  grid points.}
  \label{fig:sod.1d}
\end{figure*}
 
Figure~\ref{fig:sod.1d} shows the analytic solution and the results from
\fornax{} using a coarse mesh of 64 points. The code correctly captured
all of the waves. The shock wave is sharply captured within ${\sim}2$
grid points, while the contact wave is smeared over ${\sim}4$ grid
points.

\subsection{Double Mach Reflection}
\label{mach}

To test the sensitivity of the \fornax{} code to ithe numerical
diffusion of contact waves, we perform the classic double Mach
reflection test of Woodward \& Collela (1984).  This test consists of a
Mach-10 oblique shock through air ($\gamma=1.4$), inclined with respect
to a reflecting boundary in which the incident and reflected shocks
then interact to produce a triple point.  In the space between the
reflected shock and the contact discontinuity, an upward-directed jet
should form along the slip surface, but if the numerical dissipation
of the contact wave is too high, the formation of this jet will be
suppressed.  We use a two-dimensional grid of $520 \times 160$ zones
over the domain $(x,y) \in [0,3.25] \times [0,1]$ and position an
oblique shock at $x_0=1/6$ inclined at an angle $\alpha=\pi/3$ with
respect to the $x$-axis.  The pre-shock state is given by
$(\rho,P,v_x,v_y)_\mathrm{R}=(1.4,1,0,0)$ and the post-shock state by
$(\rho,P,v_x,v_y)_\mathrm{L}=(8,116.5,8.25\sin \alpha,-8.25\cos
\alpha$).  The left $x$-boundary is held fixed at the post-shock state
and outflow conditions are imposed at the right $x$-boundary.  At the
lower $y$-boundary, the post-shock state is fixed for $x \le x_0$ and
reflecting boundary conditions are imposed for $x > x_0$.  Meanwhile,
at the upper $y$-boundary, the time-dependent shock position, $x_s =
x_0 + y/\tan \alpha + 10t/\sin \alpha$, is used to set either the
post-shock state ($x \le x_s$) or the pre-shock state ($x>x_s$).

\begin{figure}
\includegraphics[width=\columnwidth]{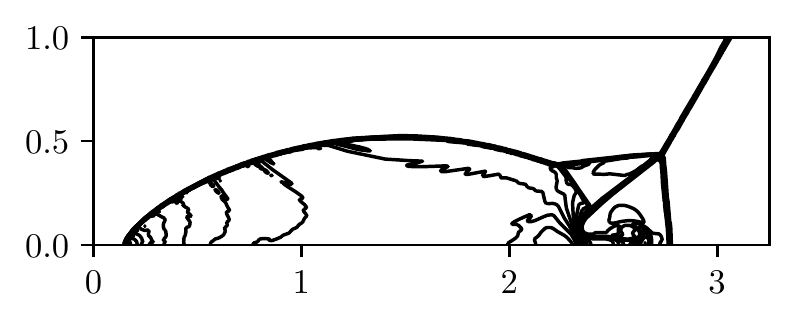}
\caption{Results at time $t_\mathrm{final}=0.2$ of the double Mach
reflection test.  We use a resolution of $520 \times 160$ Cartesian
grid zones and plot 30 linearly-spaced contours of the density.  On
the right-hand side, the incident and reflected shocks form a triple
point, below which an upward-directed jet is formed along the slip
surface at $y=0$.}
\label{fig:double_mach_rho}
\end{figure}

Figure~\ref{fig:double_mach_rho} shows the evolved system at
$t_\mathrm{final} = 0.2$ using 30 linearly-spaced contours of the
density.  The features at the various shock fronts remain sharp, and
in the region below the triple point a small upward-directed jet is
indeed produced along the slip surface.  This indicates that numerical
dissipation of the contact wave is well-controlled in \fornax{}.

\subsection{Rayleigh-Taylor Instability}
\label{rayleigh}

\subsubsection{Two-Dimensional Rayleigh-Taylor Test}
\label{rt_2d}

Here, we study the linear and non-linear development of the Rayleigh-Taylor
instability in 2-d. The initial data describes an unstably stratified
fluid with
\begin{equation}
  \rho = \begin{cases}
    \rho_h & \textrm{if } z \geq 0\,, \\
    \rho_l & \textrm{otherwise}\,. \\
  \end{cases}
\end{equation}
The gravitational acceleration $g = 1/2$ is parallel to the $z$ axis,
and $\rho_h = 2$, $\rho_l = 1$, so that the Atwood number $A = (\rho_h -
\rho_l)/(\rho_h + \rho_l)$ is $1/3$. The boundary conditions are
reflective for $z = \pm 1$ and periodic for $x = \pm 1/2$. The initial
pressure is chosen to ensure hydrostatic equilibrium:
\begin{equation}
  P(z) = P(-1) + \int_{-1}^z \rho\, g\, \mathrm{d} z\,,
\end{equation}
with $P(-1) = 10/7 + 1/4$. For this test we adopt an adiabatic equation
of state with index \mbox{$\gamma = 1.4$}.

At time $t = 0$ the interface between the two fluid components is
perturbed to be
\begin{equation}
  h(x) = h_0 \cos(\kappa x)\,,
\end{equation}
with $h_0 = 0.01$ and $\kappa = 4\pi$. The sharp interface is then smoothed
into an hyperbolic tangent profile with characteristic length $0.005$.

According to analytic theory $h$ should evolve according to the
following equation:
\begin{equation}\label{eq:rt.linear.theory}
  h(x,t) = h_0\, \cosh(\sqrt{A \kappa g}\, t)\, \cos(\kappa x)\,.
\end{equation}

\begin{figure}
  \includegraphics[width=\columnwidth]{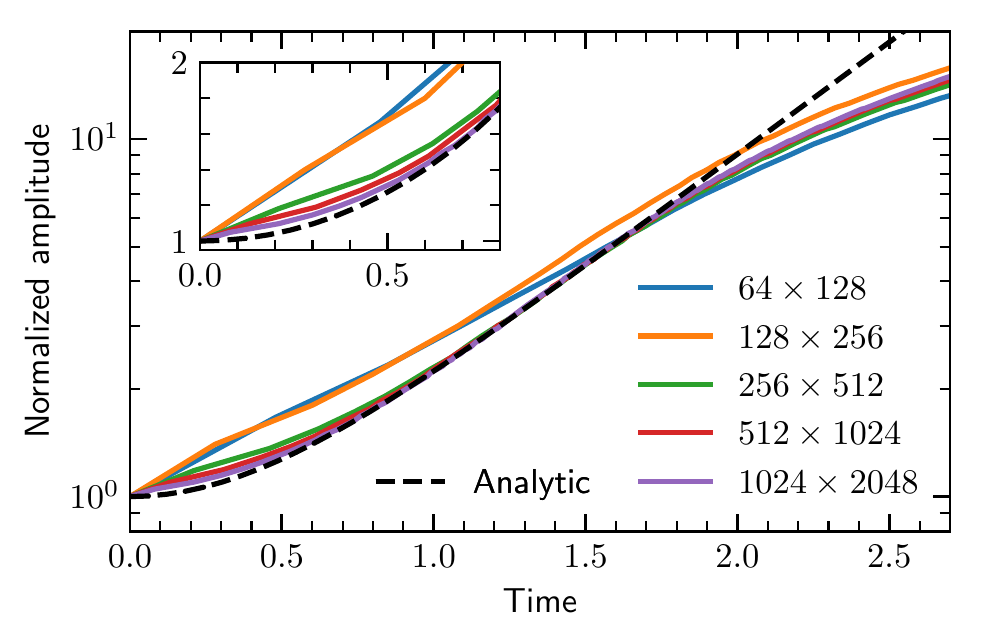}
  \caption{Amplitude of the interface perturbation for the single mode
  2-d Rayleigh-Taylor test. We show results from different resolutions
  and the expected growth rate from analytic theory. We find perfect
  agreement with linear theory up to $t \simeq 1.75$, when secondary
  instabilities start to appear in the flow and the dynamics becomes
  fully nonlinear.}
  \label{fig:rt.linear}
\end{figure}
 
We perform simulations with resolutions ranging from $64\times 128$ to
$1024\times 2048$. We track the mixing of the two fluid components by
means of a passive scalar tracer. This is initialized to be $\rho -
\rho_l$, so that the value $1$ corresponds to the heavy fluid, while the
value $0$ corresponds to the light fluid. We track the growth of the
initial perturbation by locating the $0.99$ isocontour of the tracer.
Figure~\ref{fig:rt.linear} shows the perturbation amplitude, normalized by
its initial value, and the prediction from analytic theory
(Eq.~\ref{eq:rt.linear.theory}).

At low resolution, the perturbations grow exponentially from $t = 0$,
but the growth rate is somewhat smaller than that predicted by analytic
theory. The slower growth rate at low resolution is not unexpected and
is due to the numerical viscosity, which modifies the Rayleigh-Taylor
dispersion relation (Dimonte et al. 2004; Murphy \& Burrows 2008a). We speculate
that the reason why the initial perturbations grow exponentially at low
resolution is that they are not well resolved. At higher resolution the
correct $\cosh$ time dependence is recovered and the agreement with
linear theory is excellent up to time $t \simeq 1.75$, when the
Rayleigh-Taylor plumes start to break and rolls develop.

\begin{figure*}
  \includegraphics[width=\textwidth]{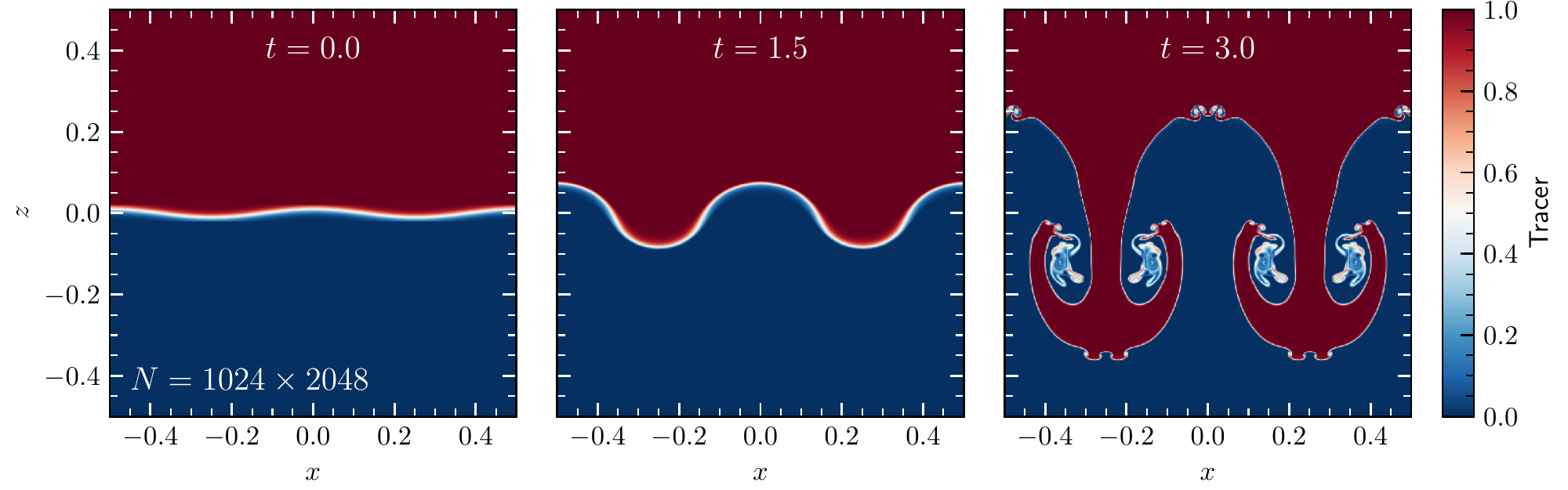}
  \caption{Passive tracer concentration for the single mode 2-d
  Rayleigh-Taylor test at three representative times. The resolution is
  $1024 \times 2048$. During the linear phase of the instability, up to
  $t \simeq 1.75$, the contact discontinuity is well preserved and there
  is no spurious mixing of the two fluid phases. Secondary
  instabilities, seeded by the numerical noise, appear at later times,
  during the non-linear phase of the evolution.}
  \label{fig:rt.2d}
\end{figure*}

During the non-linear phase of the evolution, secondary Rayleigh-Taylor
and Kelvin-Helmholtz instabilities appear (Fig.~\ref{fig:rt.2d}). These
are seeded by the numerical noise and their detailed morphology, in
the absence of explicit dissipation, is known to be dependent on the details
of the numerical scheme (Liska \& Wendroff 2003a). Despite the
presence of these features, \fornax{} is able to preserve the sharp
discontinuity in the fluid tracer. Artificial mixing between the two
fluid components is only present in the Rayleigh-Taylor rolls, where the
flow develops features on scales comparable to that of the grid.

\subsubsection{Three-Dimensional Rayleigh-Taylor Test}
\label{rt_3d}

Neutrino-driven convection plays a central role in the explosion
mechanism of CCSNe (Burrows 2013; Radice et al. 2017; Burrows 
et al. 2018; Vartanyan et al. 2018). In this section, we
benchmark \fornax{} for the modeling of convective flows by studying the
nonlinear development of the Rayleigh-Taylor instability in 3-d. We
consider the setup introduced by Dimonte et al. (2004), for which
the dynamics is dominated by mode couplings. 

The computational domain is a box with $-L/2 < x,y < L/2$, $-L < z < L$,
where $L = 10$ in arbitrary units. We consider a constant vertical
gravitational acceleration $g := - g_z = 2$, and we prepare initial
conditions with density stratification
\begin{equation}
  \rho(z) = \rho_0 \left(1 - \frac{\gamma - 1}{\gamma}\frac{\rho_0 g
  z}{P_0}\right)^{\frac{1}{\gamma - 1}}\,
\end{equation}
where $\gamma = 5/3$,
\begin{equation}
\rho_0 = \begin{cases}
  \rho_h & \textrm{if } z \geq 0\,, \\ 
  \rho_l & \textrm{otherwise}\,,
\end{cases}
\end{equation}
and $p_0 = 2\pi(\rho_h + \rho_l) g L$. We set $\rho_h = 3$ and $\rho_l =
1$, so that the Atwood number $A = (\rho_h - \rho_l)/(\rho_h + \rho_l)$
is $1/2$. The initial pressure is set to
\begin{equation}
  P = P_0 \left(\frac{\rho}{\rho_0}\right)^{\gamma}\,,
\end{equation}
which ensures that the initial conditions are in hydrostatic
equilibrium. We track the concentration of the ``heavy'' fluid by
evolving a passive scalar field $f_h$ initialized to be one for positive
$z$ and zero otherwise. We assume periodicity in the $x$ and $y$
direction and hydrostatic boundary conditions at $z = \pm L$.  We use
uniform grids and label our simulations by the number of grid points in
the $x$ direction so that, e.g., the N128 run has a resolution of $128
\times 128 \times 256$ points.

At time $t = 0$ the interface between the two fluid components is
perturbed to be
\begin{equation}\label{eq:rt3d.pert}
\begin{split}
  h(&x,y) = \frac{1}{H} \sum_{8 \leq k_x^2 + k_y^2 \leq 16}
  \Big[
      a_{\mathbf{k}} \cos(k_x X) \cos(k_y Y) + \\
    & b_{\mathbf{k}} \cos(k_x X) \sin(k_y Y) +
      c_{\mathbf{k}} \sin(k_x X) \cos(k_y Y) + \\
    & d_{\mathbf{k}} \sin(k_x X) \sin(k_y Y) \Big]\,,
\end{split}
\end{equation}
where $X = 2\pi x/L$ and $Y = 2\pi y/L$. The
coefficients $a_{\mathbf{k}}$, $b_{\mathbf{k}}$, $c_{\mathbf{k}}$, and
$d_{\mathbf{k}}$ are sampled from a uniform distribution taking values
between $-1$ and $1$, boundary excluded. $H$ is a normalization
coefficient that we adjust after having sampled  $a_{\mathbf{k}}$,
$b_{\mathbf{k}}$, $c_{\mathbf{k}}$, and $d_{\mathbf{k}}$ so that $h_{\rm
rms} =  3\times 10^{-4} L$. We remark that we use the same initial
conditions for all of our simulations, while Dimonte et al. (2004) used
a different number of modes in their initial perturbation depending on
the resolution. We also use the same realization of the coefficients in
Equation~\eqref{eq:rt3d.pert} for all the simulations presented in this
section. In this way, the convergence of our numerical results can be
better assessed. 

When constructing the perturbed initial data, we compute the cell-averaged 
density, internal energy, and heavy-fluid concentration, i.e.,
the quantities evolved by the finite-volume scheme in \fornax{}, by
numerically integrating the profiles discussed above on an auxiliary
mesh with 50 times higher resolution than the base grid in each
direction.

\begin{figure*}
  \includegraphics[width=\textwidth]{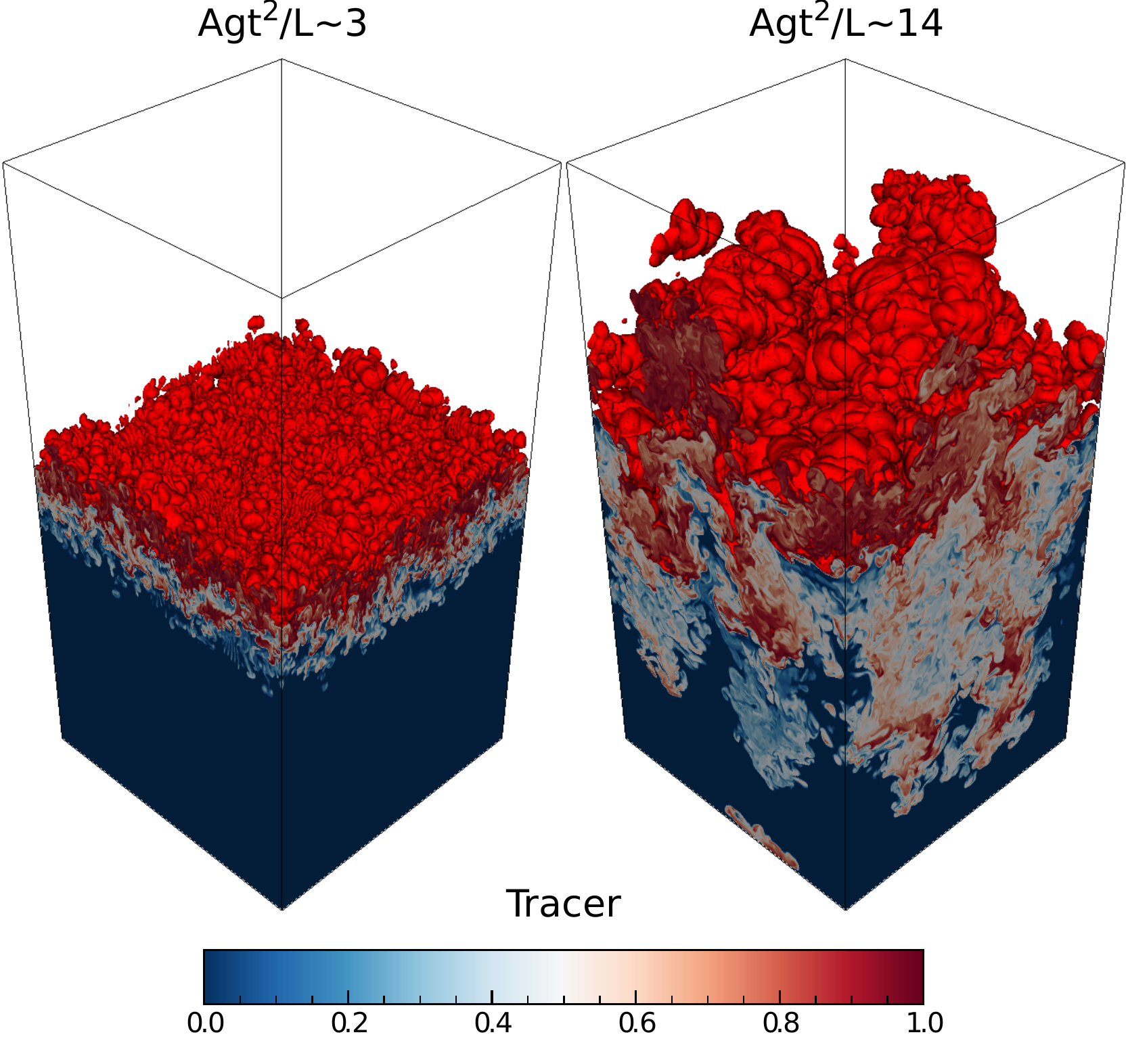}
  \caption{Passive tracer concentration for the multi-mode 3-d
  Rayleigh-Taylor test at two representative times. Red values indicate
  higher concentration for the ``heavy'' fluid. The red surface is the
  99\% isocontour of the concentration. The resolution is $512
  \times 512 \times 1024$.}
  \label{fig:rt.3d}
\end{figure*}
 
The initial perturbations generate small buoyant bubbles that interact
and grow into larger bubbles as the system evolves, as shown in
Figure~\ref{fig:rt.3d}. We find that there is significant mixing between
the two fluids. This is evidenced by the intermediate values taken by
$\langle f_h \rangle$ (see Eq.~\ref{fh} below), especially in the buoyant bubbles which entrain a
significant amount of the heavy fluid. At late times, the topology of the
interface between the two fluids becomes non-trivial and we observe the
detachment of some of the bubbles.

\begin{figure*}
  \begin{center}
    \includegraphics[width=0.24\textwidth]{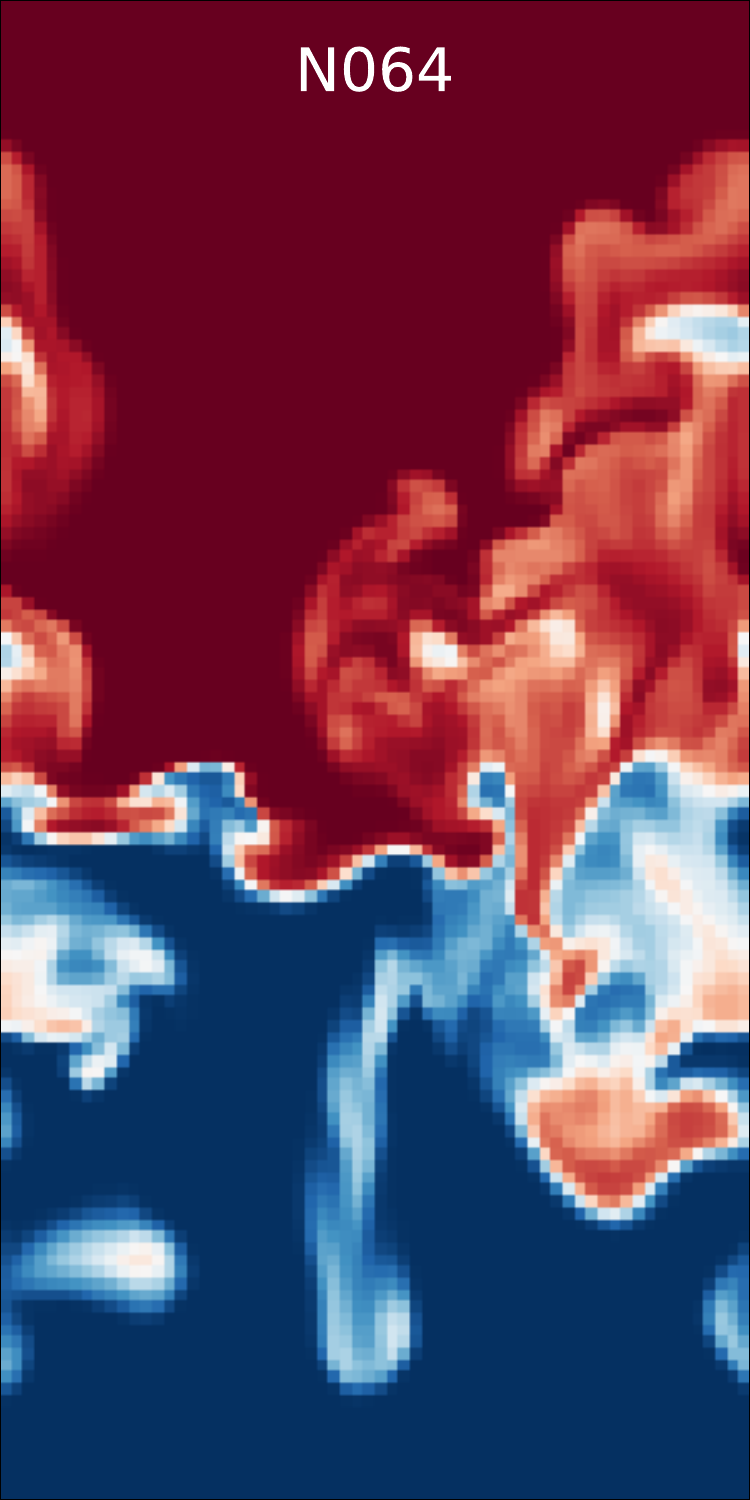}
    \includegraphics[width=0.24\textwidth]{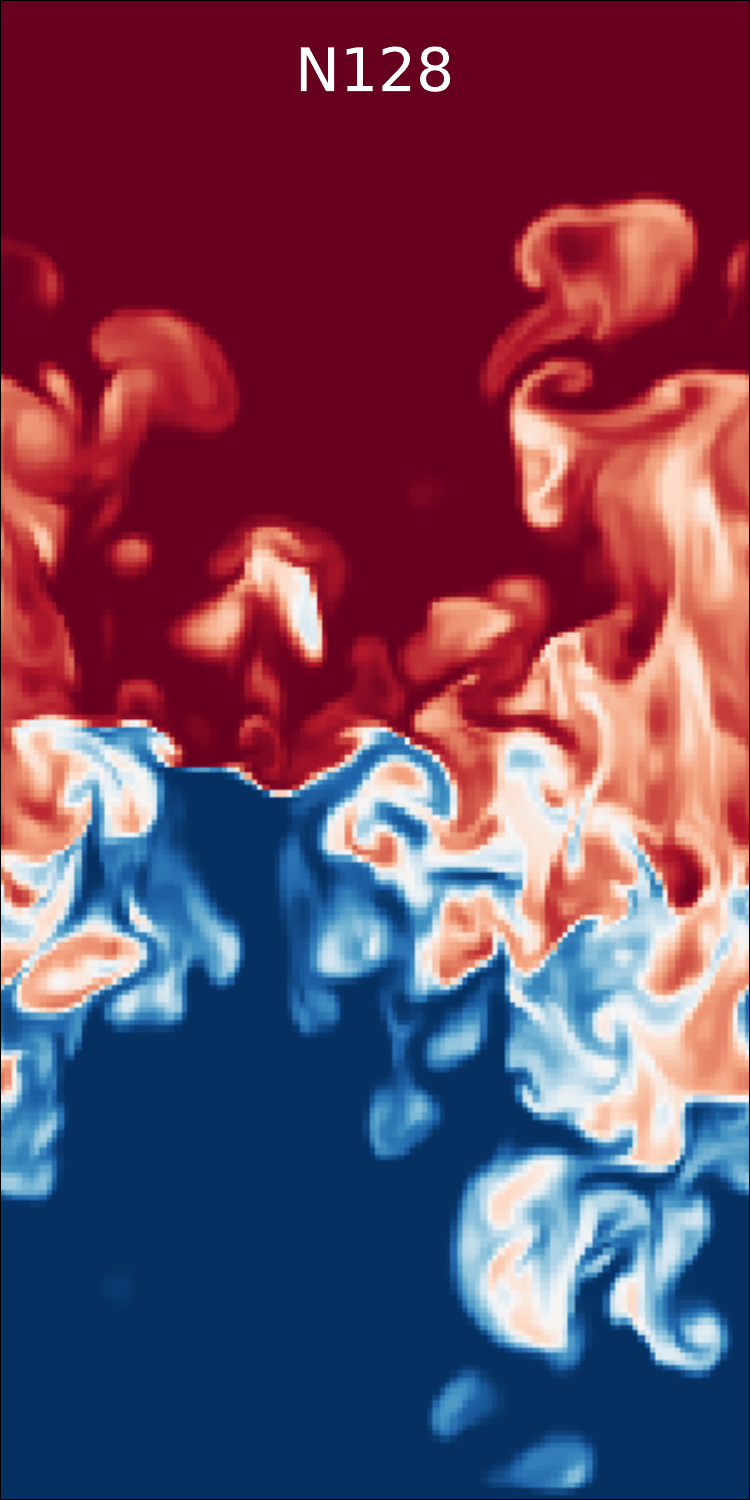}
    \includegraphics[width=0.24\textwidth]{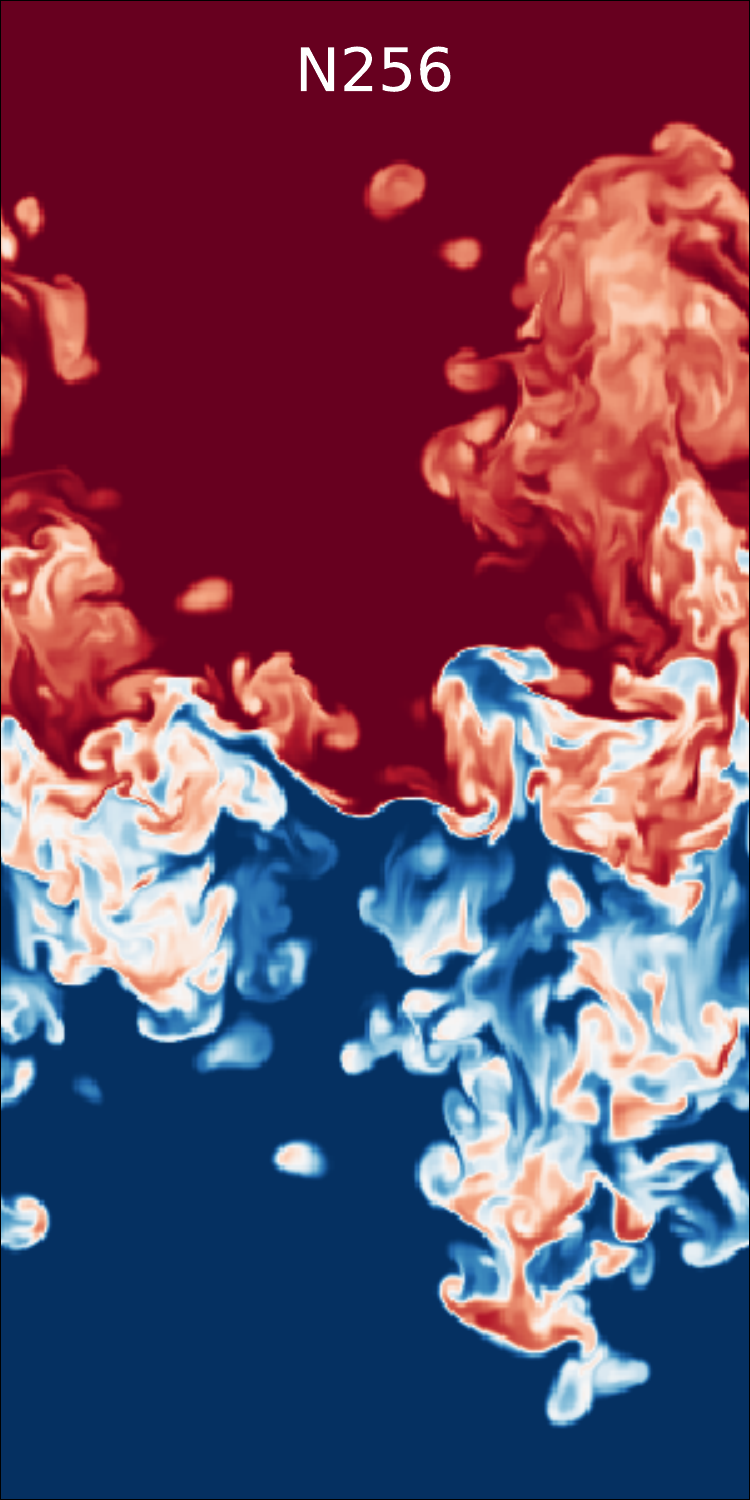}
    \includegraphics[width=0.24\textwidth]{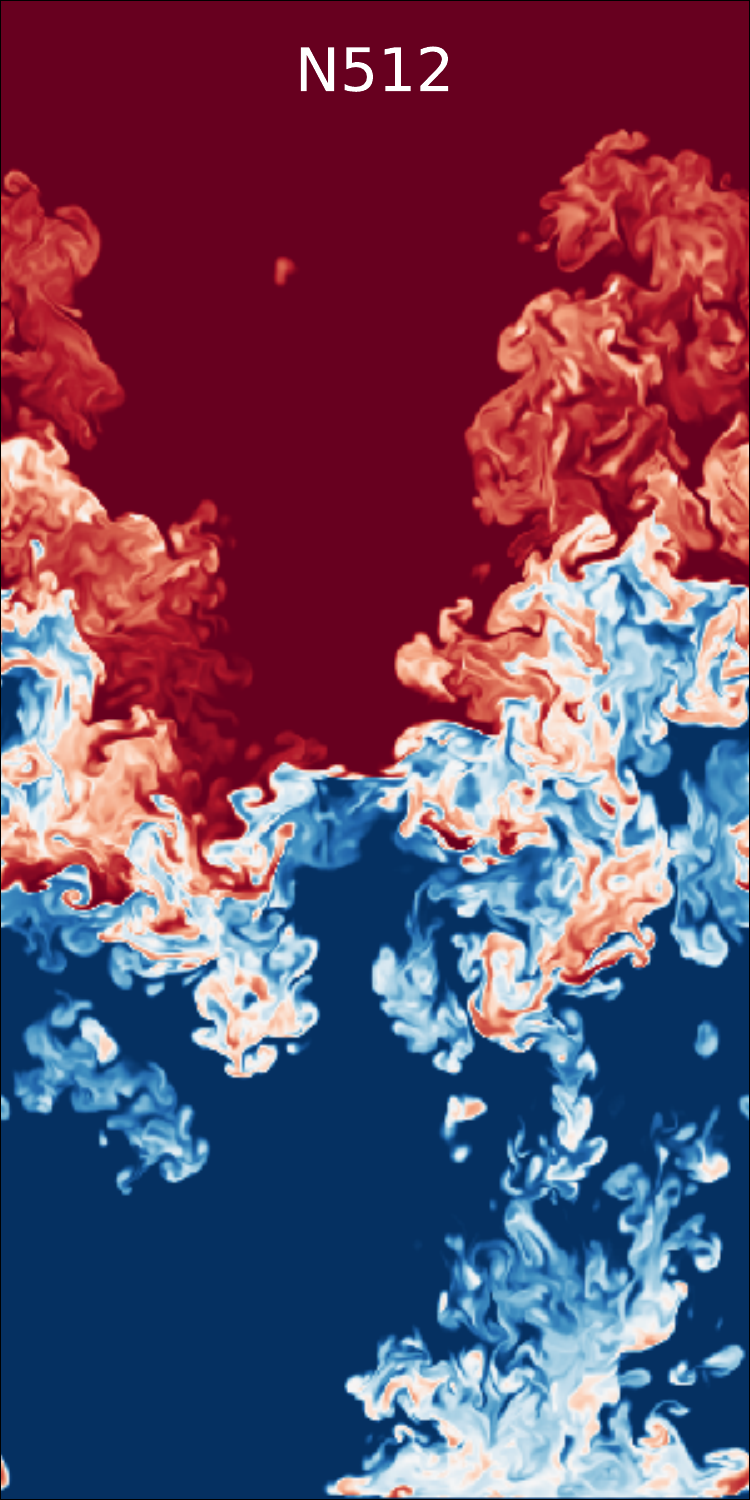}
  \end{center}
  \caption{Passive tracer concentration in the $xz$-plane for the
  multi-mode 3-d Rayleigh-Taylor test at time $A g t^2/L \simeq 14$ for
  different resolutions. The dominant features of the flow are in
  qualitative agreement between the different resolutions, despite the
  highly non-linear nature of the flow at this time.}
  \label{fig:rt.3d.xz}
\end{figure*}
 
The entrainment of the heavy fluid in the buoyant bubbles and of the
light fluid in the sinking plumes is not strongly dependent on the
resolution, but they appear to grow slightly with resolution, as shown
in Figure~\ref{fig:rt.3d.xz}. This is presumably due to the higher
effective Reynolds numbers achieved in the best resolved simulations. Note 
that we do not include an explicit viscosity in these simulations,
consequently the dissipation scale is a multiple of the grid scale. At
high resolutions secondary instabilities with smaller wave number and
faster growth rates are allowed to develop. Their presence might explain
the increase in the entrainment with resolution. Despite these
quantitative differences we find a remarkable qualitative agreement
between the different resolutions with the gross flow feature captured
even at the lowest resolution of $64\times 64\times 128$.

\begin{figure*}
  \includegraphics[width=\columnwidth]{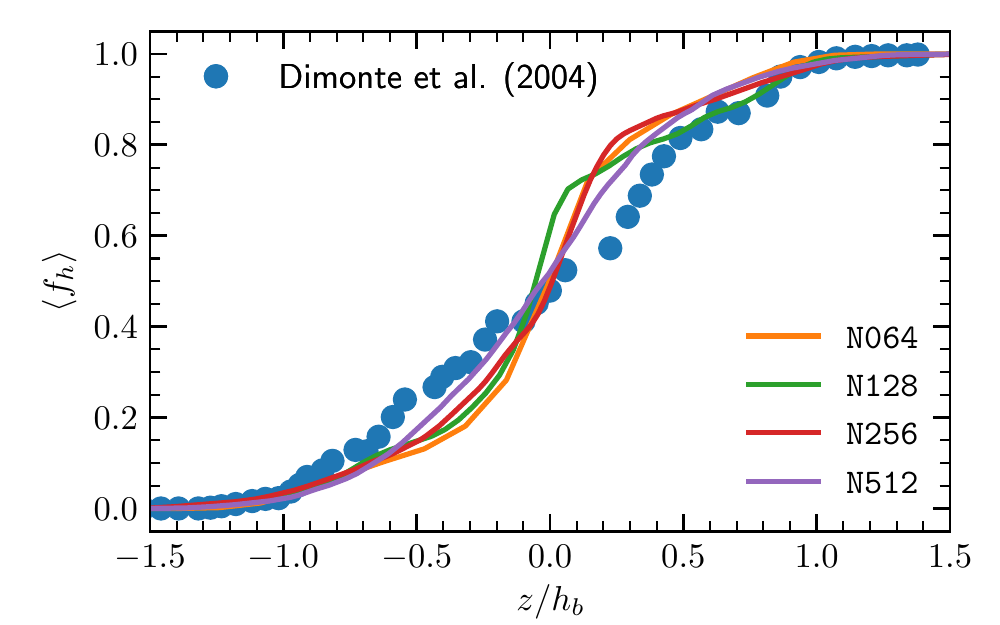}
  \includegraphics[width=\columnwidth]{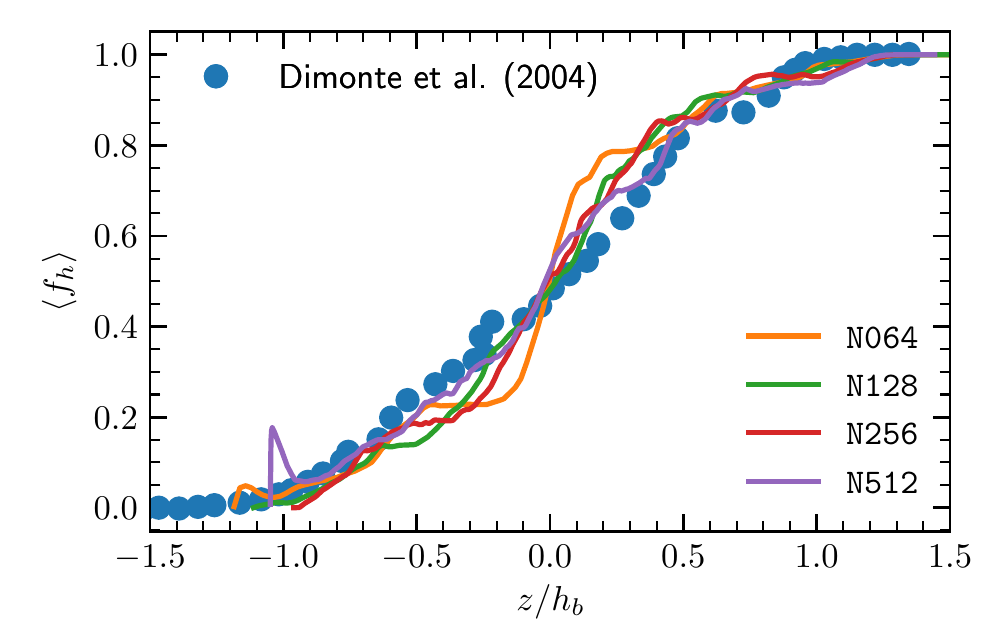}
  \caption{Horizontally-averaged concentration of the heavy fluid at
  an early time ($Agt^2/L \simeq 3$; left panel) and at a late time ($A
  g t^2/L \simeq 14$; right panel). The blue points are the experimental
  data extracted from Dimonte et al. (2004), the solid lines are
  \fornax{} calculations at different resolutions. We find good
  agreement between the numerical and the experimental data.}
  \label{fig:rt.3d.fh}
\end{figure*}
 
Following Dimonte et al. (2004) we compute the horizontally-averaged
concentration of the heavy fluid
\begin{equation}
  \langle f_h \rangle = \frac{1}{L^2} \iint f_h\,
    \mathrm{d}x\, \mathrm{d}y
\label{fh}
\end{equation}
and we define as the bubble penetration depth $h_b$ the $z$ value at
which $\langle f_h \rangle$ reaches 99\%. The profile of $\langle f_h
\rangle$ is observed to be self-similar during the non-linear
development of the Rayleigh-Taylor instability. We show the profile of
$\langle f_h \rangle$ obtained with \fornax{} at two representative
times in Figure~\ref{fig:rt.3d.fh}. Also shown are the experimental
results obtained by Dimonte et al. (2004), as reported in
that paper. This figure should be contrasted with Figure~7 of
Dimonte et al. (2004). We find good agreement between the \fornax{}
results and the experimental data. The agreement improves as the
resolution increases, with the exception of a spike appearing at late
times in the N512 data around $z/h_b \simeq -1$. This is caused by the
interaction of some of the down-falling plumes with the boundary at the
bottom of the simulation box. 

\begin{figure}
  \includegraphics[width=\columnwidth]{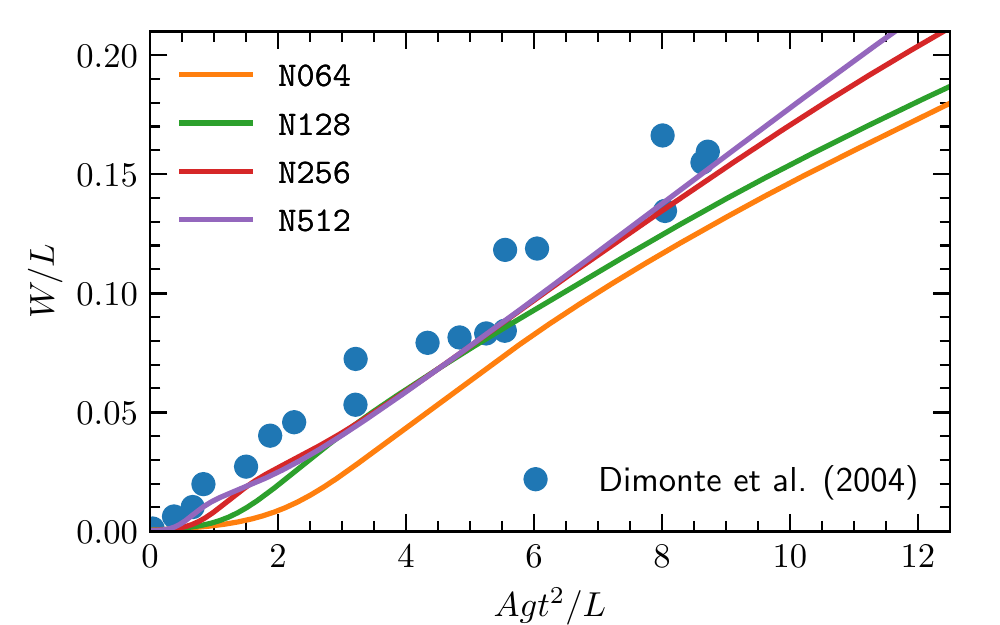}
  \caption{Mixing degree as a function of the dimensionless time $A g
  t^2/L$. The blue points are the experimental data extracted from
  Dimonte et al. (2004). The other lines are results from \fornax{}
  calculations. We find that \fornax{} agrees well with the experimental
  data predicting the right growth rate of the instability in the
  non-linear regime.}
  \label{fig:rt.3d.W}
\end{figure}
 
Following Andrews \& Spalding (1990) and Dimonte et al. (2004), we quantify
the mixing degree between the heavy and the light fluid with the metric
\begin{equation}
  W = \int \langle f_h \rangle \left(1 - \langle f_h \rangle \right)\,
    \mathrm{d} z\,.
\end{equation}
Our results are shown in Figure~\ref{fig:rt.3d.W}. The growth rate of the
instability in the nonlinear phase is proportional to $A g t^2/L$, as
predicted by theoretical models and confirmed by experimental results
(Dimonte et al. 2004). The progression of the mixing between the two
fluids as computed by \fornax{} agrees well with the experimental
results reported in Dimonte et al. (2004).
We find that \fornax{} compares favorably to the other Eulerian codes
presented in Dimonte et al. (2004), such as \texttt{FLASH}. \fornax{}
yields results of comparable in quality to those of Arbitrary
Lagrangian-Eulerian (ALE) codes with interface reconstruction methods
(cf. Fig.~8 of Dimonte et al. 2004). That said we want to remark
that, as documented in Dimonte et al. (2004) and as we have confirmed in
preliminary calculations, the exact rate of mixing between the two
fluids depends to some extent on the spectrum and the character of the
initial perturbations. We have not fine-tuned the initial conditions of
our simulations to match the experimental data.\footnote{We adopted the
initial conditions of Dimonte et al. (2004) for the N064 resolution, and
we have kept them fixed as we increased the resolution.}  Nevertheless,
we cannot exclude the possibility that the better agreement with the experimental data
of \fornax{} compared to the results from the other Eulerian codes
presented in Dimonte et al. (2004) is due partly to the different choice
of initial conditions.

\begin{figure}
  \begin{center}
    \includegraphics[width=0.3\textwidth]{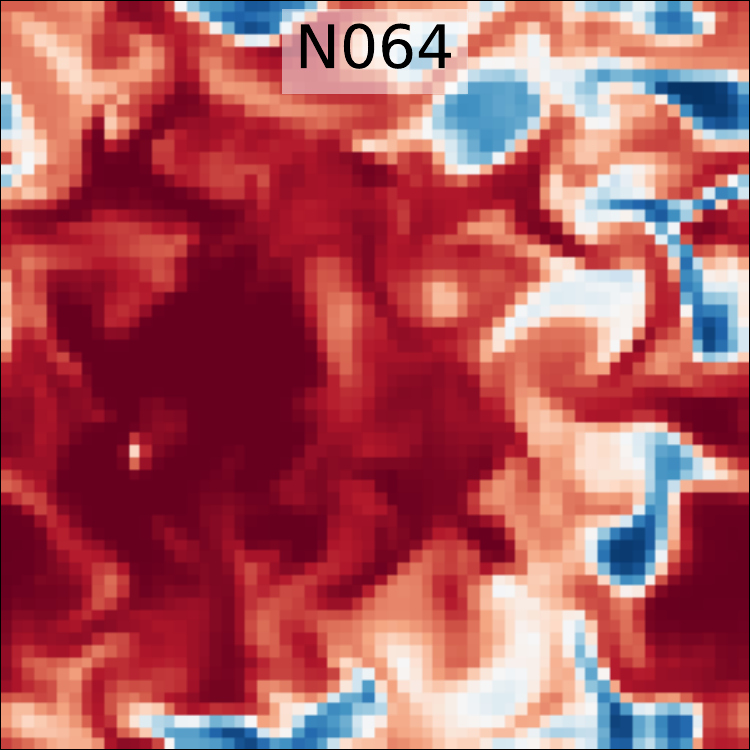}
    \includegraphics[width=0.3\textwidth]{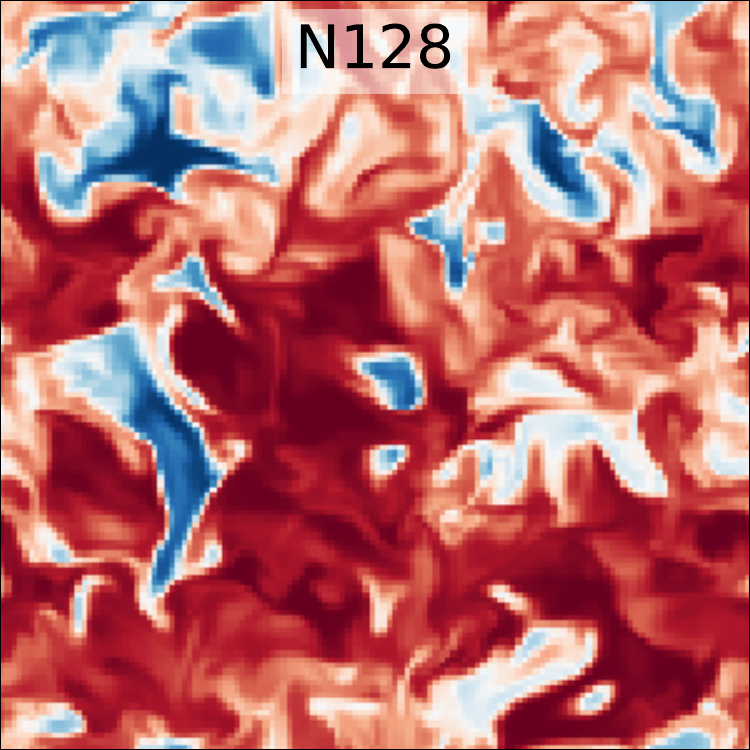}
    \includegraphics[width=0.3\textwidth]{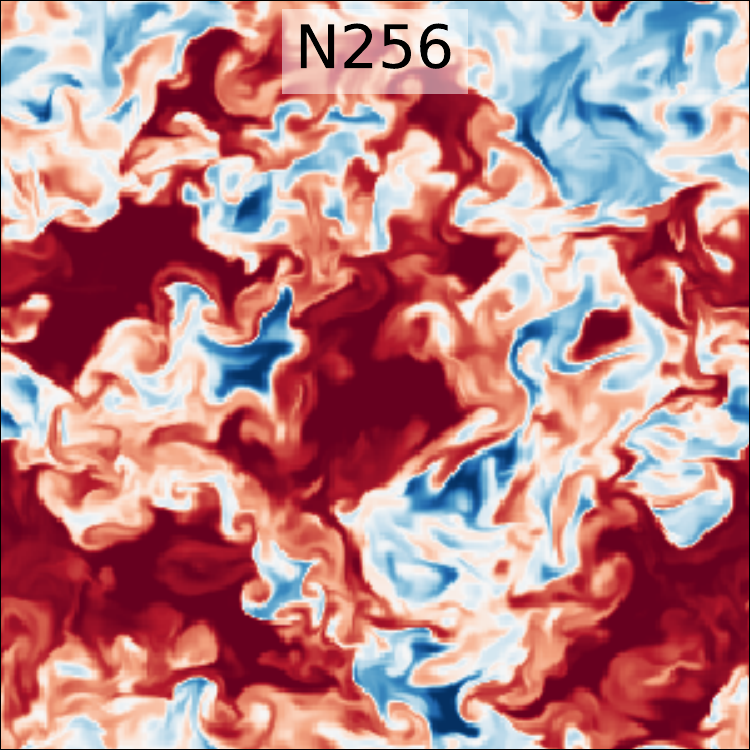}
    \includegraphics[width=0.3\textwidth]{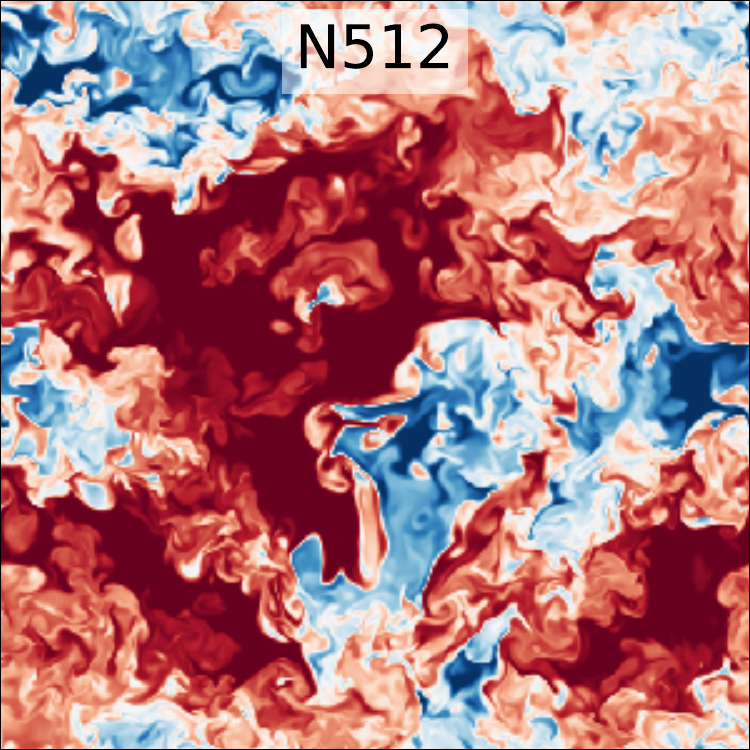}
  \end{center}
  \caption{Passive tracer concentration in the $xy$-plane for the
  multi-mode 3-d Rayleigh-Taylor test at time $A g t^2/L \simeq 14$ for
  different resolutions. Note the appearance of small-scale flow
  structures as the resolution is increased.}
  \label{fig:rt.3d.xy}
\end{figure}
 
We analyze the character of the fluctuations in $f_h$ on the original
interface at $z = 0$ at an advanced time $A g t^2/L \simeq 14$
(Fig.~\ref{fig:rt.3d.xy}). We find good qualitative agreement between
the two highest resolution simulations. However, there are quantitative
differences. Particularly noticeable is the appearance of smaller scale
fluid features as the resolution is increased. This is expected since we
do not include an explicit viscosity in our simulations. Instead, the
dissipation scale is related to the grid resolution, so that higher
resolution simulations have a larger effective Reynolds number.

\begin{figure}
  \includegraphics[width=\columnwidth]{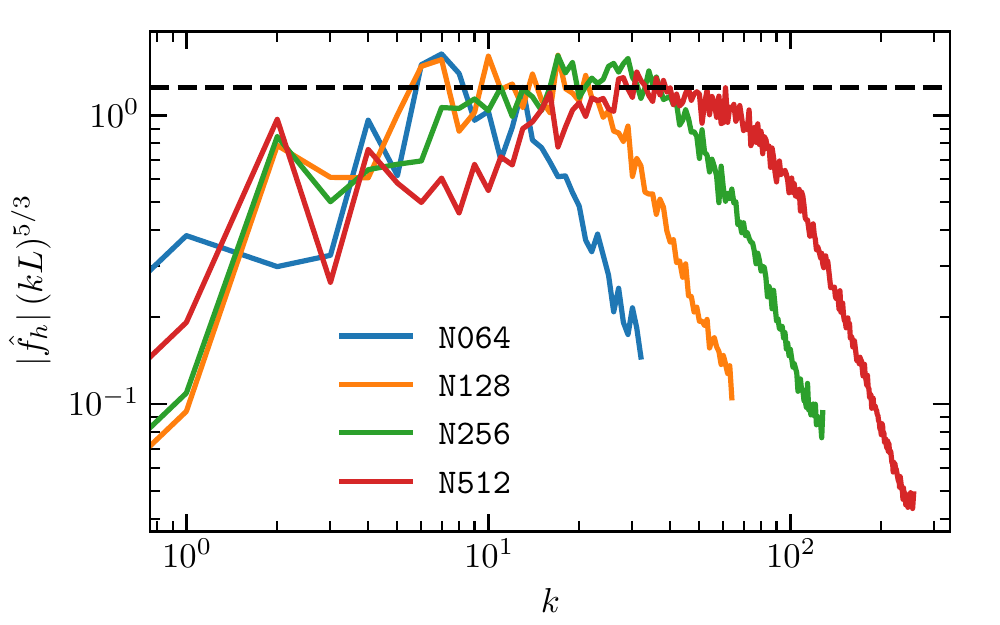}
  \caption{Power-spectra of the heavy fluid concentration on the initial
  interface $z = 0$ at $A g t^2/L\simeq 14$. The spectra are normalized
  to have unit norm and then compensated by $k^{5/3}$ so that regions
  with Kolmogorov scaling appear flat. As the resolution is increased,
  a progressively larger extent of the inertial range is recovered.}
  \label{fig:rt.3d.spec}
\end{figure}

We quantify this observation by computing the power spectrum of
$f_h$ on the original interface at $z = 0$ at $A g t^2/L\simeq
14$. We define the 2-d spectrum of $f_h$ as
\begin{equation}\label{eq:rt.3d.spec2d}
  \hat{f}_h(\mathbf{k}) = \iint 
    f_h(\mathbf{x})|_{z = 0} \exp\left(2 \pi \mathrm{i}
    \mathbf{k}\cdot\frac{\mathbf{x}}{L}\right) \mathrm{d}^2 \mathbf{x}
\end{equation}
and the 1D spectrum as
\begin{equation}\label{eq:rt.3d.spec1d}
  \hat{f}_h(k) = \iint \delta(\mathbf{k} - k)\, \hat{f}_h(\mathbf{k})\,
    \mathrm{d}^2\mathbf{k}\,.
\end{equation}
Because of the periodic boundary conditions, the 2-d spectrum is
nontrivial only for $\mathbf{k} = (k_x, k_y)$ with $k_x$, $k_y$
integers. When computing the 1D spectrum, we convert the integral into a
weighted summation following Eswaran \& Pope (1988), i.e., we set
\begin{equation}
  \hat{f}_h(k) = \frac{2 \pi k}{N_k} \sum_{\|\mathbf{k}\| = k}
    \hat{f}_h(\mathbf{k})\,,
\end{equation}
where $N_k$ is the number of modes with $\kappa - 1/2 < \|\mathbf{k}\| < \kappa + 1/2$. Our
results are shown in Figure~\ref{fig:rt.3d.spec}.

We find that the spectrum appears converged for $k \sim 2$, as could
have been anticipated by looking at the dipolar component of the
oscillations in Figure~\ref{fig:rt.3d.xy}. However, we find that
fluctuations with scales $k \sim 10$ are progressively suppressed as the
resolution increases. The lack of convergence at these intermediate
scales can be attributed to nonlinear mode coupling becoming stronger as
the effective Reynolds number of the simulations increases. At smaller
scales, the flow becomes self-similar and develops an inertial range.
We find that the power spectrum follows the scaling expected from
Kolmogorov's theory of turbulence for a passively advected scalar field,
i.e., $|\hat{f}_h| \propto k^{-5/3}$ (Pope 2000). We also find
that, as the resolution increases, a progressively larger part of the
inertial range is uncovered. At scales of less than ${\sim}5$ grid
points the power spectrum drops due to the direct effect of
numerical viscosity.

\subsection{Kelvin-Helmholtz Instability}
\label{kh}

The Kelvin-Helmholtz (KH) instability is a two-dimensional instability arising 
from a velocity shear within a fluid. The instability results in vorticity, which 
can cascade into turbulence. As such, it is of particular interest in the study of 
core-collapse supernova, where cascading turbulence can contribute to shock revival.

We consider a fluid over the rectangular domain within $x \in [0,4]$ and $z \in [0,2]$,
with a setup similar to that found in Lecoanet et al. (2016) and 
with a density jump characterized by the initial conditions below. We conduct tests at 
low, medium, high resolutions, defined as $512 \times 256$, $1024 \times 512$, and $2048 \times 1024$,
respectively.

\begin{subequations}
\begin{align}
\rho & = 1 + \frac{\delta\rho}{2\rho_0}\,\left[\mathrm{tanh}\left(\frac{z-z_1}{a}\right)-\mathrm{tanh}\left(\frac{z-z_2}{a}\right)\right]\\
v & = u_0\times\left[\mathrm{tanh}\left(\frac{z-z_1}{a}\right)-\mathrm{tanh}\left(\frac{z-z_2}{a}\right)\right] \\
w & = \mathrm{A}\,\mathrm{sin}(\pi x)\times\left[\mathrm{exp}\left(-\frac{(z-z_1)^2}{\sigma^2}\right)+\mathrm{exp}\left(-\frac{(z-z_2)^2}{\sigma^2}\right)\right]\, ,
\end{align}
\end{subequations}
with $a = 0.05$, $\sigma=0.2$, $z_1 = 0.5$, and $z_2 = 1.5$. We define $v$ as the velocity in the 
$x$-direction and $w$ as the velocity in the $z$-direction.

The initial amplitude, $A$, of the perturbation in the vertical ($z$-direction) velocity ($w$) is set to 
0.01, and the initial density perturbation, $\delta \rho/\rho_0$, is set to 1.0. The initial 
pressure $P_0$ is 10, the initial lateral velocity constant, $u_0$, is set to 1.0, and the initial 
density $\rho$ is set to 1.0. We assume an adiabatic gas with adiabatic index $\gamma = 5/3$. 
The resulting Mach number ($M$) is ${\sim}0.25$, consistent with quasi-incompressible flow. 
As is often done in KH instability tests, we ignore gravity.

\begin{figure}
\centering
\includegraphics[width=\columnwidth]{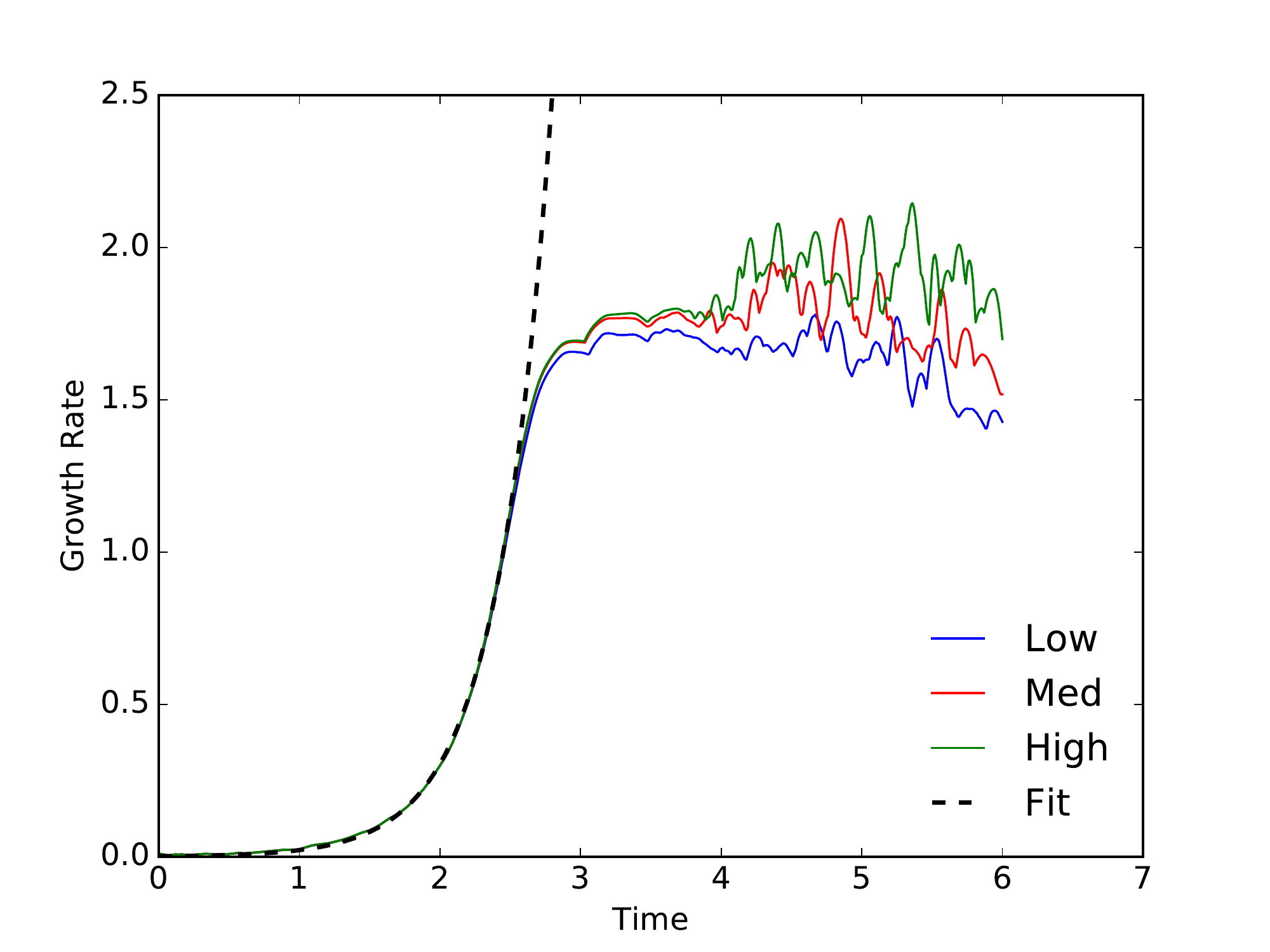}
\caption{In the context of the Kelvin-Helmholtz study, the growth rate of the maximum of $w$, the velocity in the $z$-direction as a
function of time for three different resolutions: low (blue), medium (red), and high
(green). In thick dashed black we plot the best fitting exponential to the growth
rate in the linear regime, which breaks down at $t \sim 2.6$.}
\label{fig:rho_test}
\end{figure}
 
In Figure~\ref{fig:rho_test}, we plot as a function of time the growth rate of the 
maximum of $w$, the velocity in the $z$-direction. The linear growth phase of the 
Kelvin-Helmholtz instability lasts until $t \sim 2.6$ (in our dimensionless units) 
for all models, which evolve virtually identically, independent of resolution, until the nonlinear regime. 

\begin{figure}
\centering
\includegraphics[width=0.9\columnwidth]{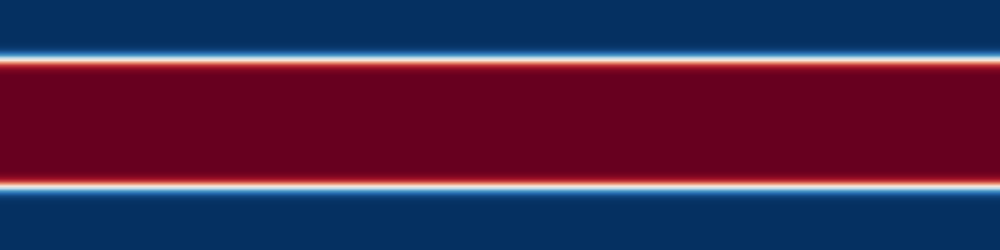}
\includegraphics[width=0.9\columnwidth]{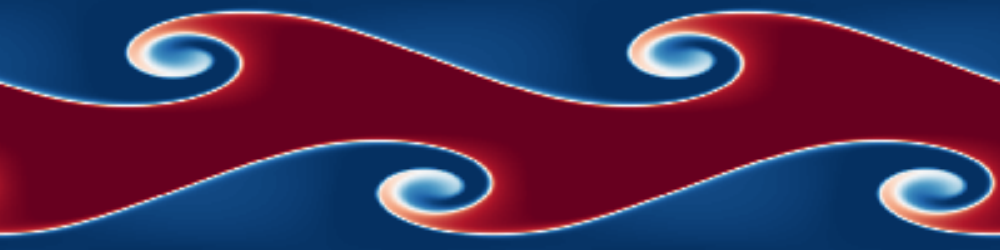}
\includegraphics[width=0.9\columnwidth]{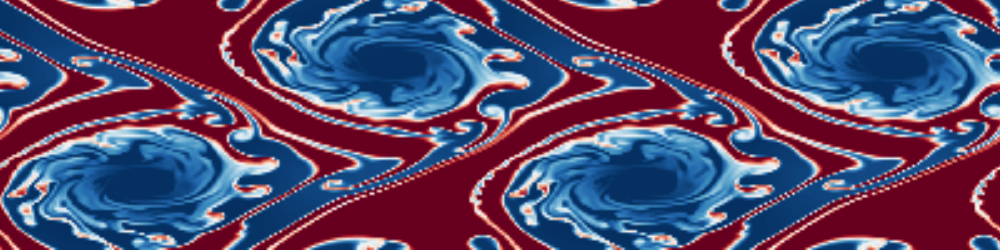}
\caption{Two-dimensional density evolution of the Kelvin-Helmholtz instability over the
grid at 3 different times: $t=0$, our initial condition (top), $t=3$ (middle), and $t=6$ (bottom).
The vertical velocity perturbations translate to density perturbations, forming vortices at later times.}
\label{fig:KHI}
\end{figure}

In Figure~\ref{fig:KHI}, we plot the 2-d density evolution at three times 
($t=0,3,6$ in dimensionless units) over our grid. Initial density 
perturbations evolve to form characteristic vortices at later times.

\subsection{Liska-Wendroff Implosion Test}
\label{lw_implosion}

\begin{figure*}
  \includegraphics[width=\textwidth]{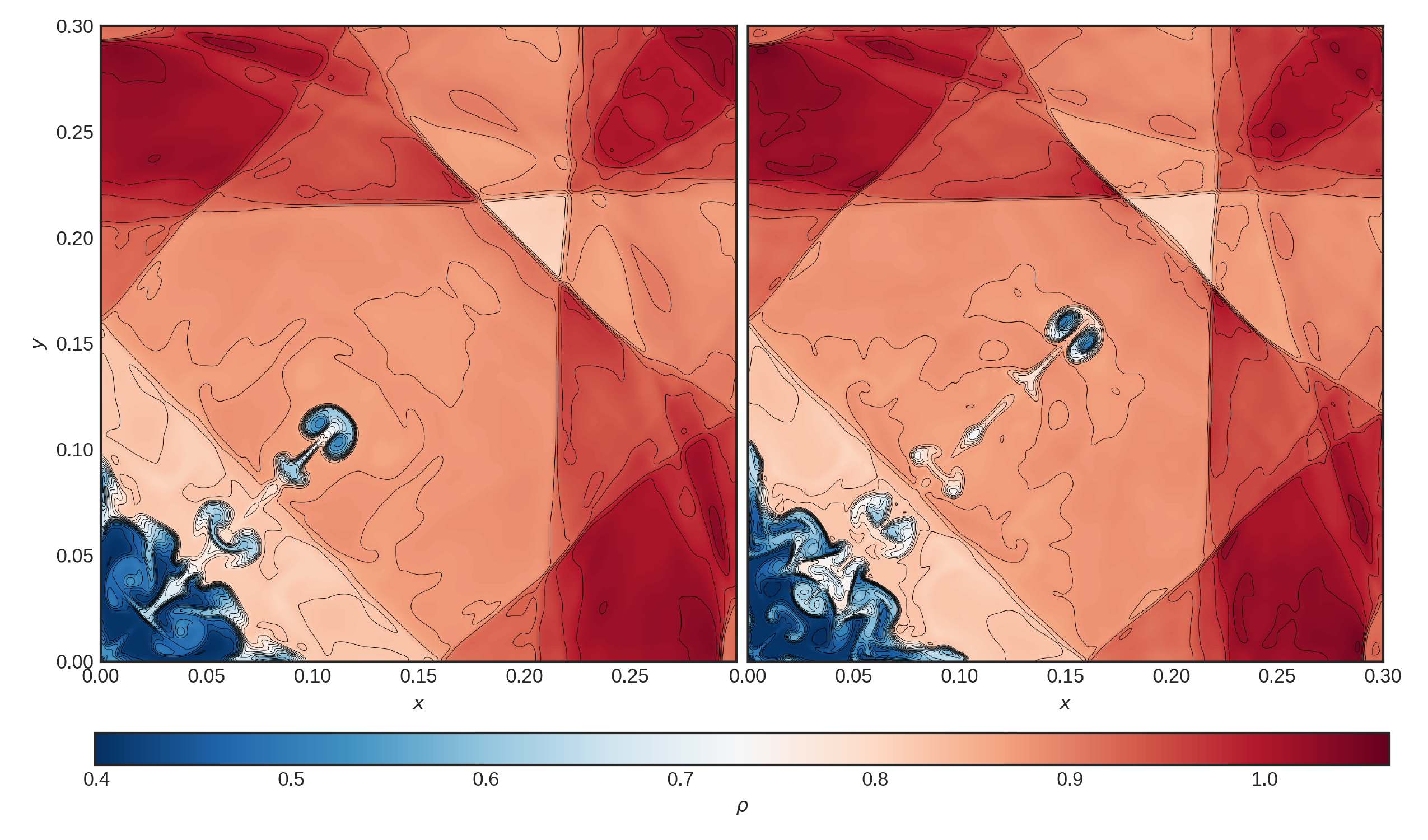}
  \caption{Results of the 2-d implosion problem of Liska \& Wendroff (2003b)
at $t=2.5$ on a $400\times400$ mesh using linear (left) and
parabolic (right) reconstruction methods.  Contours are shown
for $\rho=[0.35,1.1]$ at 31 equidistant values and are
exactly symmetric about $y=x$.  The longer jet in the
results at right reflect the lower numerical diffusion
characteristic of parabolic relative to linear reconstruction
methods.}
  \label{fig:lw_implosion}
\end{figure*}
 
\begin{figure}
  \includegraphics[width=\columnwidth]{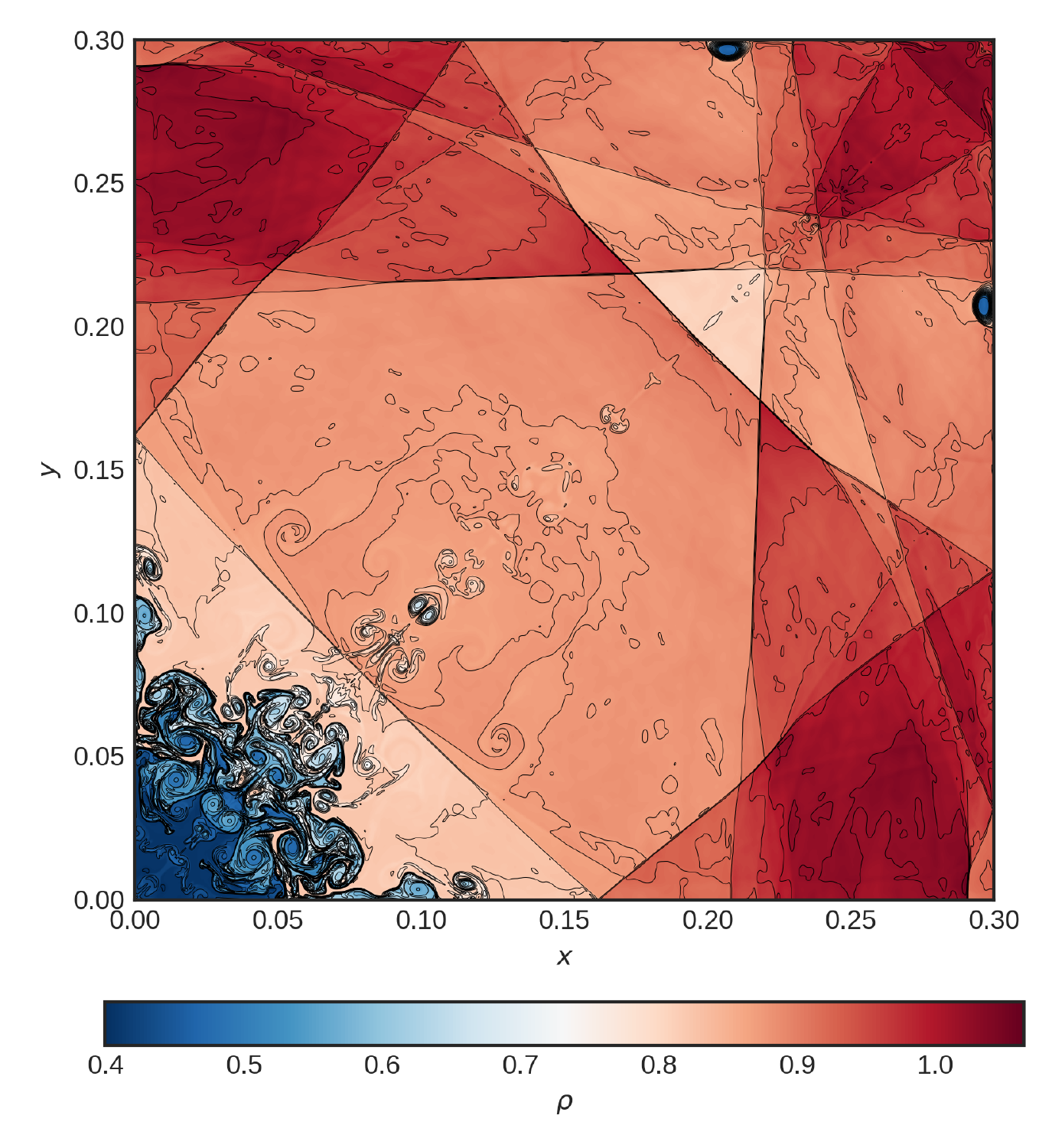}
  \caption{Results of the 2-d implosion problem of Liska \& Wendroff (2003) at $t=2.5$ on a $1600\times1600$ mesh using parabolic reconstruction. At this very high resolution, the vortices at the tip of the jet have propagated into the upper right corner and reflected, running back along the top and right sides. The overall structure of the solution, however, remains largely unchanged in comparison to the lower resolution results shown in Figure~\ref{fig:lw_implosion}.}
\label{fig:lw_implosion_hires}
\end{figure}

The two-dimensional implosion test of Liska \& Wendroff (2003b) consists of Sod-like initial conditions, but rotated by $45^\circ$ in a two-dimensional reflecting box. Specifically, the domain extends from 0 to 0.3 along both the $x$- and $y$-directions, with $(\rho,P)$ set to $(1,1)$ for $x+y>0.15$ and $(0.125,0.14)$ otherwise.  The gas is intially at rest and obeys an ideal gas equation of state with $\gamma=1.4$. The solution is evolved until $t=2.5$ on a uniform $400\times400$ mesh. This setup results in a shock, contact, and rarefaction moving from the initial discontinuity that reflect off the boundaries and interact. As highlighted by Stone et al. (2008), the correct solution to this problem includes a low-density jet that travels along the symmetry axis $y=x$. This problem is exquisitely sensitive to the preservation of this reflection symmetry, with the jet failing to form or wandering off $y=x$ if symmetry is violated even at the level of round off in the discretized update. The directionally-unsplit update in Fornax is able to maintain this symmetry, as shown in Figure~\ref{fig:lw_implosion}. Additionally, the distance traveled by the jet can be a useful measure of the numerical diffusion of contacts. This is illustrated in Figure~\ref{fig:lw_implosion} in the comparison between the results using linear and parabolic reconstructions. In fact, at much higher resolutions, the jet propagates into and interacts with the top, right corner of the domain, as shown in Figure~\ref{fig:lw_implosion_hires}.  Our results are comparable, though perhaps characteristic of slightly higher numerical diffusion, to the results presented in Stone et al. (2008).

\subsection{Pressureless Dust Collapse}
\label{Dust}

Here we investigate the nearly pressureless and homologous collapse of a uniformly dense cloud of dust under its own self-gravity.  This problem has a solution described in Colgate \& White (1966) and was used by M\"onchmeyer \& M\"uller (1989) to argue for the reconstruction of volume-averaged grid variables over the corresponding volume centroids, rather than the zone-centered coordinates.  At a time $t$, the outer radius of the cloud, $r_\mathrm{cl}(t)$, is given by the solution to the equation
\begin{equation}
\left(\frac{8\pi G}{3}\rho_0\right)t = \left[\frac{r_\mathrm{cl}}{r_0}\left(1 - \frac{r_\mathrm{cl}}{r_0}\right)\right]^{1/2} + \sin^{-1}\left(1 - \frac{r_\mathrm{cl}}{r_0}\right)^{1/2}\, ,  \label{eq:rcloud}
\end{equation}
where $\rho_0$ and $r_0$ are the cloud's initial density and radius, respectively.  Once $r_\mathrm{cl}$ is obtained, the cloud density $\rho_\mathrm{cl}(t)$ is given by
\begin{equation}
\rho_\mathrm{cl} = \rho_0 \left( \frac{r_\mathrm{cl}}{r_0}\right)^{-3}, \label{eq:rhocloud}
\end{equation}
and the radial velocity profile inside the cloud, $v_\mathrm{cl}(r,t)$, is given by
\begin{equation}
v_\mathrm{cl} = -r\left[\frac{8\pi G}{3} \rho_0 \left(\frac{r_0}{r_\mathrm{cl}} - 1\right)\right]^{1/2}\, .  \label{eq:vcloud}
\end{equation}
We use $\rho_0=10^9 \text{ g cm}^{-3}$, $r_0 = 6500 \text{ km}$, an adiabatic equation of state with $\gamma=5/3$, and evolve to $t_\mathrm{final} = 0.065 \text{ s}$.  The pressure is set to a negligibly small, constant value, $P_0$, such that $P_0 \ll (4\pi G/\gamma)\rho_0^2 r_0^2$, hence at $t_\mathrm{final}$, the outer edge of the cloud is collapsing at a Mach number of $\sim 80$.  Finally, we use a three-dimensional spherical dendritic grid with radial extent out to $r_\mathrm{max}=7000 \text{ km}$ and a resolution of $200 \times 64 \times 128$ zones.  The radial grid spacing is constant in the interior with $\Delta r \approx 0.5 \text{ km}$ out to approximately $15 \text{ km}$, then smoothly transitions to logarithmic spacing out to $r_\mathrm{max}$.

\begin{figure}
\includegraphics[width=\columnwidth]{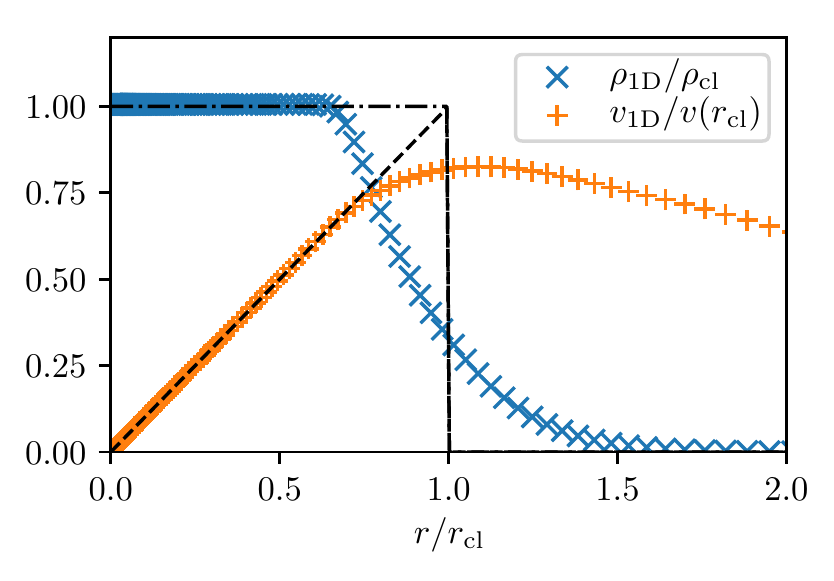}
\caption{Results of the three-dimensional pressureless dust collapse test 
at time $t_\mathrm{final}=0.065\text{ s}$.  The computed density, $\rho$ 
(blue exes), and velocity, $v$ (orange crosses) are scaled by the cloud density 
and velocity, $\rho_\mathrm{cl}$ and $v_\mathrm{cl}$, respectively, 
obtained from the semi-analytic solution (see text for details).  The 
semi-analytic solutions for the radial profiles of density (dash-dotted) 
and velocity (dashed) are also shown.  At a given radius, the data in 
all angular zones are identical, indicating perfect spherical symmetry 
is maintained on infall.  Results from one- and two-dimensional tests are also identical.}
\label{fig:dustcollapse}
\end{figure}

Figure~\ref{fig:dustcollapse} shows the density, $\rho$, and radial velocity, $v$, at 
time $t_\mathrm{final}$.  We scale these by the cloud density and radius, $\rho_\mathrm{cl}$ 
and $v_\mathrm{cl}$, respectively, as obtained from equations~\eqref{eq:rhocloud} and~\eqref{eq:vcloud}, 
which in turn depend on the numerical solution for $r_\mathrm{cl}$ obtained from equation~\eqref{eq:rcloud} 
at time $t_\mathrm{final}$.  As evident from the figure, the cloud density remains 
constant throughout the cloud, and velocity remains linear.  There is some smearing at 
the outer edge of the cloud due to finite pressure gradients, but otherwise, the results 
are comparable to those obtained by other authors (Monchmeyer \& M\"uller 1989; Mignone 2014).  
This test demonstrates the code's ability to obtain accurate fluxes from reconstruction in the volume coordinate.

\subsection{One-dimensional Core-collapse Mechanical Energy Conservation Test}
\label{1d_conservation_hydro}

\begin{figure}
\includegraphics[width=\columnwidth]{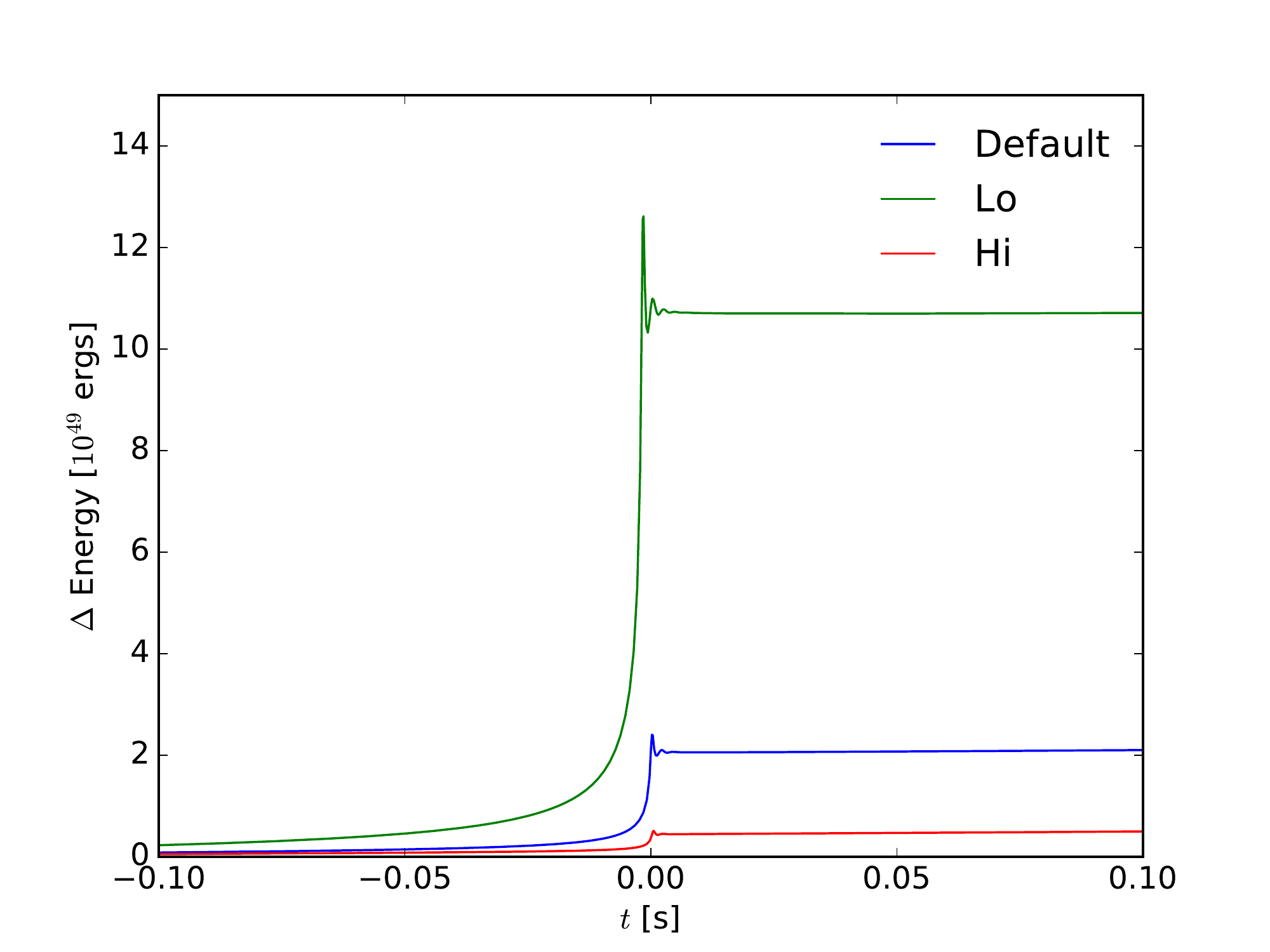}
\caption{The change in total mechanical energy (in $10^{49}$ erg), defined as the sum of gravitational potential,
internal, and kinetic energies, as a function of time after bounce (in seconds) for three different resolutions:
Lo (green, 304 radial cells), Default (blue, 608 radial cells), and Hi (red, 1216 radial cells). Energy is
conserved to better than $2\times 10^{49}$ erg at bounce for the default resolution, and ${\sim}3\times 10^{48}$ erg at the higher resolution. See text for a discussion.}
\label{energy_hydro}
\end{figure}

We perform several spherical 1-d core-collapse hydrodynamical simulations
without transport or general-relativistic corrections and with Newtonian
gravity to study energy conservation in \fornax{}. For this test, we use the SFHo
equation of state (Steiner et al. 2013).
The total mechanical energy is defined as the sum of the kinetic, internal, and
gravitational potential energies on the grid, extending out to $20,000\text{ km}$. 
The latter is calculated as:
\begin{equation}
E_\mathrm{grav} = -\frac{1}{8\pi G} \int | \nabla \Phi |^2 d^3 x\, , \label{eq:Egrav}
\end{equation}
where $\Phi$ is the gravitational potential.  
We take the jump in total energy around bounce as an apt measure, since
gravitation is not implemented in automatically conservative form (the
other terms are) and bounce is the most problematic phase of core-collapse
simulations vis-\`a-vis total energy conservation. Moreover, we continue
the calculations to ${\sim}100\text{ ms}$ after bounce for three
resolutions:  Lo (304 radial cells, green), Default (608 radial cells,
blue), and Hi (1216 radial cells, red). We find that energy is conserved
from just before bounce to just after bounce to approximately 10$^{50}$
erg, $2 \times 10^{49}\text{ erg}$, and ${\sim}3 \times 10^{48}\text{ erg}$,
respectively, for the three resolutions (Figure \ref{energy_hydro}).

Earlier generations of codes, including AGILE-BOLTZTRAN (Liebend\"orfer et
al. 2004) and BETHE-hydro (Murphy \& Burrows 2008b), saw total energy
shifts from just before bounce to over 100 milliseconds after bounce of
on the order of 10$^{51}$ erg. For comparison, CHIMERA (Bruenn et al., 2009; Bruenn
et al., 2016) conserves energy to 0.5 Bethe ($1\text{ B} = 10^{51}\text{ erg}$) over ${\sim}1\text{ s}$ post-bounce. Our default resolution from before
bounce to ${\sim}1\text{ s}$ after bounce conserves total energy to $0.05\text{ B}$,
an order of magnitude better. Importantly, we find that the total energy 
is conserved to ${\sim} 10^{47}\text{ erg}$ from $10\text{ ms}$ post-bounce
to $100\text{ ms}$ post-bounce.  By comparison, using CoCoNUT for a Newtonian core-collapse
simulation, and subtracting out neutrino losses, M\"uller et al. (2010)
report energy conservation of ${\sim}2 \times 10^{49}\text{ erg}$ from just before
bounce to just after bounce, and subsequent non-conservation for the
following tens of milliseconds at the ${\sim} 10^{48}$ level. In \S\ref{multid_conservation_total} we perform an analogous core-collapse test using multi-dimensional, multi-species transport.

\section{Radiation Tests}
\label{rad_tests}

\subsection{Multi-dimensional Core-collapse Total Energy Conservation Test}
\label{multid_conservation_total}

\begin{figure}
\includegraphics[width=\columnwidth]{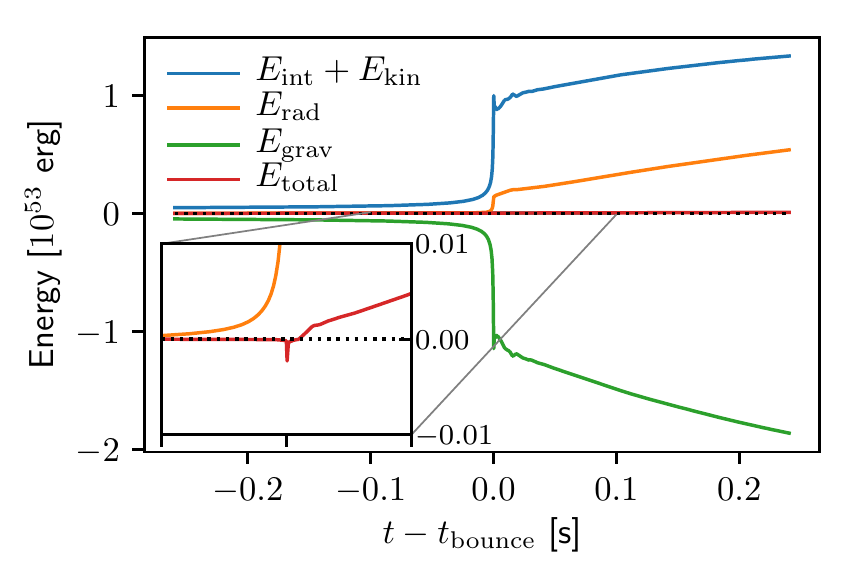}
\caption{We plot the gravitational potential (green), internal and kinetic (blue), and lab-frame radiation (orange) energies (in 10$^{53}$ erg), as well as the total energy (red) defined as their sum, versus time after bounce (in seconds) for a 2-d, multi-species core-collapse simulation at our default resolution (608 radial cells by 128 polar-angle cells).  In this run, we use the 16-$M_\odot$ progenitor model of \cite{Woos07} (which was studied in \cite{vart2018}) with Newtonian gravity and our dendritic grid decomposed over 1024 processors.  At each step of the calculation, we account for the flux of each energy component onto and off of the grid through the outer radial boundary.  The inset zooms in on the total energy 100 ms before and after bounce.  The total energy is conserved to within a few $\times$${\sim}10^{51}\text{ erg}$ (to within 0.5\% relative to $E_\mathrm{grav}$) during the entire 500 ms of the run.}
\label{fig:energy_radhydro}
\end{figure}

{
As a first test of our full radiation-hydrodynamics code, we test the conservation of total energy by performing a 2-d core-collapse calculation on a grid with a resolution of 608 radial cells by 128 polar-angle cells.  We use Newtonian monopolar gravity, the SFHo equation of state, 12 energy groups per species, and our dendritic grid decomposed over 1024 processors.  We track the gravitational potential energy (defined as in Eq.~\ref{eq:Egrav}), internal plus kinetic energy, and the total lab-frame radiation energy at each time step, accounting for the flux of each energy component through the outer radial boundary.  

The results of this test are shown in Figure~\ref{fig:energy_radhydro}.  As previously 
demonstrated in \S\ref{1d_conservation_hydro}, we experience a glitch in 
the total energy at bounce, since gravity is treated in non-conservation form.  
With radiation transport, there are other additional sources of energy 
non-conservation, e.g., due to the finite tolerance of the implicit solver 
or due to truncation of the comoving-to-lab frame transformation in powers 
of $v/c$.  However, as Figure~\ref{fig:energy_radhydro} demsontrates, the 
total energy remains well controlled to within a few$\times$${\sim}10^{50}\text{ erg}$ 
(or to within {0.5\%} relative to $E_\mathrm{grav}$) throughout the $500\text{ ms}$ 
of the simulation run.\footnote{At the same time, the total mass is conserved 
to within a relative error of ${\sim}2\times 10^{-8}$.}  Energy conservation 
within such low levels is especially important considering the high-velocity 
energy fluxes crossing the internal refinement boundaries of our dendritic grid 
in both the radial and angular directions and across a distribution of many processors.
}

\subsection{Doppler Shift Test}

To test the advection of radiation energy in frequency space caused by strong 
spatial variations in the gas velocity, we perform the Doppler shift test of Vaytet et al. (2011).  
We use a one-dimensional domain with 50 equally-spaced zones between $x=0\text{ cm}$ and $x=10\text{ cm}$ 
with constant background density of $\rho = 1\text{ g cm}^{-3}$ and zero opacity.  The velocity 
varies smoothly between $v=0\text{ cm s}^{-1}$ at the left edge of the domain and $v=A$ in the 
center, where $A=5\times 10^7\text{ cm s}^{-1}$.  At the left boundary, we introduce a radiation 
field having a Planck spectrum at a temperature $T=1000\text{ K}$, which we resolve using 20 
equally-spaced frequency groups from $\nu=0\text{ Hz}$ to $\nu=2\times 10^{14}\text{ Hz}$, with 
two additional groups to account for radiation in the frequency range $2\times 10^{14}\text{ Hz} \to \infty$.  
We evolve the system for 10 light-crossing times in order to reach a steady state.

\begin{figure}
\includegraphics[width=\columnwidth]{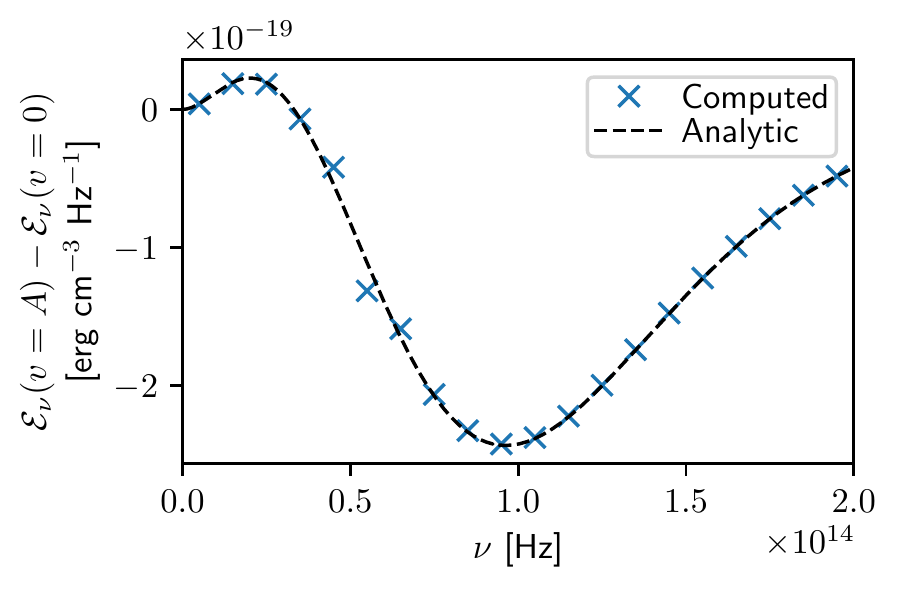}
\caption{Difference between the shifted and unshifted energy spectra for the Doppler shift test.  
The energy difference in the computed solution (blue exes) and the analytic solution (black dashed line) 
are plotted.  The relative error in the solution is less than $6.5\%$ except near the spectral peak 
at $\sim 5\times10^{13}\text{ Hz}$ where it jumps up to $23\%$ due to slope limiting of the 
frequency-advected spectrum, which results in a local first-order error.}
\label{fig:doppler}
\end{figure}

In Figure~\ref{fig:doppler}, we plot the difference between the Doppler-shifted spectral 
energy distribution at the center of the domain, $\mathcal{E}_\nu(v=A)$, and the unshifted 
distribution at the left boundary, $\mathcal{E}_\nu(v=0)$.  We also plot the analytic solution, 
obtained from the difference between the relativistic Doppler shift of the incident spectrum 
into the frame comoving at velocity $v=A$ and the unshifted spectrum.  To do this, we first 
multiply the lab-frame frequencies by the relativistic Doppler factor,
\begin{equation}
\sqrt{\frac{1+\beta}{1-\beta}}\, ,
\end{equation}
where $\beta = A/c$, to obtain the corresponding comoving-frame frequencies.  Then, we 
set $\mathcal{E}_\nu = (4\pi/c)B_\nu$, with frequencies evaluated in the comoving frame, and 
finally divide by the relativistic Doppler factor to transform the resulting spectrum back 
into the laboratory frame.

The computed solution agrees well with the analytic solution at almost all frequencies.  The 
relative error is slightly larger near $\nu \sim 5\times 10^{13}\text{ Hz}$, the peak of 
the Planck spectrum at $T=1000\text{ K}$, where TVD slope limiting in our frequency advection 
step results in a local first-order error.  With 20 groups, the relative error is less than 
$\sim 6.5\%$ away from the spectral peak, where it jumps up to $\sim 23\%$.  Running the 
same test with 40 groups reduces the maximum relative error to $0.85\%$ away from the 
spectral peak, where it jumps to $5.6\%$.

\subsubsection{Continuity of Lab-Frame Flux at Transparent Shocks}
\label{continuity}

\begin{figure*}
\includegraphics[width=\columnwidth]{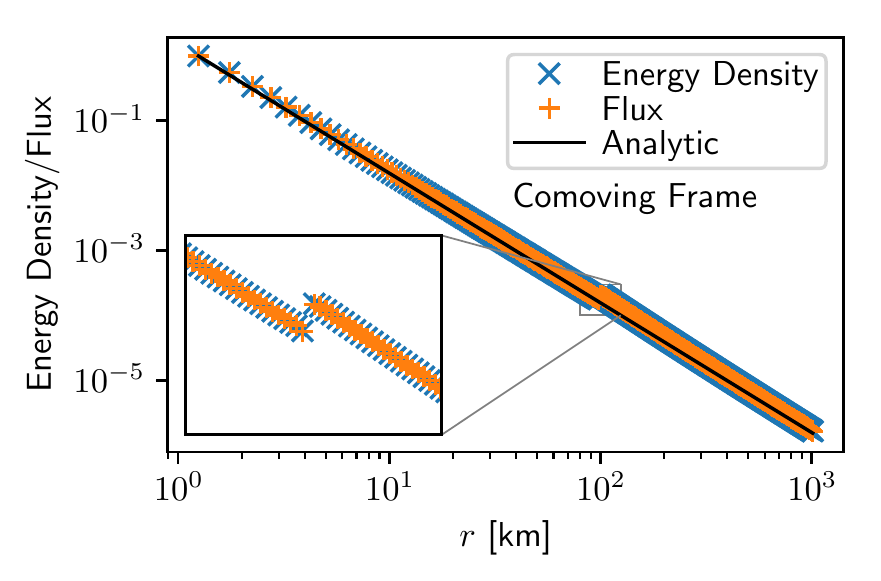}
\hfill
\includegraphics[width=\columnwidth]{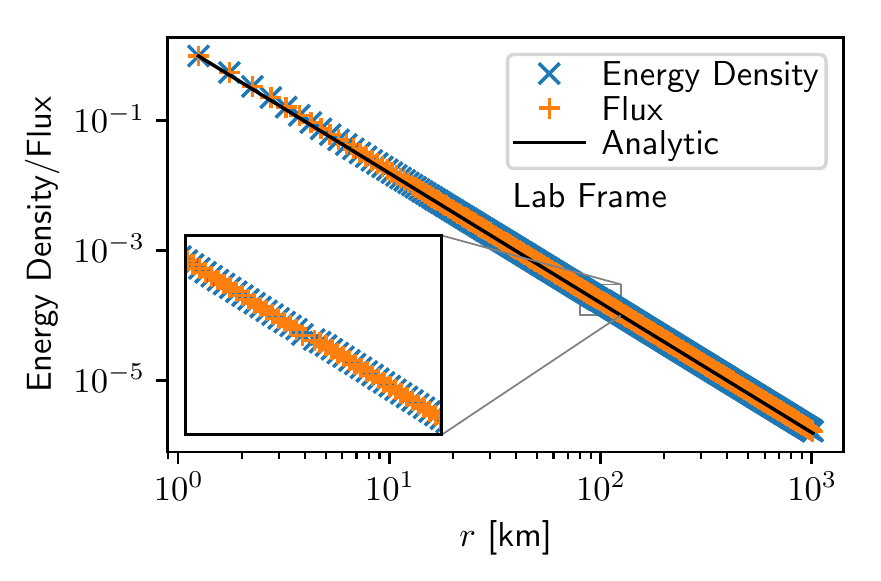}
\caption{{\bf Left:} The profiles of the radiation energy density and flux, scaled to their values in the first radial zone, in the comoving frame versus radius for a steady-state constant point luminosity source.  A non-trivial radial velocity field has been imposed that includes a strong standing shock (see Eq.~\ref{eq:vprofile}). {\bf Right:} The corresponding scaled radiation energy density and flux profiles Lorentz-transformed into the laboratory frame. We note that the matter is assumed to be transparent, hence the laboratory-frame profiles of the scaled radiation quantities should go as $1/r^2$, which we indicate by the solid line. See text for a discussion.}
\label{jump_plot}
\end{figure*}

To demonstrate further that \fornax{} accounts properly for frame effects, we show a comparison of the radial profile of the total radiation energy density and flux calculated in the steady state for a constant point luminosity source at the spherical grid center.  We set the material opacity to zero and impose a core-collapse-like radial velocity field given by
\begin{equation}
v(r) = \left\{ \arraycolsep=1.4pt\def\arraystretch{2.2} \begin{array}{ll}
-0.1 \, v_\mathrm{shock} \frac{r}{r_\mathrm{shock}}, & r < r_\mathrm{shock} \\
-v_\mathrm{shock} \frac{\sqrt{r_\mathrm{max}/r} - 1}{\sqrt{r_\mathrm{max}/r_\mathrm{shock}} - 1}, & r \ge r_\mathrm{shock}
\end{array} \right. , \label{eq:vprofile}
\end{equation}
where $r_\mathrm{shock} = 100\text{ km}$ is the location of a strong standing shock of maximum velocity $v_\mathrm{shock} = 0.1 c$ and $r_\mathrm{max} = 1000\text{ km}$ is the maximum radius.  

Since \fornax{} calculates radiation quantities in the comoving frame, the corresponding energy density and flux profiles will have discontinuities at shocks.  However, since the radiation is uncoupled from the matter, the radiation quantities should possess a smooth profile that goes as $1/r^2$ in the laboratory frame.  Thus, performing a simple Lorentz transformation back to the laboratory frame should eliminate the discontinuity.  Figure~\ref{jump_plot} depicts the radial profiles of the radiation energy density and flux, each scaled to their values in the first radial zone, with the left and right panels depicting the profile in the comoving and laboratory frames, respectively.  The right-hand-side panel shows that the Lorentz transformation has indeed removed the discontinuity.  This simple test is one way of demonstrating that \fornax{} handles the velocity-dependent terms in radiation Equations~\eqref{eq:rad_E} and~\eqref{eq:rad_mom2} properly.

\subsection{Frequency-Dependent Opacity Test}

\begin{figure}
\includegraphics[width=\columnwidth]{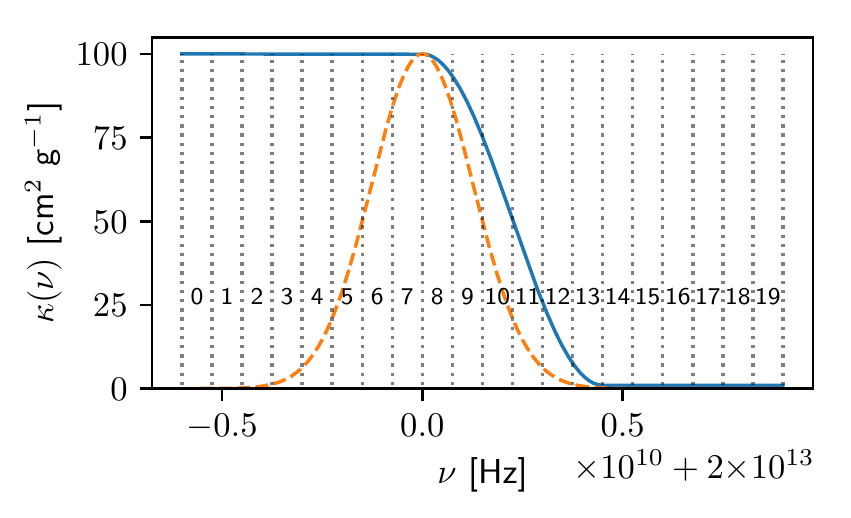}
\caption{Opacity function (solid blue line) for the frequency-dependent opacity test.  
The opacity changes from $\kappa_\nu = 100\text{ cm}^2\text{ g}^{-1}$ for $\nu<2\times10^{13}\text{ Hz}$ to $\kappa_\nu=1\text{ cm}^2\text{ g}^{-1}$, 
transitioning smoothly over a frequency range of width $\Delta \nu=4.5\times 10^9\text{ Hz}$.  
Overplotted is a scaled representation of the incident energy spectrum (dashed orange line), 
which peaks at $\nu=2\times 10^{13}\text{ Hz}$ and has FWHM equal to $2/3 \Delta\nu$.  The numbered 
frequency groups are indicated by vertical dotted lines.  Groups 8-13 comprise the opacity transition region.}
\label{fig:freq_dep_opac_kappa}
\end{figure}

To test the effect of a strong velocity gradient on a frequency-dependent opacity 
function, we perform a modified version of a test by Vaytet et al. (2011).  We use a 
one-dimensional domain with $x \in [0,1]\text{ cm}$ resolved over 100 zones and set a 
fixed velocity profile $v = \mathcal{D}x$, where $\mathcal{D}$ is either $0\text{ s}^{-1}$ 
or $10^7\text{ s}^{-1}$.  We use a density profile of $\rho = 1/(\mathcal{C}x)$ with $\mathcal{C} = 0.2\text{ cm}^2\text{ g}^{-1}$ 
and set the gas temperature to $T_0 = 3\text{ K}$ everywhere.  We use a frequency-dependent opacity 
with $\kappa_\nu = 100\text{ cm}^2\text{ g}^{-1}$ for $\nu < 2\times 10^{13}\text{ Hz}$, transitioning 
smoothly to $\kappa_\nu = 1\text{ cm}^2\text{ g}^{-1}$ over a region of width $\Delta \nu = 4.5\times 10^9\text{ Hz}$ 
as shown in Figure~\ref{fig:freq_dep_opac_kappa}.  At the left boundary, we inject a Gaussian 
radiation spectrum that is normalized to have total energy $a_\mathrm{R} T_1^4$ with $T_1 = 1000\text{ K}$, 
is peaked at $\nu = 2\times 10^{13}\text{ Hz}$,  has FWHM equal to two-thirds the width of the transition 
region, and is resolved over 20 equally-spaced frequency groups (also shown in Figure~\ref{fig:freq_dep_opac_kappa}).  
For each case, we evolve only the radiation system for one light crossing time, keeping the hydrodynamics frozen.

\begin{figure}
\includegraphics[width=\columnwidth]{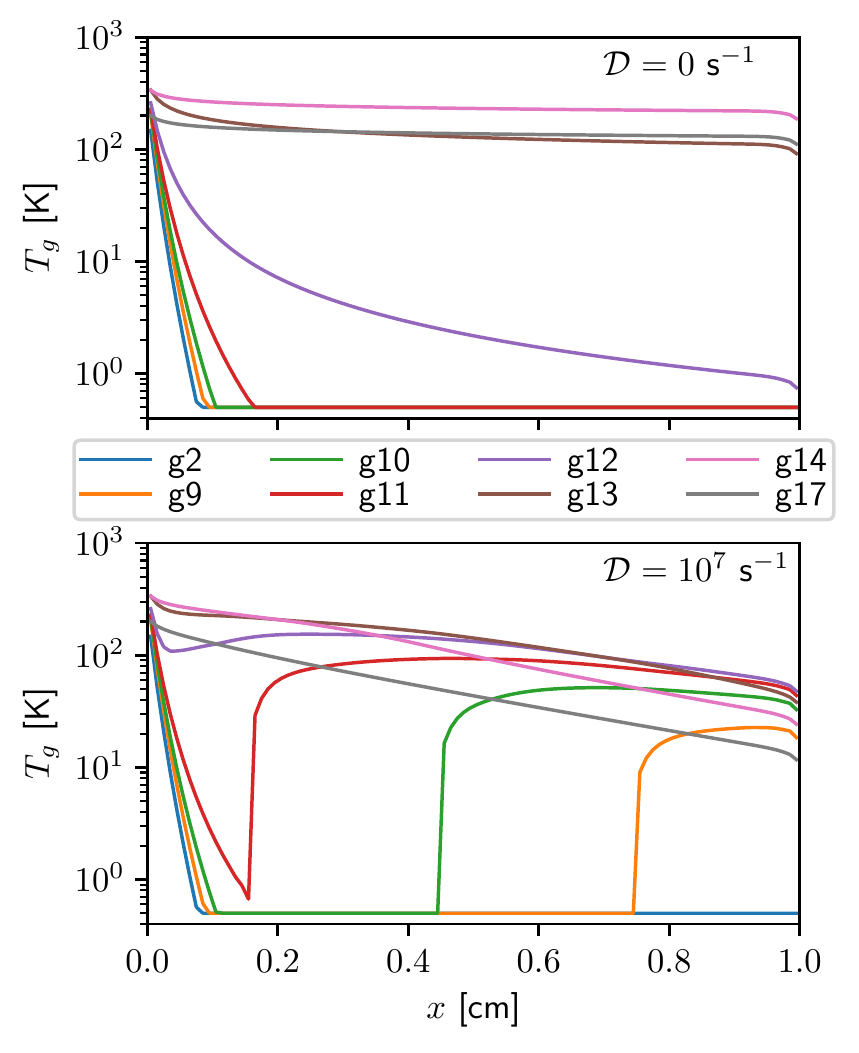}
\caption{Selected group temperatures, $T_g \equiv (\mathcal{E}_g/a_\mathrm{R})^{1/4}$ for the 
frequency-dependent opacity test.  The velocity profile is given by $v=\mathcal{D}x$, 
where either $\mathcal{D}=0\text{ s}^{-1}$ for the zero-gradient case (top) or 
$\mathcal{D} = 10^7\text{ s}^{-1}$ for the strong velocity gradient case (bottom).  For 
the zero-velocity case, only radiation in groups with very low opacities can stream 
freely across the domain, but for the case with a strong velocity gradient, energy in 
groups 9-12 inside the opacity transition region are shifted to higher frequencies where 
the opacities are smaller.  This allows these groups to stream more freely once optically 
thin conditions are reached.}
\label{fig:freq_dep_opac}
\end{figure}

Figure~\ref{fig:freq_dep_opac} shows selected group temperatures, $T_g \equiv (\mathcal{E}_g/a_\mathrm{R})^{1/4}$, 
for the cases with $\mathcal{D} = 0\text{ s}^{-1}$ (top) and $\mathcal{D} = 10^7\text{ s}^{-1}$ (bottom).  
In the zero-velocity case, the group energies are unshifted, and only those groups with low enough optical 
depth are able to stream freely across the domain.  By contrast, in the case with a strong velocity gradient, 
the comoving-frame frequencies are Doppler-shifted to the right where the opacities are lower.  Thus, 
the energy in groups 9-12 in the transition region of the opacity function are able to stream more freely 
once optically-thin conditions are reached.

\subsection{Multi-Group Radiation Pulse Advection}
\label{rad_pulse}

To study the behavior of \fornax{} in the strong diffusion regime, where radiation is 
transported both by diffusion through the gas and advection along with it, we perform 
a variation of the pulse advection test of Krumholz et al. (2007) (see also Zhang et al. 2013), 
modified for our explicit transport solver.  We initialize the gas temperature and density profiles as
\begin{eqnarray}
T &=& T_0 + (T_1 - T_0)\exp \left(-\frac{x^2}{2w^2}\right), \\
\rho &=& \rho_0 \frac{T_0}{T_1} + \frac{a_\mathrm{R} \mu}{3 k_\mathrm{B}} \left(\frac{T_0^4}{T} - T^3 \right),
\end{eqnarray}
where $T_0$ and $\rho_0$ are the background temperature and density, respectively, 
$T_1 = 2T_0$ is the peak of a Gaussian pulse of thermal energy of width $w$, 
and $\mu = (m_e + m_p)/2$ is the mean particle mass for ionized hydrogen.  Following Zhang et al. (2013), 
we use a temperature- and frequency-dependent absorption opacity of the form
\begin{equation}
\kappa_\nu(T) = \kappa_0 \left( \frac{T}{T_1} \right)^{-1/2} \left( \frac{\nu}{\nu_1} \right)^{-3} \left[ 1 - \exp \left(-\frac{h\nu}{k_\mathrm{B}T}\right)\right]\, ,
\end{equation}
where $\kappa_0$ is some normalization constant and we take $\nu_1 = 2.821 k_\mathrm{B} T_1/h$ 
to approximate the spectral peak of a Planck distribution at temperature $T_1$.  The problem 
dimensions are set by the choice of an arbitrary length scale, $w$, and by the values of the 
non-dimensional parameters $\beta \equiv v/c$, $\mathcal{M} \equiv v/a_0$, $\mathcal{P} \equiv a_\mathrm{R}T_0^4/(\rho_0 a_0^2)$, 
and $\tau \equiv \kappa_\mathrm{P}(T_1) w$, where $\kappa_\mathrm{P} = 3.457 \,\kappa_0 (T/T_1)^{-3.5}$ is 
the Planck-mean opacity such that $\tau$ represents the optical depth of the pulse.  For our version of
this problem, we choose $w = 1\text{ cm}$, $\beta = 0.01$, $\mathcal{M}=0.1$, $\mathcal{P} = 0.1$, and $\kappa_0 = 0.2892 \,\tau \text{ cm}^{-1}$, 
where $\tau$ is chosen to control which diffusion regime characterizes the flow.

The product $\beta\tau$ is equal to the ratio of the radiation diffusion and advection time scales.  
For $\beta\tau \la 1$, radiation diffusion dominates and the system is in the so-called ``static diffusion" regime, 
but for $\beta\tau \gg 1$, advection dominates and the system is in the ``dynamic diffusion" regime.  
In either case, although the velocity-dependent radiation work and advection terms will be very different 
depending on the frame in which the equations are solved, the results should be frame-independent.  
Therefore, for both the static and dynamic diffusion regimes, we compare runs in which the 
radiation pulse is initially at rest to runs in which the radiation pulse is advected over 
twice its initial width and shifted back to lie on top of the unadvected results.  In either case, 
with $\tau \gg 1$, the radiation is in thermal equilibrium with the gas, and the total pressure is initially constant.

\begin{figure}
\includegraphics[width=\columnwidth]{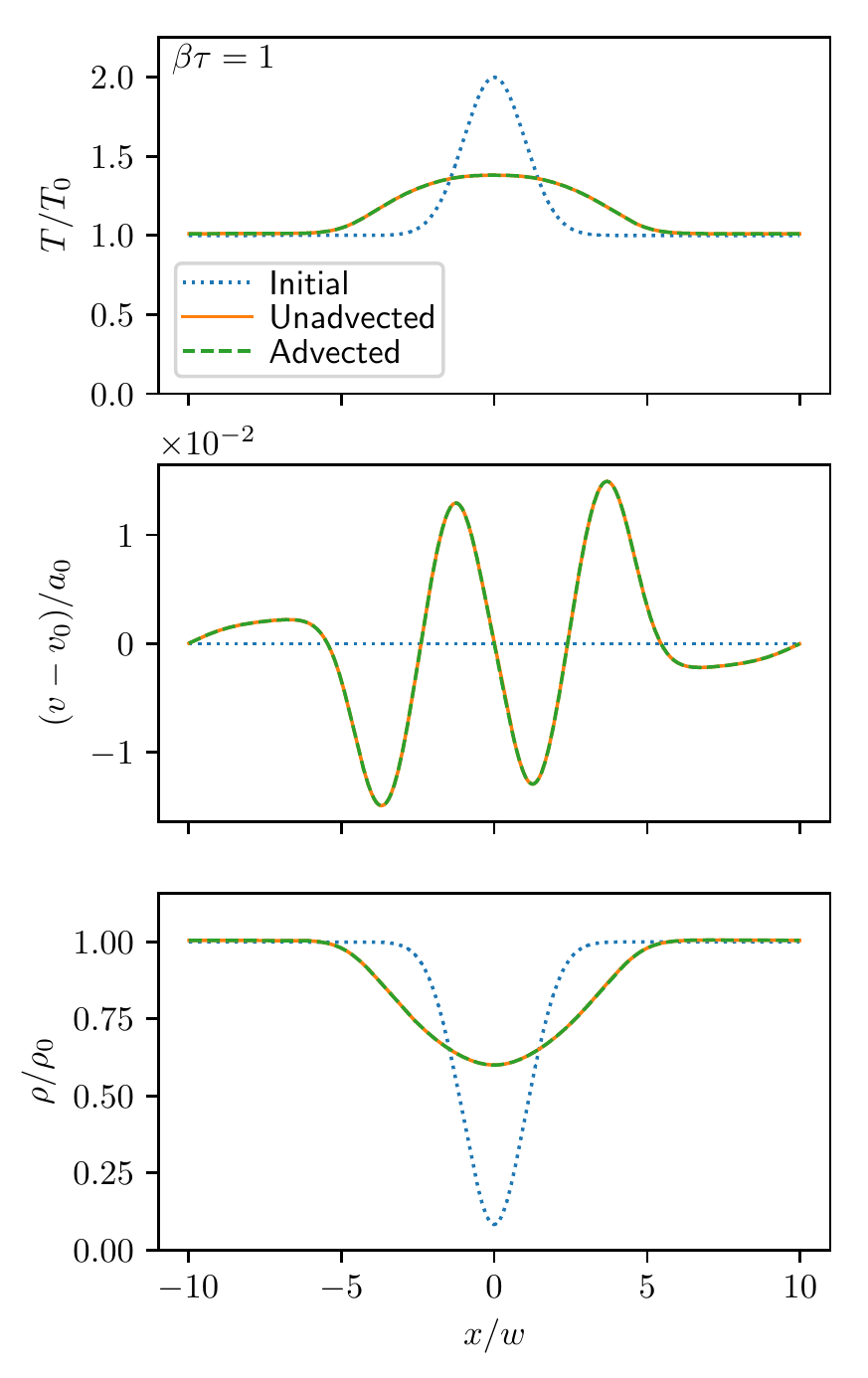}
\caption{Temperature (top), velocity (middle), and density (bottom)
for the multigroup pulse advection test with $\beta\tau=1$ in the
static diffusion regime.  Profiles at the initial time (dotted blue
line) and at time $t=2w/\beta c$ for the unadvected (solid orange
line) and advected (dashed green line) runs are shown.  In both runs,
the radiation pulse diffuses through the gas, and as pressure support
is lost, the density in the pulse increases.  Although the radiative
work and advection terms are very different between the unadvected and
advected runs, the relative error between them is so small that they
are visually indistinguishable (see
Figure~\ref{fig:pulse_advection_smallbt_relerr} for a plot of the
errors).}
\label{fig:pulse_advection_smallbt_duT}
\end{figure}
 
\begin{figure}
\includegraphics[width=\columnwidth]{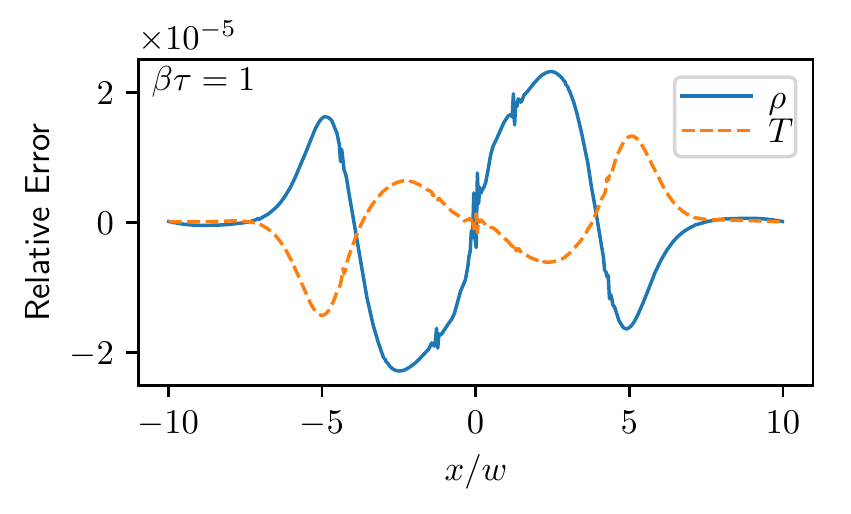}
\caption{Relative errors between the unadvected and advected runs for
the density (solid blue line) and temperature (dashed orange line) in
the multigroup pulse advection test in the static diffusion regime
with $\beta\tau=1$ (see Figure~\ref{fig:pulse_advection_smallbt_duT}
for their profiles).  The relative errors are bounded in absolute
value by $2.3 \times 10^{-5}$ and $1.4 \times 10^{-5}$ for the density
and temperature, respectively.  This indicates very good agreement
between these runs, despite the fact that the radiative work and
advection terms are very different from the vantage of the different frames.}
\label{fig:pulse_advection_smallbt_relerr}
\end{figure}
 
We use a computational grid of length $L=20w$ resolved over $512$ zones, 
with $8$ radiation groups logarithmically-spaced over the frequency range 
$[4\times 10^{18}, 4\times 10^{22}]\text{ Hz}$.  In Figure~\ref{fig:pulse_advection_smallbt_duT}, 
we show the resulting temperature, velocity, and density profiles for both the advected and 
unadvected runs with $\tau=100$ such that $\beta\tau=1$, putting the system is in the static 
diffusion regime.  Over the course of the run, the radiation has diffused through the gas, decreasing the  
temperature and thermal pressure support in the pulse, hence the density increases correspondingly.  
Since the advected and unadvected runs are visually indistinguishable, we plot the relative errors 
of the density and temperature in Figure~\ref{fig:pulse_advection_smallbt_relerr}, which are bounded 
by $2.3 \times 10^{-5}$ and $1.4 \times 10^{-5}$, respectively, in absolute value.  

\begin{figure}
\includegraphics[width=\columnwidth]{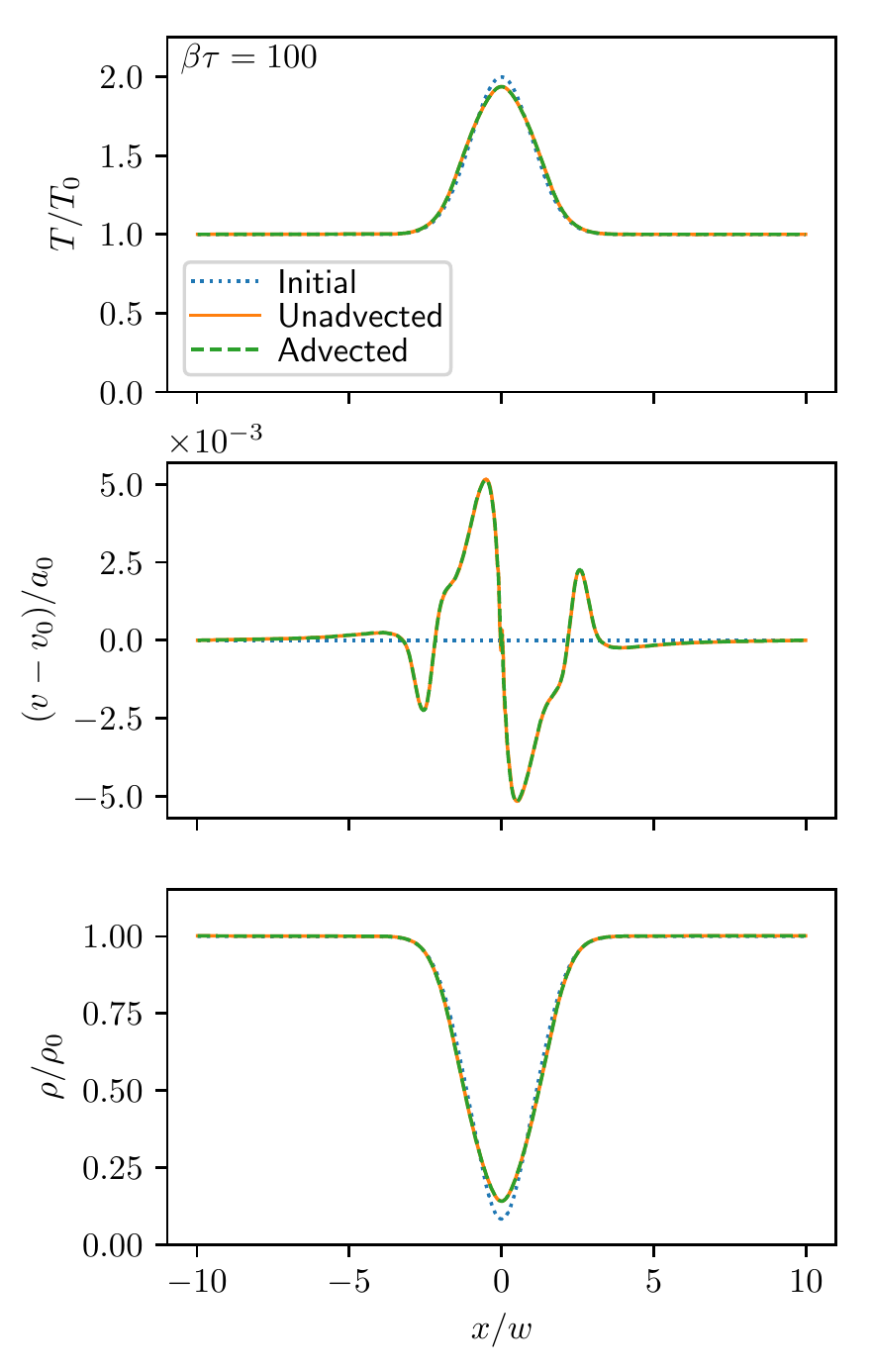}
\caption{Same as Figure~\ref{fig:pulse_advection_smallbt_duT} but for
$\beta\tau=100$ in the dynamic diffusion regime.  Here, the radiation
is trapped within the pulse and there is hardly any diffusion as it is
advected across the grid.  Again, the profiles for the unadvected
(solid orange line) and advected (dashed green line) runs are visually
indistinguishable (see Figure~\ref{fig:pulse_advection_bigbt_relerr}
for their relative errors).}
\label{fig:pulse_advection_bigbt_duT}
\end{figure}
 
\begin{figure}
\includegraphics[width=\columnwidth]{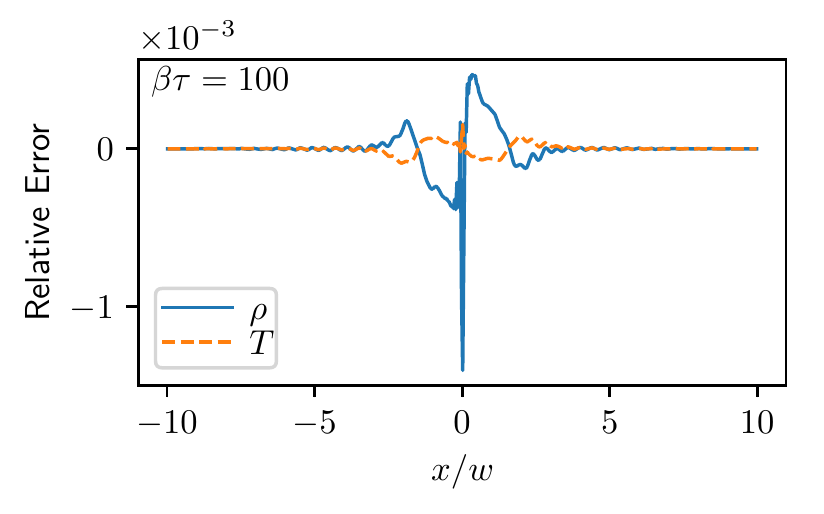}
\caption{Same as Figure~\ref{fig:pulse_advection_smallbt_relerr} but
in the dynamic diffusion regime with $\beta\tau=100$.  The relative
errors are bounded in absolute value by $1.4 \times 10^{-3}$ and $1.5
\times 10^{-4}$ for the density and temperature, respectively.  Again,
this demonstrates good agreement between the unadvected and advected
runs, although the relative sizes of the terms is very different in
each frame.}
\label{fig:pulse_advection_bigbt_relerr}
\end{figure}

Similarly, in Figure~\ref{fig:pulse_advection_bigbt_duT}, we plot the results for the case 
with $\tau=10^4$ such that $\beta\tau=100$, which puts the system in the dynamic diffusion regime.  
This time, there is hardly any diffusion as the pulse is advected, since the radiation is effectively 
trapped within the gas.  In Figure~\ref{fig:pulse_advection_bigbt_relerr}, we plot the relative 
errors of the density and temperature, which are bounded by $1.4 \times 10^{-3}$ and $1.5 \times 10^{-4}$, 
respectively, in absolute value.

Importantly, these results demonstrate the ability of \fornax{} to reproduce the correct behavior 
to a high degree of accuracy in both the static and dynamic diffusion regimes and for both 
advected and unadvected flows.  This represents a very sensitive test of the code's ability to 
model radiative trapping in the CCSN.  In the core-collapse supernova context, trapping of $\nu_e$s by infalling 
matter and their subsequent compression produces a spectrum of degenerate $\nu_e$s whose 
energies can reach the $\sim 100-300\text{ MeV}$ range while preserving a relatively 
high lepton fraction, which is a critically important aspect that allows the core to reach nuclear densities at bounce.

\subsection{Multi-group Gray-Opacity Non-Equilibrium Radiation Shock}
\label{rad_shock}

To test the \fornax{} code's ability to couple the multigroup radiation subsystem 
to the hydrodynamics in a regime where the fluid and radiation are out of equilibrium, 
we present here the results of the classical gray non-equilibrium radiation shock described by Zel'Dovich \& Raizer (1969).
Following that work, we adopt the Eddington approximation and set the radiation pressure 
tensor to $(\mathcal{P}_\epsilon)^i_j = (\mathcal{E}_\epsilon/3) \delta^i_j$, where $\delta^i_j$ 
is the Kronecker symbol.  The shock structure has a semi-analytic solution described by Lowrie \& Edwards (2008).  
Using their parameters with an upstream Mach number of $\mathcal{M}=3$, we obtain a subcritical 
radiation shock whose temperature jumps discontinuously at the shock interface.  We set the constant, gray 
absorption opacity to $\kappa = 577\text{ cm}^{-1}$, the mean particle mass to $\mu = m_\mathrm{H}$, 
and use an adiabatic EOS with $\gamma=5/3$.  We use a one-dimensional Cartesian domain with $x \in [-0.0132,\,0.00255]$  
resolved over $N_x = 512$ zones.  The problem is set in the rest frame of the shock, which we initialize 
at $x=0$.  In the upstream state ($x<0$), we set the gas temperature $T_0 = 2.18 \times 10^6 \text{ K}$, 
density $\rho_0 = 5.69 \text{ g cm}^{-3}$, and velocity $v_0 = 5.19 \times 10^7 \text{ cm s}^{-1}$, and 
using the Rankine-Hugoniot jump conditions to determine the downstream state ($x<0$), we set 
$T_1 = 7.98 \times 10^6 \text{ K}$, density $\rho_1 = 17.1 \text{ g cm}^{-3}$, and velocity $v_1 = 1.73 \times 10^7 \text{ cm s}^{-1}$.  
Similar to Vaytet et al. (2011), we use 8 radiation groups logarithmically spaced over the 
frequency range $\nu \in [10^{15}\text{ Hz}, \,10^{19}\text{ Hz}]$, initialize the group radiation energy 
densities using a Planck spectrum at the local gas temperature with zero radiation flux, and evolve for 
3 shock-crossing times to $t_\mathrm{final} = 9.08 \times 10^{-10}\text{ s}$.  Finally, since the 
structure of the steady-state shock solution is independent of the radiation propagation speed, 
$\hat{c}$, and since we must solve the semi-explicit radiation subsystem on a hydrodynamic time scale, 
we adopt a reduced speed of light $\hat{c} = 10 (v_0 + a_0)$, where $a_0 = 1.73 \times 10^7 \text{ cm s}^{-1}$ 
is the upstream sound speed.

\begin{figure}
\includegraphics[width=\columnwidth]{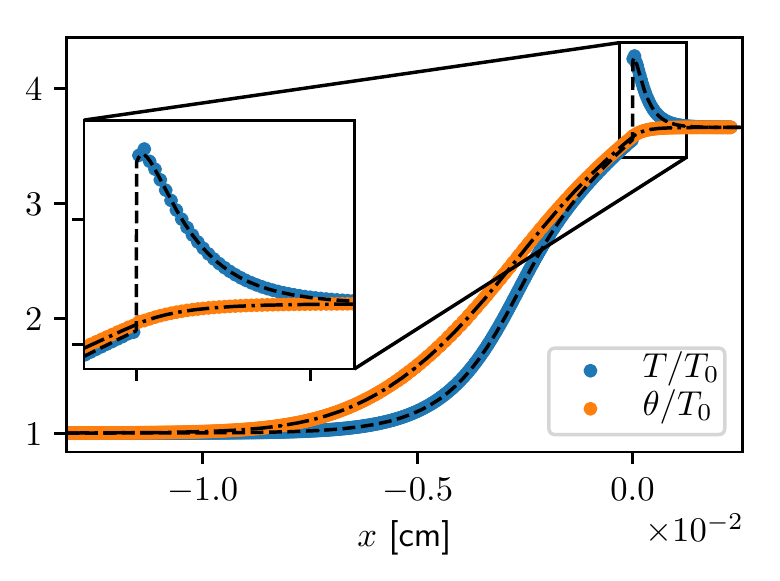}
\caption{Computed gas temperature (blue points) and radiation temperature (orange points)
for the multigroup gray radiation shock with their semi-analytic solutions (black lines)
overplotted and an inset showing the detail of the non-equilibrium Zel'Dovich spike region.}
\label{fig:radshock}
\end{figure}

Figure~\ref{fig:radshock} shows the gas temperature structure at time $t_\mathrm{final}$ with the 
semi-analytic solution from Lowrie \& Edwards (2008) overplotted and an inset showing the 
detail of the Zel'Dovich spike in the upstream temperature near the shock interface.  There 
is very good agreement between the two solutions, including in the non-equilibrium spike region 
and in the radiatively-heated shock precursor in the downstream state.  The relative error between 
the solutions is bounded by 0.8\% everywhere; the agreement in the remaining variables is 
similarly good.  Since this test simultaneously exercises the multi-group, velocity-dependent, Doppler-shift,
and matter-radiation coupling features of the \fornax{} code, it also simultaneously demonstrates the code's
ability to compute the most problematic aspects of full radiation-hydrodynamics accurately and consistently.

\section{Conclusion}
\label{conclusion}

In this paper, we have described the methods and implementation of the multi-group, 
multi-dimensional, radiation-hydrodynamic code \fornax{} and numerically 
exercised it with a variety of standard and non-standard simulation tests.  These included
tests of on- and off-center Sedov blast waves, the Sod shock tube, double Mach reflection, 
the 2-d and 3-d Rayleigh-Taylor and 2-d Kelvin-Helmholtz instabilities, the Liska-Wendroff implosion, 
pressureless dust collapse, energy conservation with gravity and a complicated EOS, 
radiation advection and Doppler ($v/c$) shifts in the context of strong velocity gradients, the 
handling of velocity- and frequency-dependent opacity, and non-equilibirum radiative shocks with 
the Zel'Dovich spike. We demonstrated that \fornax{} performs
well and accurately for all these tests, robustly handling non-trivial multi-dimensional 
hydrodynamic and radiation-hydrodynamic problems. To date, \fornax{} has been employed
to study core-collapse supernovae (Wallace et al. 2016; Radice et al. 2017; Burrows et al. 
2018; Vartanyan et al. 2018; Seadrow et al. 2018), for which extensions that include 
approximate general relativity and inelastic scattering processes are of relevance. 
The implementation of these effects is not address here, but can be found in Marek 
et al. (2006) (for the former) and Thompson, Burrows, \& Pinto (2003) and Burrows 
\& Thompson (2004) (for the latter).  In constructing \fornax{}, we endeavored to incorporate
best numerical and solution practices, with the result that the code is fast, scales well
on most modern HPC platforms, and has useful geometrical and grid flexibilities. Given this,
we anticipate its continued use to explore cutting-edge radiation-hydrodynamical
challenges of astrophysical import and plan to extend its reach to include magnetic fields
and neutrino oscillations in the years to come.

\acknowledgments

The authors acknowledge support under U.S. NSF Grant AST-1714267,
the Max-Planck/Princeton Center (MPPC) for Plasma Physics (NSF PHY-1144374),
and the U.S. Department of Energy, Office of Science, Office of Advanced
Scientific Computing Research, and the Scientific Discovery through
Advanced Computing (SciDAC4) program under Grant DE-SC0018297 (subaward 00009650).
DR acknowledges support as a Frank and Peggy Taplin Fellow at the Institute for Advanced Study.
The authors employed computational resources provided by the TIGRESS
high performance computer center at Princeton University, which is jointly supported by the Princeton
Institute for Computational Science and Engineering (PICSciE) and the Princeton University Office of
Information Technology, and by the National Energy Research Scientific Computing Center
(NERSC), which is supported by the Office of Science of the US Department of
Energy (DOE) under contract DE-AC03-76SF00098. The authors express their gratitude to
Ted Barnes of the DOE Office of Nuclear Physics for facilitating their use of NERSC.
This research is part of the Blue Waters sustained-petascale computing project, 
which is supported by the National Science Foundation (awards OCI-0725070 and ACI-1238993) 
and the state of Illinois. This work is also part of the ``Three-Dimensional Simulations of 
Core-Collapse Supernovae" PRAC allocation support by the National Science Foundation 
(under award \#OAC-1809073). Blue Waters is a joint effort of the University of Illinois 
at Urbana-Champaign and its National Center for Supercomputing Applications.
Under the local award \#TG-AST170045, this work used the resource Stampede2 in the 
Extreme Science and Engineering Discovery Environment (XSEDE), which is supported by 
National Science Foundation grant number ACI-1548562.
This work was performed under the auspices of the U.S. Department of Energy by 
Lawrence Livermore National Laboratory under Contract DE-AC52-07NA27344.  
This article has been assigned an LLNL document release number LLNL-JRNL-752999-DRAFT.  
In additoon, this work was performed in the auspices of the U.S. Department of Energy 
by Los Alamos National Laboratory under Contract DE-AC52-06NA25396.  This article has 
been assigned an LANL document release number LA-UR-18-25082.

\appendix

\section{General Spherical Metric}
\label{metric_tech}

We use the general spherical coordinates $(x^1,x^2,x^3)$, where the physical spherical 
coordinates are related via the mappings
\begin{subequations}  \label{eq:coords}
\begin{align}
r(x^1) &= r_t \sinh\left(\frac{|x^1|}{r_t}\right)\, , \\
\theta(x^2) &= \frac{\pi}{2} + C x^2 \left[1 + \frac{1}{1+\alpha} \left(\frac{x^2}{x_t}\right)^\alpha\right]\, , \\
\phi(x^3) &= x^3\, .
\end{align}
\end{subequations}
The invariant proper interval for regular spherical coordinates is given by
\begin{equation}
ds^2 = dr^2 + r^2 \,d\theta^2 + r^2 \sin^2\theta \,d\phi^2 \,.
\end{equation}
For general spherical coordinates, such as those defined by Equations~\eqref{eq:coords}, 
the invariant proper interval is given via the Chain Rule by
\begin{align}
ds^2 &= (\partial_1 r)^2 (dx^1)^2 \notag \\
& + \, r^2 (\partial_2 \theta)^2 (dx^2)^2 \notag \\
& + \, r^2 \sin^2\theta (dx^3)^2 \,.
\end{align}
Thus, with $ds^2 \equiv g_{\mu\nu} \,dx^\mu\,dx^\nu$, we identify the metric components as
\begin{subequations} \label{eq:metric}
\begin{align}
g_{11} &= (\partial_1 r)^2 \,, \\
g_{22} &= r^2 (\partial_2 \theta)^2 \,, \\
g_{33} &= r^2 \sin^2\theta \,.
\end{align}
\end{subequations}
Since the metric $g_{\mu\nu}$ is orthogonal (diagonal), the contravariant components 
are simply the inverses of their covariant counterparts, i.e., $g^{ii} = (g_{ii})^{-1}$ for $i=1,2,3$.  
The contravariant and covariant components of the metric can be used to raise and lower indices, respectively, i.e.,
\begin{align}
v^\mu &= g^{\mu\nu} v_\nu \,, \\
v_\nu &= g_{\mu\nu} v^\mu \,.
\end{align}
Finally, the determinant of the metric, $|g|$, is defined as
\begin{equation}
\sqrt{|g|} = r^2 (\partial_1 r) \sin \theta \,(\partial_2 \theta) \,. \label{eq:rootg}
\end{equation}

\section{Covariant Form of the Momentum Equation}

In Equation~\ref{eq:mom_covderiv}, the covariant derivative of the tensor is defined as
\begin{equation}
\tensor{T}{^\mu_{\nu;\sigma}} \equiv \tensor{T}{^\mu_{\nu,\sigma}} + \Gamma^\mu_{\sigma\lambda} \tensor{T}{^\lambda_\nu} - \Gamma^\lambda_{\sigma\nu} \tensor{T}{^\mu_\lambda} \,, \label{eq:covderiv}
\end{equation}
where
\begin{equation}
\Gamma^\lambda_{\mu\nu} \equiv \frac{1}{2} g^{\lambda\sigma} \left( g_{\nu\sigma,\mu} + g_{\sigma\mu,\nu} - g_{\mu\nu,\sigma} \right)  \label{eq:connection}
\end{equation}
is the connection coefficient (Christoffel symbol).  Note that connection coefficients are symmetric in their lower indices, i.e., $\Gamma^\lambda_{\mu\nu} = \Gamma^\lambda_{\nu\mu}$.  Note further that the Einstein summation convention \emph{does not apply} to connection coefficients.

From Equation~\eqref{eq:covderiv} and a well-known identity (Weinberg 1972, pp. 106-107),
\begin{equation}
\Gamma^i_{i\lambda} = \frac{1}{\sqrt{|g|}} \left( \sqrt{|g|} \right)_{,\lambda} \,,
\end{equation}
it follows that
\begin{align}
\tensor{T}{^i_{j;i}}
&= \frac{1}{\sqrt{|g|}} \left( \sqrt{|g|} \, \tensor{T}{^i_{j}} \right)_{,i} - \Gamma^\lambda_{ij} \tensor{T}{^i_\lambda} \,,
\end{align}
where we identify the first and second terms on the right-hand side as the ``flux term'' and ``geometric source term'', respectively.
Thus, Equation~\eqref{eq:mom_covderiv} can be rewritten in the equivalent form of Equation~\eqref{eq:gas_mom}.

\section{Connection Coefficients for Orthogonal Metrics}

Using the definition in Equation~\eqref{eq:connection}, it can be shown that the only non-zero connection coefficients are given for $i \ne j$ by
\begin{subequations} \label{eq:Gamma}
\begin{align}
\Gamma^i_{ii} 
&= \left(\ln \sqrt{|g_{ii}|}\right)_{,i}, \label{eq:Gamma:iii} \\
\Gamma^i_{ij} 
&= \left(\ln \sqrt{|g_{ii}|}\right)_{,j} \,, \label{eq:Gamma:iij} \\
\Gamma^i_{jj} 
&= - \frac{1}{2} \frac{g_{jj,i}}{g_{ii}} \,. \label{eq:Gammai:jj}
\end{align}
\end{subequations}

\section{Connection Coefficients for the General Spherical Metric} \label{conn_gen_sph}

Using Equations~\eqref{eq:metric} in Equations~\eqref{eq:Gamma}, we derive here the connection coefficients for the general spherical metric and their volume averages, defined as
\begin{equation}
\langle \bullet \rangle \equiv \frac{1}{\Delta V} \iiint \bullet \,\sqrt{|g|} \,dx^1 \,dx^2 \,dx^3 \,,
\end{equation}
where $\sqrt{|g|}$ is as given in Equation~\eqref{eq:rootg} and 
\begin{equation}
\Delta V \equiv \iiint \sqrt{|g|} \,dx^1 \,dx^2 \,dx^3
\end{equation}
is a given control volume.  The coefficients are of only three types:  $\Gamma^i_{ii}$, $\Gamma^i_{ij}$, and $\Gamma^i_{jj}$.

The coefficients of Type $\Gamma^i_{ii}$ are given by
\begin{subequations}  \label{eq:Gamma_iii}
\begin{align}
\langle \Gamma^1_{11} \rangle &= \Bigg\langle \frac{(\partial^2_1 r)}{\partial_1 r} \Bigg\rangle \,,\notag \\
&= \frac{1}{\Delta (r^3/3)} \int (\partial^2_1 r) \,r^2 \,dx^1 \,, \\
\langle \Gamma^2_{22} \rangle &= \Bigg\langle \frac{(\partial^2_2 \theta)}{\partial_2 \theta} \Bigg\rangle \,,\notag \\
&= \frac{1}{\Delta (-\cos\theta)} \int (\partial^2_2 \theta) \,\sin\theta \,dx^2 \,, \\
\langle \Gamma^3_{33} \rangle &= 0 \,.
\end{align}
\end{subequations}
The coefficients of Type $\Gamma^i_{ij}$ are given by
\begin{subequations}  \label{eq:Gamma_iij}
\begin{align}
\langle \Gamma^1_{12} \rangle &= 0 \,. \\
\langle \Gamma^1_{13} \rangle &= 0 \,. \\
\langle \Gamma^2_{21} \rangle &= \Bigg\langle \frac{(\partial_1 r)}{r} \Bigg\rangle \,,\notag \\
&= \frac{1}{\Delta (r^3/3)} \int (\partial_1 r)^2 \,r \,dx^1 \,. \\
\langle \Gamma^2_{23} \rangle &= 0 \,. \\
\langle \Gamma^3_{31} \rangle &= \Bigg\langle \frac{(\partial_1 r)}{r} \Bigg\rangle \,,\notag \\
&= \langle \Gamma^2_{21} \rangle \,, \\
\langle \Gamma^3_{32} \rangle &= \Bigg\langle \cot \theta \,(\partial_2 \theta) \Bigg\rangle \,,\notag \\
&= \frac{1}{\Delta(-\cos\theta)} \int \cos\theta \,(\partial_2 \theta)^2 \,dx^2 \,.
\end{align}
\end{subequations}
The coefficients of Type $\Gamma^i_{jj}$ are given by
\begin{subequations}  \label{eq:Gamma_ijj}
\begin{align}
\langle \Gamma^1_{22} \rangle &= \Bigg\langle -\frac{r (\partial_2 \theta)^2}{(\partial_1 r)} \Bigg\rangle \,,\notag \\
&= -\frac{1}{\Delta(r^3/3)\,\Delta(-\cos\theta)} \notag \\
&\times \int r^3 \,dx^1 \int \sin\theta \,(\partial_2 \theta)^3 \,dx^2 \,, \\
\langle \Gamma^1_{33} \rangle &= \Bigg\langle -\frac{r \,\sin^2 \theta}{(\partial_1 r)} \Bigg\rangle \,,\notag \\
&= -\frac{1}{\Delta(r^3/3)\,\Delta(-\cos\theta)} \notag \\
&\times \int r^3 \,dx^1 \int \sin^3\theta \,(\partial_2 \theta) \,dx^2 \,, \\
\langle \Gamma^2_{11} \rangle &= 0 \,, \\
\langle \Gamma^2_{33} \rangle &= \Bigg\langle -\frac{\sin\theta \,\cos\theta}{(\partial_2 \theta)} \Bigg\rangle \,,\notag \\
&= -\frac{1}{\Delta(-\cos\theta)} \int \sin^2 \theta \,\cos\theta \,dx^2 \,, \\
\langle \Gamma^3_{11} \rangle &= 0 \,, \\
\langle \Gamma^3_{22} \rangle &= 0 \,.
\end{align}
\end{subequations}

In practice, we use numerical (Romberg) integration to approximate only the volume-averaged 
connection coefficients $\langle \Gamma^1_{11} \rangle$, $\langle \Gamma^2_{22} \rangle$, 
$\langle \Gamma^2_{21} \rangle$, $\langle \Gamma^3_{32} \rangle$, and $\langle \Gamma^1_{22} \rangle$.  
Since $\Gamma^3_{31} = \Gamma^2_{21}$, we can simply set $\langle \Gamma^3_{31} \rangle = \langle \Gamma^2_{21} \rangle$, and by symmetry of the connection coefficients in their lower coordinates, 
we can set $\langle \Gamma^2_{12} \rangle = \langle \Gamma^2_{21} \rangle$ and $\langle \Gamma^3_{23} \rangle = \langle \Gamma^3_{32} \rangle$.

Next, in order to get exact conservation of $\phi$-angular momentum, we wish to have 
the geometric source term for the $x^3$-momentum equation vanish, i.e., $\Gamma^\lambda_{i3} T^i_\lambda = 0$.  Expanding in repeated indices and substituting \mbox{Equations~\eqref{eq:Gamma_iii}, \eqref{eq:Gamma_iij}, and~\eqref{eq:Gamma_ijj},} it follows that
\begin{equation}
\left( \Gamma^3_{13} \,g_{33} + \Gamma^1_{33} \, g_{11} \right) v^1 + \left( \Gamma^3_{23} \, g_{33} + \Gamma^2_{33} \, g_{22} \right) v^2 = 0 \,.  \label{eq:angmom}
\end{equation}
Since $v^1$ and $v^2$ in Equation~\eqref{eq:angmom} are completely arbitrary, it must be that
\begin{equation}
\Gamma^3_{13} \,g_{33} + \Gamma^1_{33} \,g_{11} = \Gamma^3_{23} \,g_{33} + \Gamma^2_{33} \,g_{22} = 0 \,.
\end{equation}
Thus, having calculated $\langle g_{11} \rangle$, $\langle g_{22} \rangle$, $\langle g_{33} \rangle$, $\langle \Gamma^3_{13} \rangle$, and $\langle \Gamma^3_{23} \rangle$, we can set
\begin{subequations}
\begin{align}
\langle \Gamma^1_{33} \rangle &= -\frac{\langle g_{33} \rangle}{\langle g_{11} \rangle} \,\langle \Gamma^3_{13} \rangle \,, \\
\langle \Gamma^2_{33} \rangle &= -\frac{\langle g_{33} \rangle}{\langle g_{22} \rangle} \,\langle \Gamma^3_{23} \rangle \,,
\end{align}
\end{subequations}
to ensure numerical conservation of $\phi$-angular momentum to machine precision.

Finally, to ensure that $\partial_2 P = 0$ numerically when it ought to be, we require that the flux term 
and the geometric source term in the 2-momentum equation cancel identically in a finite-volume sense.  
Assuming $v_j =0$, only $\tensor{T}{^i_i}=P$ and $\Gamma^i_{i2}$ survive.  It follows that
\begin{equation}
\frac{1}{\sqrt{|g|}} \left( \sqrt{|g|} \,\tensor{T}{^2_2} \right)_{,2} = \Gamma^2_{22} \tensor{T}{^2_2} + \Gamma^3_{32} \tensor{T}{^3_3} \,.  \label{eq:grad2P}
\end{equation}
Taking the volume average of each side of equation~\eqref{eq:grad2P}, it follows that
\begin{equation}
\langle\Gamma^2_{22}\rangle = \frac{\Delta A_2}{\Delta V} - \langle\Gamma^3_{32}\rangle \,,
\end{equation}
where $A_2 \equiv \int \sqrt{|g|} \,dx^1 dx^3$ and $\Delta A_2$ denotes the area difference 
at upper and lower 2-faces.

\section{Some Strong Scaling Test Results}
\label{scaling}

\begin{figure}
\includegraphics[width=\columnwidth]{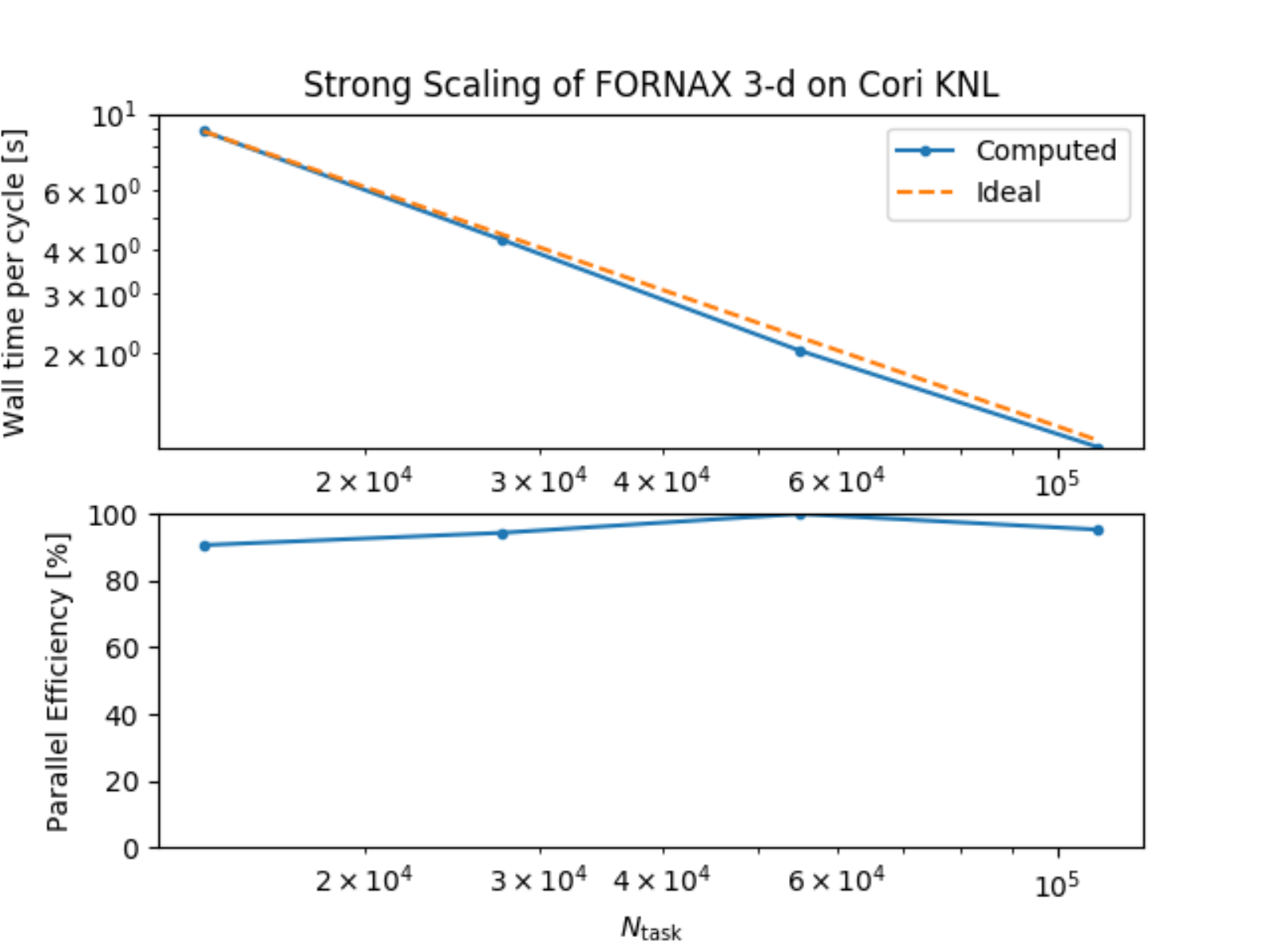}
\caption{
This figure demonstrates our code's raw per-cycle timing (top) and excellent strong scaling
efficiency (bottom) out to 110k MPI tasks.  Due to the greedy algorithm used to load-balance the
decomposition of our dendritic mesh, efficiency can improve with increasing task count as
demonstrated here.}
\label{fig:fornax_scaling1}
\end{figure}
 
\begin{figure}
\includegraphics[width=\columnwidth]{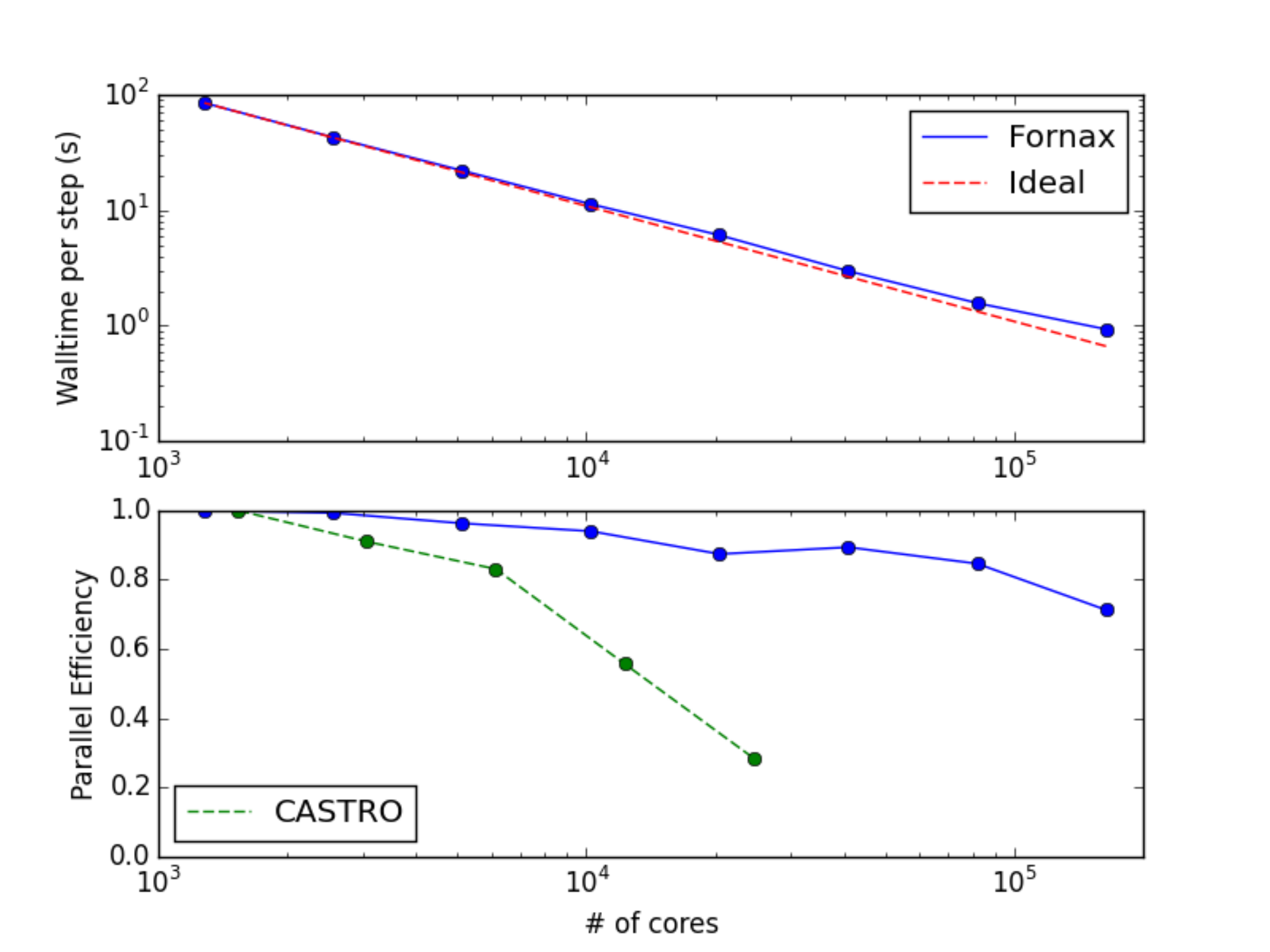}
\caption{
{\bf Top:} Wall clock time per time step (in seconds) versus core count
for 3-d radiation-hydrodynamic runs.  This is a {\bf strong} scaling test on Blue Waters. Red-dashed is the ideal expectation, and blue is the realization using \fornax{}.  {\bf Bottom:}  The corresponding parallel efficiency
versus core count for \fornax{} (blue) and CASTRO (green).  Note that beyond ${\sim}10,000$ cores,
the efficiency of CASTRO plummets, while that of \fornax{} maintains high values.  At ${\sim}160,000$
cores the efficiency of \fornax{} is still ${\sim}75$\%. Our efficiencies have improved slightly
since this test on Blue Waters was performed (see Fig.~\ref{fig:fornax_scaling1}).
Perfect scaling would be a flat curve.}
\label{fig:fornax_scaling2}
\end{figure}

To benchmark the parallel performance of \fornax{} under production-run 
conditions, we ran a full radiation-hydrodynamic core-collapse 3-d simulations 
with 20 energy groups, 12th-order multipole gravity with GR corrections
to the monopole component, and a resolution of $608 \times 256 \times 512$ (radial, poloidal, toroidal) for
10 cycles with an increasing number of pure MPI tasks out to 110k tasks.  The results shown
in Figure~\ref{fig:fornax_scaling1} demonstrate that the code runs extremely fast on NERSC/Cori II,
with excellent strong scaling efficiency over 90\%.  In fact, because of the way
we use a greedy algorithm to load-balance the decomposition of our dendritic mesh,
the efficiency peaks at larger task count.

Figure~\ref{fig:fornax_scaling2} demonstrates strong scaling results for \fornax{} on
the Cray/XE6 on Blue Waters for full-star radiation-hydrodynamic simulations
using 20 energy groups per neutrino species and $1000 \times 256 \times 512$
spatial gridding in spherical coordinates.  We see excellent scaling
results to ${\sim}130,000$ cores.  The lowest efficiency
achieved was for ${\sim}130,000$ cores and was ${\sim}75$\%.  A comparison was
made to the corresponding results for our previous supernova code CASTRO,
that requires an iterative transport solution and Krylov subspace methods.
After ${\sim}30,000$ cores, \fornax{} is five times more efficient than CASTRO, and after
${\sim}100,000$ cores \fornax{} is approximately ten times as efficient.
Moreover, the wallclock and CPU-hour per time step comparisons have revealed that
\fornax{} is also ten times more favorable by these metrics than CASTRO.
This comparision is not meant as a criticism of CASTRO, which has many state-of-the-art
features and capabilities (Zhang et al. 2011,2013).  It merely highlights the
differences between an implicit and an explicit treatment of radiative transport
in the context of current computer architectures and parallelism modalities.


\end{document}